\documentclass[12pt]{article}
\usepackage[centertags]{amsmath}
\usepackage{amssymb}
\usepackage{graphicx}
\usepackage{epsfig}
\usepackage{ulem}
\usepackage[english]{babel}
\usepackage{array}
\usepackage{esint}
\usepackage{amsthm}
\usepackage{latexsym}
\usepackage{mathrsfs}
\usepackage{amsbsy}
\usepackage[mathcal]{euscript}
\pdfoutput=1
\usepackage{epsfig}
 \usepackage{jheppub}
\usepackage{mathdots}
\usepackage{MnSymbol}
\usepackage{multirow}

\newcommand{\nn}{\nonumber}
\newcommand{\be}{\begin{equation}}
\newcommand{\ee}{\end{equation}}
\newcommand{\bea}{\begin{eqnarray}}
\newcommand{\eea}{\end{eqnarray}}

\def\Tr{{\rm Tr }}
\def\hat{\widehat}

\def\cS{{\mathcal S}}

\def\cE{{\mathcal E}}

\def\cZ{{\mathcal Z}}

\def\tilde{\widetilde}

\newcommand{\e}{\mathrm{e}}

\newcommand{\fG}{\mathfrak{G}}

\newcommand{\fZ}{\mathfrak{Z}}
\newcommand{\fR}{\mathfrak{R}}

\title{
Gas of baby universes in JT gravity and  matrix models}

\author{Irina Aref'eva,  Igor Volovich}

\affiliation{Steklov Mathematical Institute, Russian Academy of Sciences,\\Gubkina str. 8, 119991, Moscow, Russia}
\emailAdd{arefeva@mi-ras.ru}
\emailAdd{volovich@mi-ras.ru}
\abstract{It has been shown recently by Saad, Shenker and Stanford that  the genus expansion of a certain matrix integral generates  partition functions of Jackiw-Teitelboim (JT) quantum gravity on  Riemann surfaces of arbitrary
genus with any fixed number of boundaries. 
 We use an extension of this integral for  studying  gas of  baby 
universes or wormholes in JT gravity. To investigate the gas  nonperturbatively  we explore the generating functional of baby universes in the  matrix model.    The simple particular case when the matrix integral includes  the exponential potential  is discussed in some detail.
We argue that there is a phase transition in the gas of 
baby universes.
}

\begin{document}
\maketitle
%\tableofcontents
%\usepackage{mathrsfs}
%\usepackage{amsbsy}

\section{Introduction}

It has been shown  by Saad, Shenker and Stanford \cite{SSS} that  the genus expansion of a certain matrix integral generates the partition functions of Jackiw-Teitelboim (JT),
\cite{Jackiw:1984je,Teitelboim:1983ux}, quantum gravity on  Riemann surfaces of arbitrary
genus with an arbitrary fixed number of boundaries. It is shown in \cite{SSS} that  an important part of JT quantum gravity is reduced to computation of the Weil-Petersson volumes of the moduli space  of hyperbolic Riemann surfaces with various genus and number of boundaries for which Mirzakhani \cite{Mir} established  recursion relations.  Eynard and Orantin \cite{eynard2007invariants,EO} proved that Mirzakhani's relations are a special case of random matrix recursion relations with the spectral curve $y=\sin (2\pi z)/4\pi$. 
This is a natural extension  of   results on  topological gravity \cite{Witten-2gr,MK,ManinZograf,DW}. Relation of random matrices and gravity, including black hole description, has a long history, see \cite{Cotler:2016fpe,Saad:2018bqo} and refs therein.

 The results of \cite{SSS}  provide a nonperturbative approach to JT quantum gravity on Riemann surfaces of various genus and perturbative description of boundaries. We use an extension of this result for nonperturbative studying of gas of  baby universes
 in JT gravity. To investigate the boundaries nonperturbatively  we explore the generating functional of boundaries in the  matrix model and in JT gravity.   One interprets the generating functional as the partition function of gas of baby universes in grand canonical ensemble in JT multiverse with the source function  describing the distribution of  boundaries  being treated as the chemical potential. The interaction  is presented  by splitting and joining of  baby universes\footnote{One can compare this picture with  string interactions and using this analogy closed strings describe  baby universes without boundaries, meanwhile the baby universes with boundary correspond to open strings.  An analogue of matrix theory is given by string field theory \cite{BookSFT,W-SFT,AV-MatrixSFT}. As has been noted in  \cite{SSS} there is an essential  difference in coupling constant in SFT and JT.}.

Let $Z^{grav}_{g,n}(\beta_1,...,\beta_n) $ be the JT gravity path integral for Riemann surface of   genus $g\geq 2$  with $n$ boundaries with lengths $\beta_1,...,\beta_n$.
Consider a generating function for these functions
\be
\cZ_{n}^{grav}(\beta_1,...,\beta_n;\gamma)\simeq \sum_{g=0}^\infty \gamma^{2g+n-2}
\cZ_{g,n}^{grav}(\beta_1,...,\beta_n)
\ee
 where $\gamma$ is a constant which in notations of \cite{SSS} is $\gamma=e^{-S_0}$.

The following remarkable  relation between correlation functions in  matrix model and JT gravity holds \cite{SSS}:
\be
\cZ_{n}^{matrix,d.s.}(\beta_1,...,\beta_n;\gamma)\simeq Z^{grav}_{n}(\beta_1,...,\beta_n;\gamma) 
\ee
Here $\cZ_{n}^{matrix,d.s.}(\beta_1,...,\beta_n;\gamma)$ is the double scaling (d.s.) limit of the correlation function in a matrix model with the spectral curve
mentioned above. This form of the curve was obtained in \cite{SSS} by computing the JT path integral for the disc. 

In this note we consider the generating functional for the gravitational correlation functions $Z^{grav}_{n}(\beta_1,...,\beta_n;\gamma)$
\be
{\cal Z}^{grav}(J;\gamma)=\sum_{n=0}^\infty \frac{1}{n!}\int_0^\infty d\beta_1... \int_0^\infty d\beta_n Z^{grav}_{n}(\beta_1,...,\beta_n;\gamma) J(\beta_1)...J(\beta_n)\\
\ee
where $J(\beta)$ is a source function. An appropriate generating functional in matrix theory has the form
\bea
\fZ^{matrix}(J)=<\e ^{N\int \!Z(\beta)J(\beta)d\beta}>
\\
=\sum _{n=0}^\infty \frac{N^n}{n!}\int\!\!d\beta_1...
\int\!\!d\beta_n \cZ_n^{matrix}(\beta_1,...\beta_n)J(\beta_1)...J(\beta_n)\nn\\
\label{mgen-Z}\eea
Here $Z(\beta)=\Tr e^{-\beta M}$ where $M$ is a random $N\times N$ Hermitian matrix. This amounts to shifting the potential
in the matrix model $V(x)\to V(x)-\tilde J(x)$ where $\tilde J(x)$ is the Laplace transform of $ J(\beta)$, see Sect.\ref{Sect:GFMM}.
 
We define the generating functional for connected correlation functions 
\be
\fG^{matrix}(J)=-\frac{1}{N^2}\log \fZ^{matrix}(J)
\ee
 take the double scaling limit introducing the parameter $\gamma$ and obtain the relation between JT gravity and the matrix model in terms of the generating functionals: \be
d.s. \lim \fG^{matrix}(J)\simeq \fZ^{\text{grav}}(J;\gamma)\label{SSS-dual}
\ee
The "$\simeq$" symbol indicates  the equality in the sense of formal series.

The paper is organized as follows. In Sect.\ref{Sect:GFMM} the generating functional in matrix theory $\fZ^{matrix}(J)$ is discussed.
Here $J$ is the source function. In Sect.\ref{Sect:JT} the generating functional of boundaries in JT gravity $\fZ^{grav}(J)$ is considered.  In Sect.4 we investigate the double scaling limit in matrix models with a particular choice of the source $J(\beta)$
which leads to the change of the potential $V(x)\to V(x)-Je^{\omega x}$.  In Sect.5 the matrix model with the exponential potential is investigated. In Sect.6 the matrix model with the spectral curve $y=\sin (2\pi z)/4\pi$
and the source is discussed and phase transition is observed. In Sect.7 the discussion of obtained results is presented.

\section {Generating functional in matrix models}\label{Sect:GFMM}

{\bf Generating functional.} We consider ensemble of $N\times N$ Hermitian matrices \cite{Wigner,Dyson,BIPZ,mehta} with potential $V(M)$. 
Let
\be
Z(\beta)=\Tr e^{-\beta M}=\sum _{i=1}^N\exp(-\beta\lambda _i),\,\,\,\beta >0.
\ee
 where $\lambda_i$ are eigenvalues of the matrix $M$.  The $n$-point correlation function of $Z(\beta)$
in the matrix model is given by 
\bea
&\,&\,\,\,\,\,\,\,\,\,\,\,\,\,\,\cZ^{matrix}_{n}(\beta_1,...,\beta_n)\equiv<Z(\beta_1)...Z(\beta_n)>\nn\\
&=&\int Z(\beta_1)...Z(\beta_n) \exp(-N\Tr V(M))dM/ \int  \exp(-N\Tr V(M))dM\label{Znc}\eea
Its generating functional can be presented as 
\bea
\fZ^{matrix}(J)=<\e ^{N\int \!Z(\beta)J(\beta)d\beta}>=\sum _{n=0}^\infty \frac{N^n}{n!}\int\!\!d\beta_1...
\int\!\!d\beta_n \cZ_n^{matrix}(\beta_1,...\beta_n)J(\beta_1)...J(\beta_n)\nn\\
\label{gen-Z}\eea
%$\mbox{\Large{$ \cZ $}}^{matrix}(J)\,\,\,\,$
or
\be
\fZ^{matrix}(J)=\int \exp\{N \sum _{j=1}^N\tilde J(\lambda _j)\}
 \,d\mu_N(\lambda_1,...\lambda _N)
\ee
where
\bea
d\mu_N(\lambda_1,...\lambda _N) &=&\frac{1}{Z_N}\prod_{j>k}(\lambda_j-\lambda_k)^2\prod_{j=1}^N e^{-NV(\lambda_j)},\\
\tilde J(\lambda )&=&\int d\beta J(\beta)\e^{-\beta \lambda }d\beta
\eea
This amounts to shift the potential
$V(x)\to V(x)-\tilde J(x)$.

One expands $\fG(J)=\log \fZ^{matrix}(J)$ to get the connected correlation functions
\bea
 \fG^{matrix}(J)=\sum _{n=0}^\infty \frac{1}{n!}\int\!\!d\beta_1...
\int\!\!d\beta_n \cZ_{n,conn}^{matrix}(\beta_1,...\beta_n)J(\beta_1)...J(\beta_n)\nn\\
\label{genconn-Z}\eea

{\bf Particular case.}
We consider a special case
\be
J(\beta)=-J\delta(\beta+\omega)\ee
In this case the consideration of $\fZ^{matrix}(J)$ is equivalent to dealing with 
the matrix model with a deformed potential
\be
U(x)=V(x)+J\e^{\omega x}\ee
In this case the singular integral equation defining the eigenvalues distribution 
has the form\footnote{See \cite{georgia,Gakhov} for  the theory of singular integral equations.}
\be
 U'(\lambda)=2\fint _{\cS}d\mu \frac{\rho(\mu)}{\lambda-\mu }\label{V'},\ee
and the spectral density is
\be
\rho(\mu)=\lim_{N\to \infty}\frac1N <\sum _{i=1}^N\delta (\mu-\lambda_i) \exp\{-NJ \sum _{j=1}^N e^{\omega\lambda _j}\}>\ee
{\bf Double scaling limit.}\footnote{Double scaling limit in matrix models has been introduced  in \cite{doublescaling}, see \cite{ DFGZ,MM,Ey5Lectures} for review and refs therein} All the correlation functions $\cZ^{matrix}_{n}(\beta_1,...,\beta_n)$ in principle could be derived if the potential $V(x)$ or the spectral density/spectral curve $\rho (\mu)$ is known.
To get a connection of the matrix model with JT gravity one has to go  to the double scaling limit, see \cite{SSS}. Consider a matrix model with a non-normalized spectral density
\be\label{nnsd}
\rho^{\text{nnorm}}(E) = \frac{e^{S_0}}{(2\pi)^2}\sinh\left(2\pi\sqrt{\frac{a^2-E^2}{2a}}\right), \hspace{20pt} -a<E<a.
\ee
where $S_0$ is a constant. Now, shifting $E\to E-a$ and sending $a\to\infty$ we get the spectral density of the double-scaled matrix model
\be\label{dsl}
\rho_{d.s.}^{\text{nnorm}}(E) = \frac{e^{S_0}}{(2\pi)^2}\sinh(2\pi\sqrt{E}), \,\,\,\,\,\,\,\,\,\, E>0.
\ee
and the correlation functions in the double scaled limit
\be
d.s. \lim \cZ_{n,conn}^{matrix}(\beta_1,...\beta_n)=\cZ_{n}^{matrix,d.s.}(\beta_1,...,\beta_n)
\ee
The limiting correlation functions $\cZ_{n}^{matrix,d.s.}(\beta_1,...,\beta_n)$ have an expansion of the form
\be
\cZ_{n}^{matrix,d.s.}(\beta_1,...,\beta_n)\simeq \sum_{g=0}^\infty (e^{-S_0})^{2g+n-2}
\cZ_{g,n}^{matrix,d.s.}(\beta_1,...,\beta_n)
\ee
The double scaling limit of the generating functional is
\bea
d.s.\lim \log\fZ^{matrix}(J)=\sum _{n=0}^\infty \frac{1}{n!}\int\!\!d\beta_1...
\int\!\!d\beta_n \cZ_{n}^{matrix,\,d.s.}(\beta_1,...\beta_n)J(\beta_1)...J(\beta_n)\nn\\
\label{genconn-Z}\eea

The correlation functions $\cZ_{g,n}^{matrix,\,d.s.}(\beta_1,...,\beta_n)$ and the constant $S_0$ will be used in the next section to describe the 
connection of the matrix model with JT gravity. The double scaling limit will be discussed also in Sect.4. 

{\bf Resolvents.} Similarly one has generating functional for correlation functions  of resolvents
\be
 \fR^{(matrix)}(f)=<\e^{N\int R(z)f(z)dz}>\ee
 where $f(z)$ is a test function and
 \be
R(z) =\Tr (z-M)^{-1}.\ee
Expanding $\log \fR^{(matrix)}(f)$ in the series on $f$ one gets connected correlation functions $<R(z_1)...R(z_n)>_c$. We perform the double scaling limit
\be
d.s.\lim <R(z_1)...R(z_n)>_c=<R(z_1)...R(z_n)>_c^{d.s.}\ee
which admits an expansion
\be
<R(z_1)...R(z_n)>_c^{d.s.}\simeq \sum_{g=0}^\infty \frac{1}{(e^{S_0})^{2g+n-2}}R_{g,n}(z_1,...,z_n) \ee
One defines the correlation functions
\be\label{wng}
W_{g,n}^{matrix}(z_1,...,z_n) = (-1)^n2^n \,z_1...z_n\,R_{g,n}(-z_1^2,...,-z_n^2) 
\ee
which satisfy the loop equations \cite{migdal,AJM,eynard} and will be used in the next section, and the generating functional
\be
{\cal W}_{g}(f)=\sum_{n=0}^\infty \frac{1}{n!}\int dz_1... \int dz_n W_{g,n}^{matrix}(z_1,...,z_n) f(z_1)...f(z_n).
\ee

\section {Generating functional in JT gravity}\label{Sect:JT}
 The Euclidean action of JT gravity \cite{Jackiw:1984je,Teitelboim:1983ux,Almheiri:2014cka} has the form
\be\label{JTa}
I_{JT} = -\frac{S_0}{2\pi}\left[\frac{1}{2}\int_{\mathcal{M}}\sqrt{g}R + \int_{\partial\mathcal{M}}\sqrt{h}K\right] -\left[ \frac{1}{2}\int_{\mathcal{M}}\sqrt{g}\phi(R+2) +\int_{\partial\mathcal{M}}\sqrt{h}\phi (K-1)\right].
\ee
Here $g_{\mu\nu}$ is a metric on a two dimensional manifold $\mathcal{M} $, $\phi$ is a scalar field (dilaton) and the constant $S_0$
was mentioned in the previous section. The path integral for Riemann surface of   genus $g\geq 2$  with $n$ boundaries with lengths $\beta_1,...,\beta_n$ reads 
\be
Z^{grav}_{g,n}(\beta_1,...,\beta_n) =
 e^{-S_0 \chi } \int \frac{\mathcal{D}g_{\mu\nu}\mathcal{D}\phi}{\text{Vol}(\text{diff})}\, e^{-{\hat I}_{JT}[g_{\mu\nu},\phi]}\ee\\
 where $\chi$ is the Euler characteristic $\chi = 2-2g-n$ and ${\hat I}_{JT}$ is the JT action with the first $S_0$ term left out.

The following   relation between the matrix model and JT gravity holds \cite{SSS}:
\be
\cZ_{g,n}^{matrix,d.s.}(\beta_1,...,\beta_n)=Z^{grav}_{g,n}(\beta_1,...,\beta_n) 
\ee
or
\be
\cZ_{n}^{matrix,d.s.}(\beta_1,...,\beta_n)\simeq Z^{grav}_{n}(\beta_1,...,\beta_n) 
\ee
where
\be
\cZ_{n}^{grav}(\beta_1,...,\beta_n)\simeq \sum_{g=0}^\infty (e^{-S_0})^{2g+n-2}
\cZ_{g,n}^{grav}(\beta_1,...,\beta_n)
\ee

It was found in \cite{SSS}  that the  partition function  has the form $(g\geq 2)$
\be
Z^{grav}_{g,n}(\beta_1,...,\beta_n) = \int_0^\infty b_1 db...\int_0^\infty b_n db_n  V_{g,n}(b_1,...,b_n) Z_{\text{Sch}}^\text{trumpet}(\beta_1,b_1)...Z_{\text{Sch}}^\text{trumpet}(\beta_n,b_n)
\ee
where $V_{g,n}(b_1,...,b_n)$ is the Weil-Petersson volume of the moduli space of a genus $g$ Riemann surface with $n$ geodesic boundaries of lengths $b_1,...,b_n$ and
\begin{align}
Z_{\text{Sch}}^{\text{trumpet}}(\beta,b) = \frac{1}{2\pi^{1/2}\beta^{1/2}}e^{-\frac{b^2}{4\beta}}.\label{Ztrum}
\end{align}
From this we get
\bea
{\cal Z}^{grav}_g(J)=\sum_{n=0}^\infty \frac{1}{n!}(e^{-S_0})^{2g+n-2}\int_0^\infty d\beta_1... \int_0^\infty d\beta_n Z^{grav}_{g,n}(\beta_1,...,\beta_n) J(\beta_1)...J(\beta_n)\\
=\sum_{n=0}^\infty \frac{1}{n!}(e^{-S_0})^{2g+n-2}\int_0^\infty b_1 db...\int_0^\infty b_n db_n  V_{g,n}(b_1,...,b_n) {\hat J}(b_1)...{\hat J}(b_n),
\eea
where
\be
{\hat J}(b)=\int_0^\infty d\beta\frac{1}{2\pi^{1/2}\beta^{1/2}}e^{-\frac{b^2}{4\beta} }J(\beta)
\ee
and the generating functional
\be
\fZ^{grav}(J) \simeq \sum_{g=0}^\infty      {\cal Z}^{grav}_g(J)
\ee
Finally, one obtains the  relation
\be
d.s. \lim\log \fZ^{matrix}(J) \simeq \fZ^{\text{grav}}(J)\label{SSS-dual}
\ee
Similarly, the correlation functions $W_{g,n}^{matrix}$ are related with volumes of the moduli spaces as
\be
W_{g,n}^{matrix}(z_1,...,z_n) = \int_0^\infty b_1 db_1 e^{-b_1 z_1}...\int_0^\infty b_n db_n e^{-b_nz_n}V_{g,n}(b_1,...,b_n).
\ee
The generating functional is 
\bea
&\,&{\cal Z}_{R,g}(f)=\sum_{n=0}^\infty \frac{1}{n!}\int dz_1... \int dz_n W_{g,n}^{matrix}(z_1,...,z_n) f(z_1)...f(z_n)\nn\\
&=&\sum_{n=0}^\infty \frac{1}{n!}  \int_0^\infty b_1 db_1 ...\int_0^\infty b_n db_n V_{g,n}(b_1,...,b_n) \tilde f(b_1)...\tilde f(b_n)\nn
\eea
where
\be
\tilde f(b)=\int dz e^{-b z}f(z).\ee

\begin{figure}[t!]
\begin{center}
\includegraphics[width=0.3\textwidth]{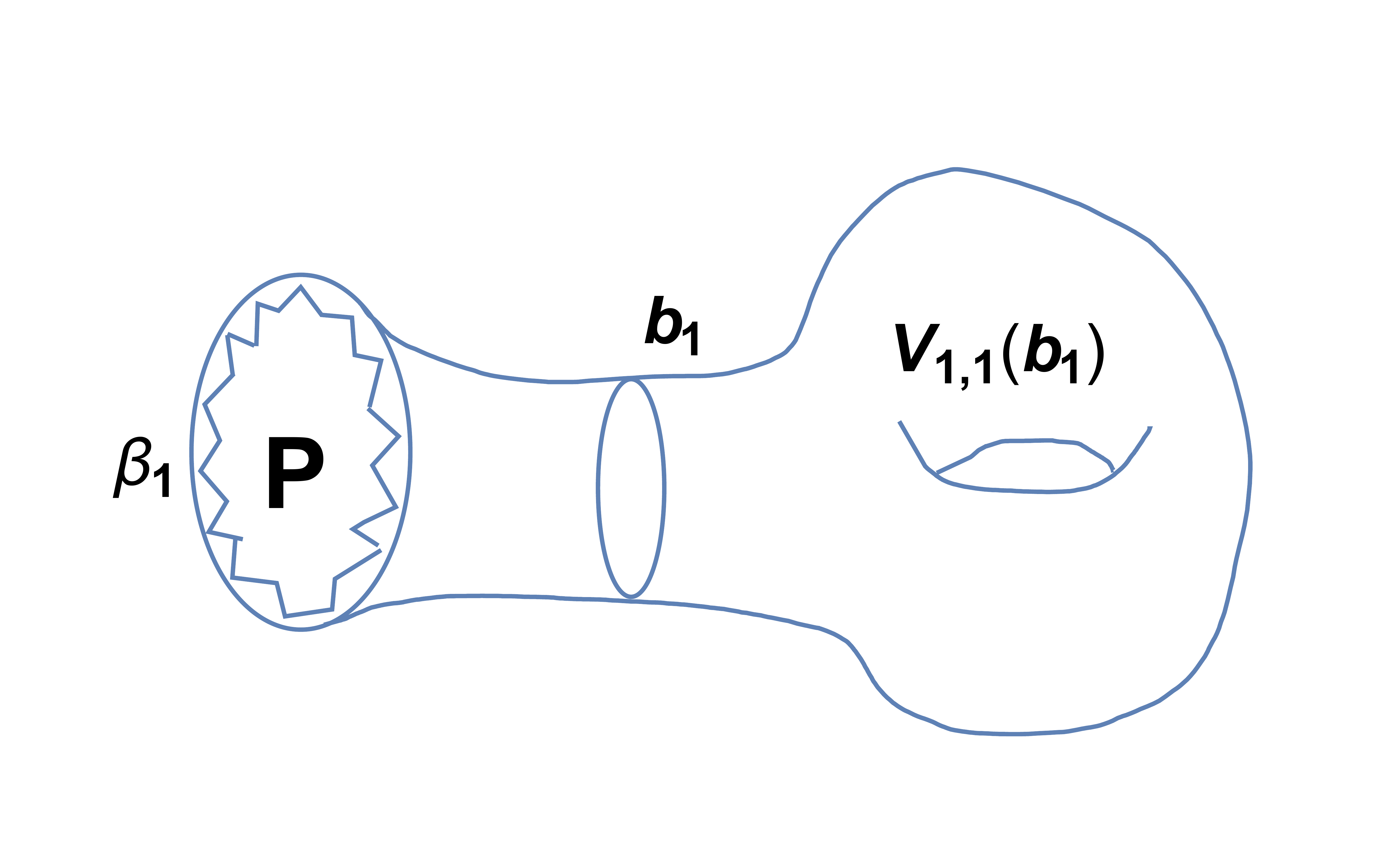}
\includegraphics[width=0.3\textwidth]{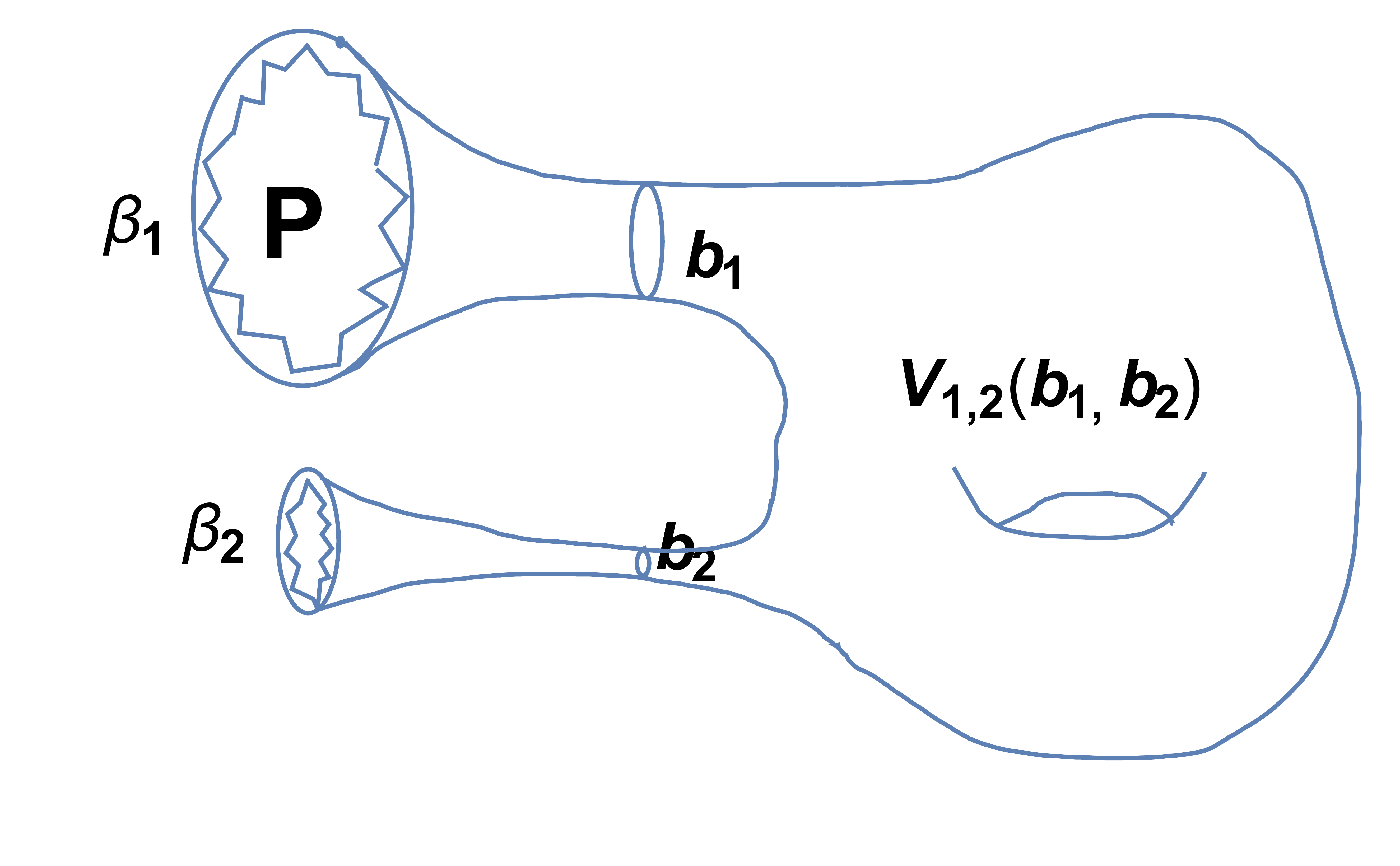}
\includegraphics[width=0.3\textwidth]{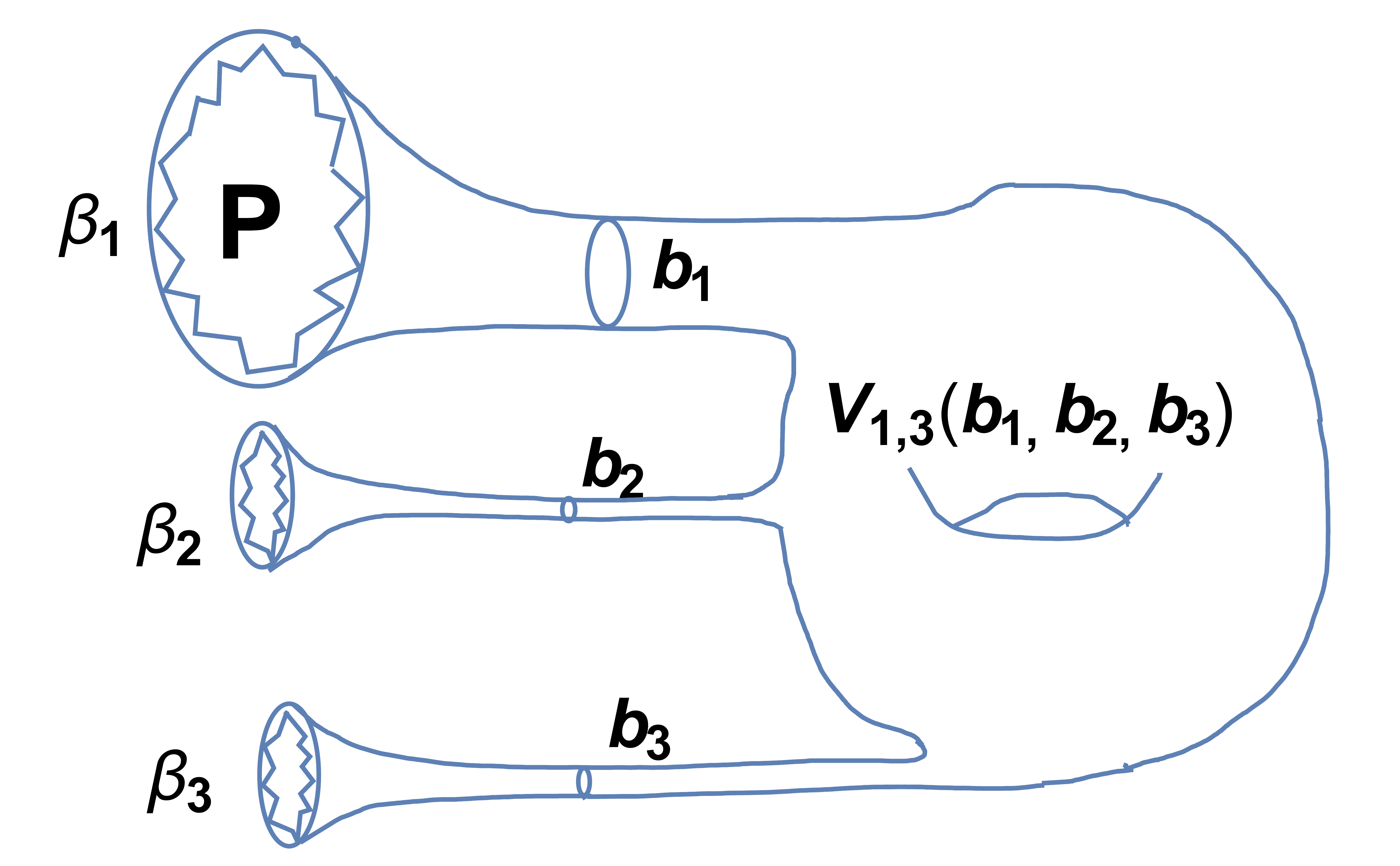}\,+\,...
(a) $\qquad \qquad \qquad \qquad$(b)$\qquad \qquad\qquad  \qquad$ (c)
\caption {Gas of baby universes. Here $|\beta_1|>>|\beta_i|$ for  $i\geq 2$ and $b_i\leq b_c$ for  $i\geq 2$}\label{Fig:P}
%{\bf Math. 2P.nb}
\end{center}
\end{figure}

\begin{figure}[h!]
\begin{center}
\includegraphics[width=0.33\textwidth]{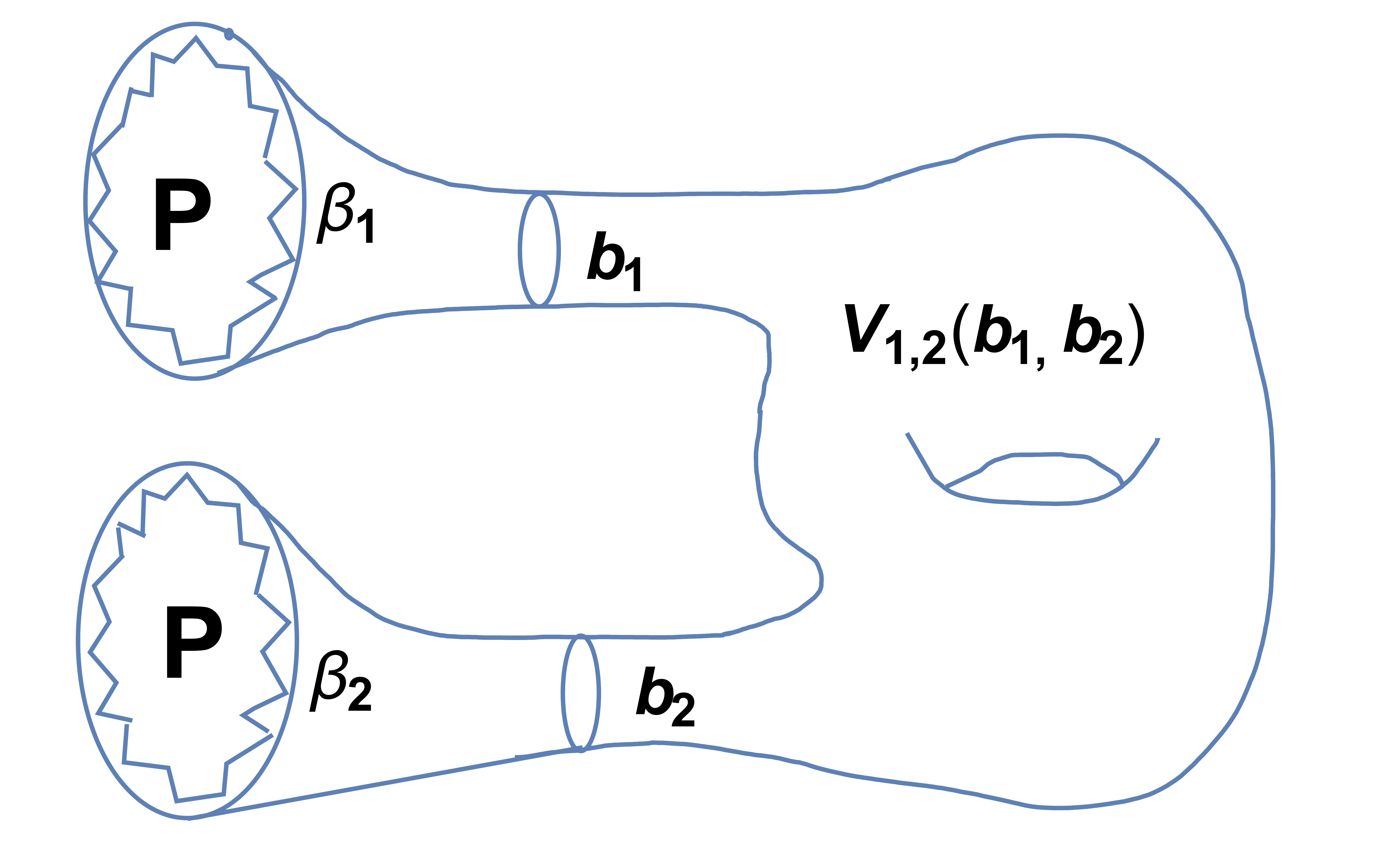}\\
\caption {Two parents connected by the wormhole }\label{Fig:WH}
%{\bf Math. 2P.nb}
\end{center}
\end{figure}

{\bf Baby universes.}  In cosmology \cite{HL,LRT,GS,Coleman,IV,1807.00824}, one usually deals with baby universes that  branch off from, 
or join onto, the parent(s) Universe(s).
 In matrix theories one parent  is a connected Riemann surface with arbitrary number of handles and at least one boundary.  We  assume that the lengths of boundaries of baby universities are small as compare with the   boundary length of the parent, see Fig.\ref{Fig:P}.   Baby universes are attached to the parent  by necks  that have restricted lengths of geodesics at which the neck is attached to the parent, 
 We assume that the lengths of the boundaries of baby universes are small compared to the length of the boundary of the parent, see Fig. \ref {Fig:P}. Baby universes are attached to the parent with the help of thin necks. Thickness of the neck is defined as the  geodesic length
 of the loop located at the thinness point of the neck, and this length is assumed to be  essentially smaller than the length of theboundary of the parent, see Fig.\ref{Fig:P}.b and Fig.\ref{Fig:P}.c. There are also restrictions on the  area of the surface of baby universes, see \cite{Mathur,Renata} for more precise definitions. Cosmological baby universes   in the parent-baby universe
approximation interact only via coupling to the parent universes, that themselves  interact via wormholes.   In matrix models the baby universes always interact via 
their  parents too and  parents  interact via wormholes, Fig.\ref{Fig:WH}. 
 One can expect that at large number of baby universes  interaction between different parts of the system increases  and this leads to phase transition (an analog of the the nucleation of a baby universe in \cite{GS}).
We interpret the matrix partition function $\fZ$, defined by equation \eqref{genconn-Z} as a partition function of the gas of baby universes.

\newpage
\section{Double scaling limit}
\subsection{Double scaling limit for the GUE }
There are various notions of the double scaling limit in matrix theory \cite{doublescaling,DS,GM,DFGZ,MM,Ey5Lectures,bz1,BB,BI,Pastur,Widom,KM,DKMVZ}. A special double scaling limit was  considered in \cite{SSS} at the level of spectral density. Here we discuss it at the level of the potentials. 
We will see that the linear term in the potential plays a special role.

Let us start with the Wigner distribution for the Gaussian Unitary Ensemble (GUE) \cite{Wigner,Dyson,BIPZ,mehta}. The ordinary Wigner distribution $2\sqrt{a^2-\lambda ^2}/\pi a^2$ is supported on the interval $[-a,a]$
and is obtained from a matrix model with the  potential $V(x)=m^2 x^2/2$, where $m^2=\textcolor{red}{4}/a^2$. We want to make a shift and get a distribution on the interval $[0,\Lambda]$, $\Lambda>0.$ To this end we consider  the gaussian model with an external source  \footnote{Note that  the Gaussian matrix model with an arbitrary matrix source has been studied in \cite{BH}. Here we consider the case  corresponding in notations \cite{BH}  to $A=j I$, $I$ is the $N\times N$ 
unit matrix}
\be
V(x)=m^2\left(\frac 12 \,  x^2+j   x\right)\ee
and we make parameters $m$ and $j$ depending on $\Lambda$ to put the measure support on $[0,\Lambda]$.
The singular integral equation  defining the density  takes the form
\be
m^2(\lambda+j)=2\fint _{\,_{[0,\Lambda]}}d\lambda' \frac {\rho_{W}(\lambda')}{\lambda-\lambda'}\label{main5m}\ee
Here $\fint$ means the Cauchy principal value of the integral. The solution of \eqref{main5m} is given by the shifted Wigner distribution
\bea
\rho_{W}(\lambda)&=&\frac{8}{\pi\Lambda^2}S_{\Lambda}(\lambda),\,\,\,\,\,\,S_{\Lambda}(\lambda)=\sqrt{\lambda(\Lambda-\lambda)}\label{SW}
\eea
Note that the constant $j$ from the linear term in the potential does not enter into the expression for the spectral density. This is valid for any potential. The constant appears only throughout    the normalization and consistency conditions that in this case read 
\bea\label{NN-L-mm}
m^2\Lambda^2=16,\,\,\,\,\,\,\,\,
\Lambda=-2j\eea
We get that the eigenvalue density \eqref{SW} is supported by  the potential 
\be
V_\Lambda(x)=\frac{4}{\Lambda ^2}(\frac12  x^2 -\Lambda  x)\ee
in the sense that
\bea
\rho_{W}(\lambda)&=&\lim _{N\to\infty}\,\rho^{V_\Lambda}_{N}(\lambda)
\eea
where
\be\rho^{V_\Lambda}_{N}(x)=\frac{1}{N}<\Tr \delta(x-M)>_{\!_{V_\Lambda}}\label{DV}\ee
  $<f>_{\!_{V}}$ in \eqref{DV} means averaging with potential 
$
<f>_{\!_{V}}=\int \!\!f e^{-N\Tr V}dM/\int\! \! e^{-N\Tr V}dM.
$

Now we take the limit $\Lambda\to\infty$ and write
\be
\lim _{\Lambda\to\infty}\,\Lambda ^\eta\,\rho_{W}(\lambda)=\frac{8}{\pi} \sqrt{\lambda},\,\,\,\,\lambda\in [0,\Lambda], \ee
where  $\eta=3/2$. So, we get an expected result in two steps. First we send $N\to\infty$ and then $\Lambda\to\infty$.

One can use also another procedure. Set $\Lambda =t\,N$ and in this case one has
\be
\lim _{N\to\infty}\,(N) ^{3/2}\,\rho^{V_{tN}}_{N}(\lambda)=\frac{8 }{ t^{3/2}\pi} \sqrt{\lambda}\
\ee
where
\be V_{tN}(x)=\frac{2}{t^2 N ^2 } x^2 -\frac{4}{t N} x
\ee

\subsection{Double scaling limit for the cubic interaction}

Let us consider the matrix model with the potential
\be
V(x)=m^2(\frac 12 x^2+j x+\frac 13 g x^3)\ee
and we will make parameters $m,j,g$ depending on $\Lambda$ to put the support of the spectral density on $[0,\Lambda]$.
Again first we take the limit $N\to\infty$ and then the limit $\Lambda\to\infty$. In the large $N$ limit one gets the singular integral equation that defines the spectral density 
\be
m^2(\lambda+j+g\lambda^2)=2\fint _{[0,\Lambda]}d\lambda' \frac {\rho(\lambda')}{\lambda-\lambda'}\label{main5}\ee
We write solution in the form
\bea
\rho(\lambda)&=&\frac{m^2\,S_{\Lambda}(\lambda)}{2\pi ^2}\int _0^\Lambda\,\frac{1+g (\lambda+\lambda')}{S_{\Lambda}(\lambda')}\,d\lambda',\eea
$
S_{\Lambda}(\lambda)$ is given  by \eqref{SW}, and we get 
\bea
\rho(\lambda)&=&
\frac{m^2\,S_{\Lambda}(\lambda)}{2\pi ^2}\Big(1+g \lambda +g\frac{\Lambda}{2}\Big).\label{rho-3-Lambda}
\eea
The normalization condition is achieved by a suitable choice of $ m ^ 2 $,
\bea
m^2\Lambda^2\Big(1+g \Lambda \Big)&=&16\label{NN-L-m}
\eea 
and the consistency  condition by  the suitable choice of j\bea
4\Lambda+g3\Lambda^2+8j&=&0\label{EC-3}.\eea
The solution of these equations is evidently
\bea
m^2=16/\Lambda^2\Big(1+g \Lambda \Big),\,\,j=-(4\Lambda+g3\Lambda^2)/8
\eea
which should be substituted into the potential.
Now we take the large $\Lambda$ limit and get
\be
\lim_{\Lambda\to\infty}\Lambda^{3/2}\rho (\lambda)=\frac{4}{\pi^2}\sqrt{\lambda}.
\ee

More reach picture appears for the quartic interaction with negative mass square. In this case for suitable choice of parameters there is a double cut solution \cite{AIM,CCM}, universality of the double scaling limit in this case has been proved in \cite{BI}, see also \cite{CM} for more general multi-cut solutions.

\newpage
\section{Deformation by an exponential potential}
 %[[{\bf File: Airy-test, density-for-exp-pot-test-IA.nb}]]
 \subsection{Exponential potential}
\begin{figure}[h!]
\begin{center}
\includegraphics[width=0.3\textwidth]{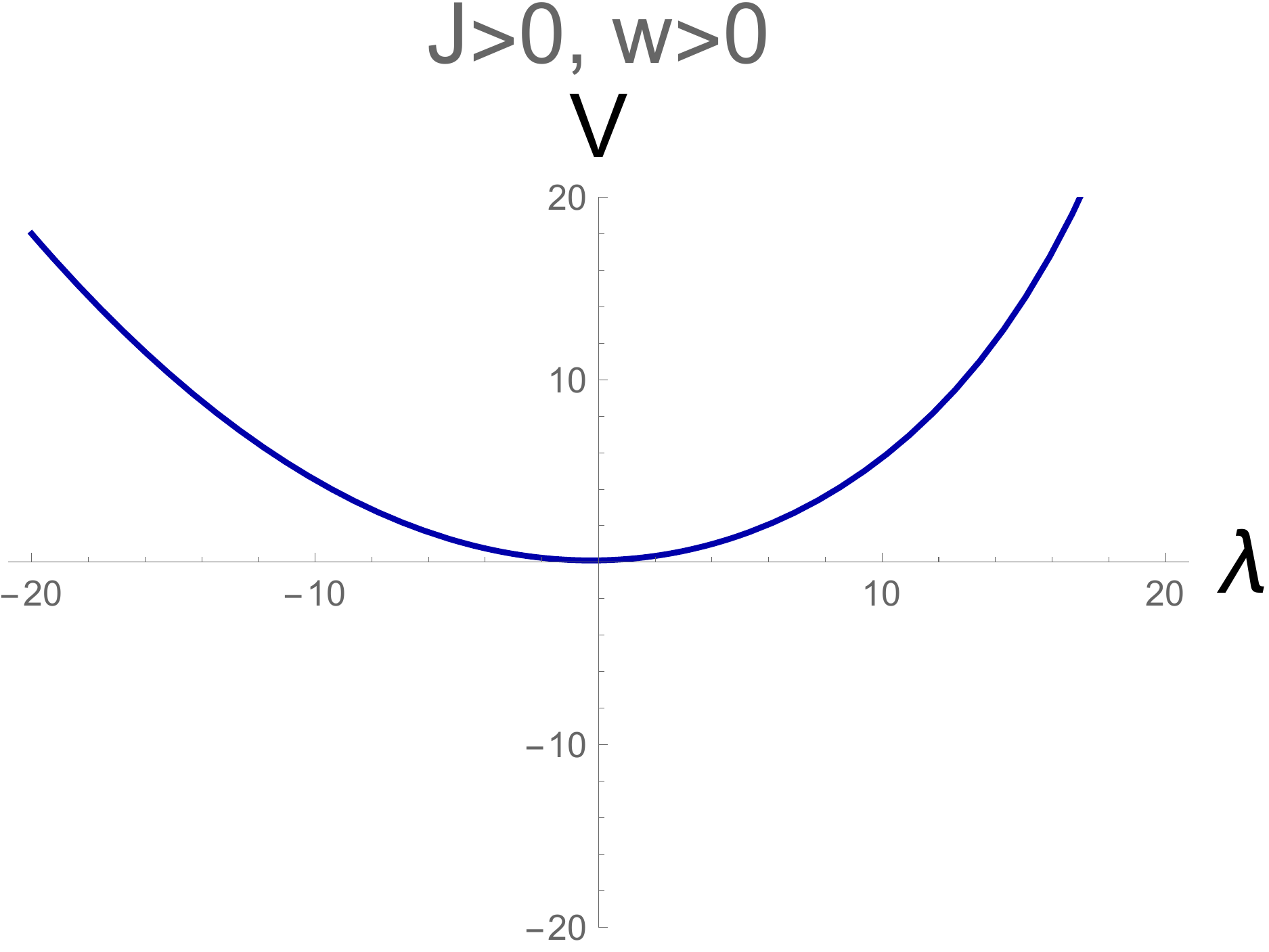}$\,\,\,\,\,\,$$\,\,\,\,\,\,$
\includegraphics[width=0.3\textwidth]{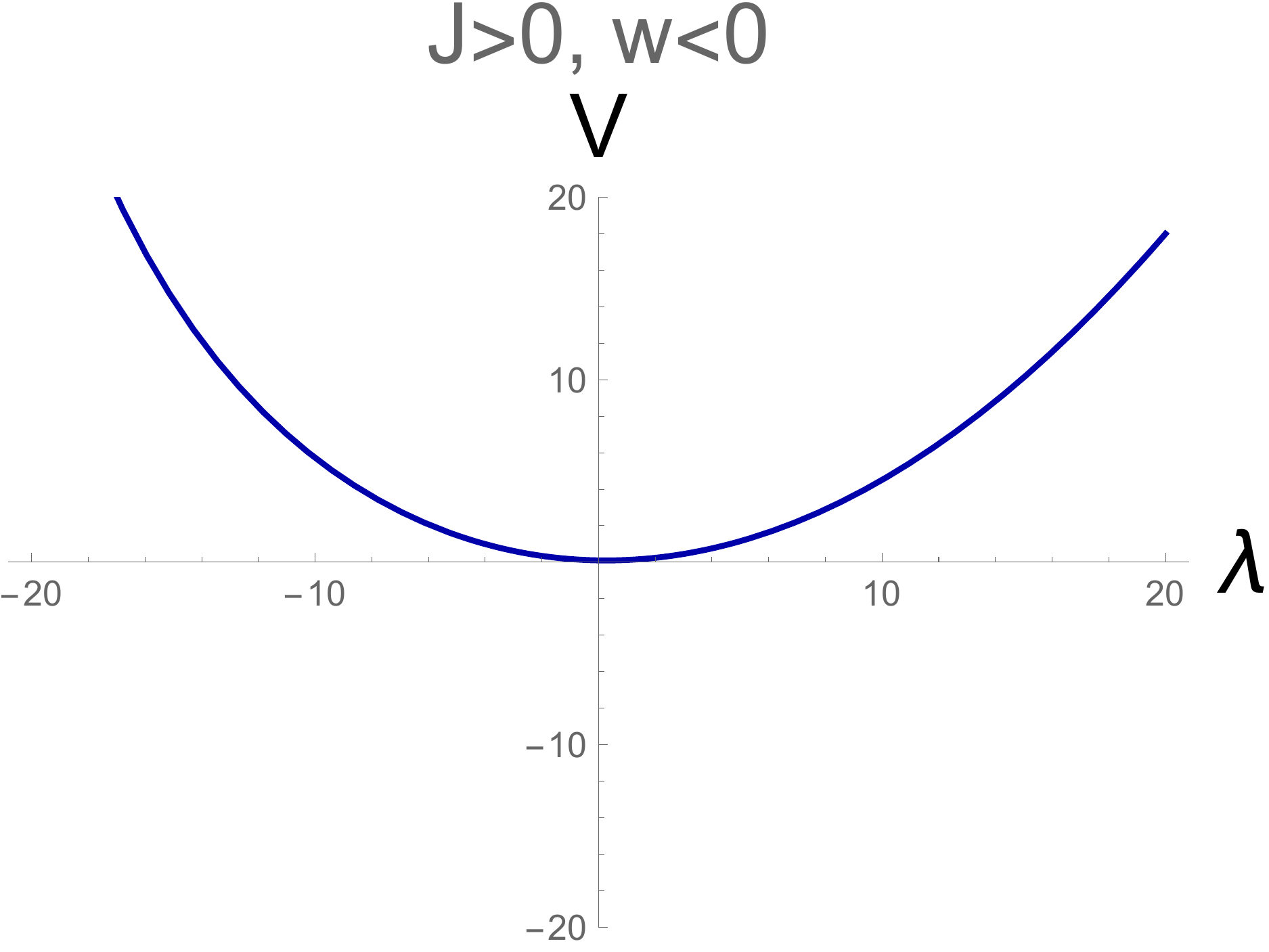}\\
(a) $\qquad \qquad \qquad \qquad\qquad \qquad\qquad  \qquad$ (b)\\ $\,$\\
\includegraphics[width=0.3\textwidth]{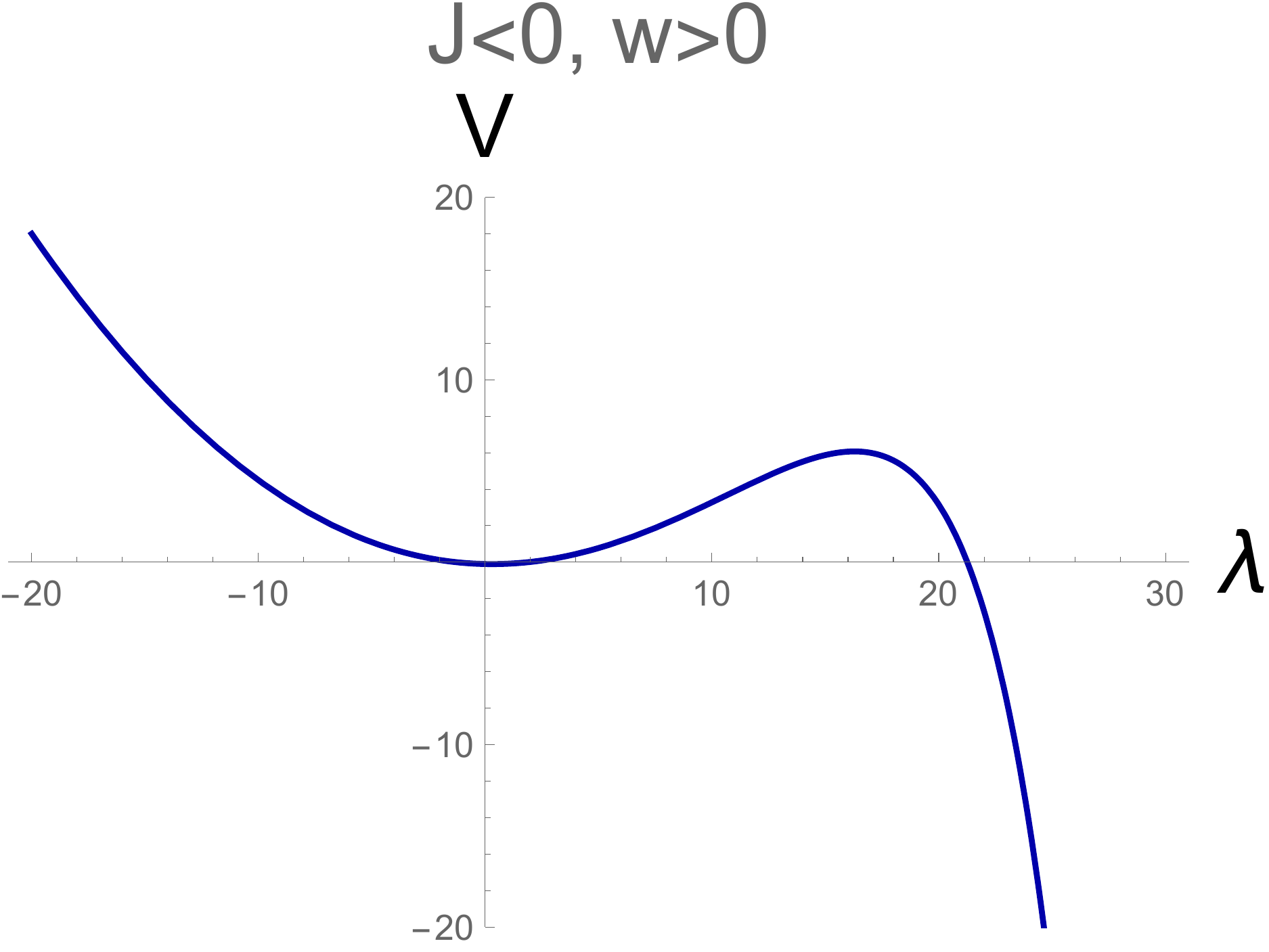}
\includegraphics[width=0.3\textwidth]{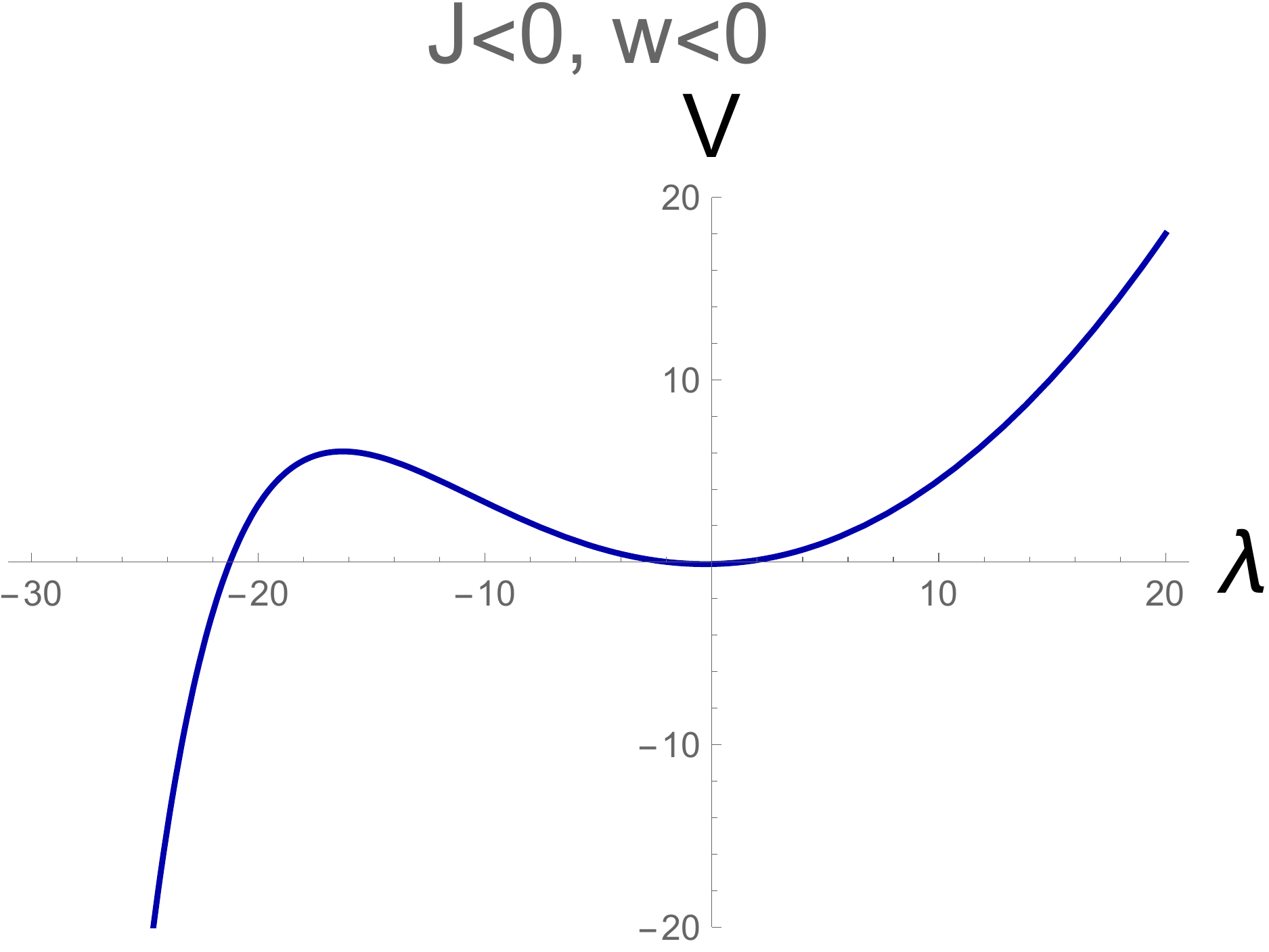}\\
(c) $\qquad \qquad \qquad \qquad\qquad \qquad\qquad  \qquad$ (d)\\ $\,$\\
\caption {Four different choices of perturbations of the gaussian ensemble by  the  exponential potential   $ V_{1}=Je^{\omega x}$.  }\label{Fig:exp-4}
%{\bf File: Deform-Gauss-model}
\end{center}
\end{figure}

 Here we  consider a
  deformation of the Wigner distribution by the insertion of the exponential potential\footnote{
It is interesting to compare this model with the  model  \cite{VK} that represents  planar graphs with dynamical holes of arbitrary sizes has been proposed. In this model there is  spontaneous tearing of the world sheet, associated with the planar graph, which gives a singularity at zero coupling constant of string interaction.}
\be
U(x)=V_0(x)+ V_{1}(x)=\frac{ m^2x^2}{2}+J e^{  \omega x}\label{2-exp}\ee
There are 4 different choices of signs of $J$ and $\omega$, see Fig.\ref{Fig:exp-4}. We see that only $J<0$ may produce some non trivial effects due to an appearance of  potential instability,  Fig.\ref{Fig:exp-4}.c and Fig.\ref{Fig:exp-4}.d. The choice of sign of $\omega$ is irrelevant.

\begin{figure}[h!]
\begin{center}
\includegraphics[width=0.45\textwidth]{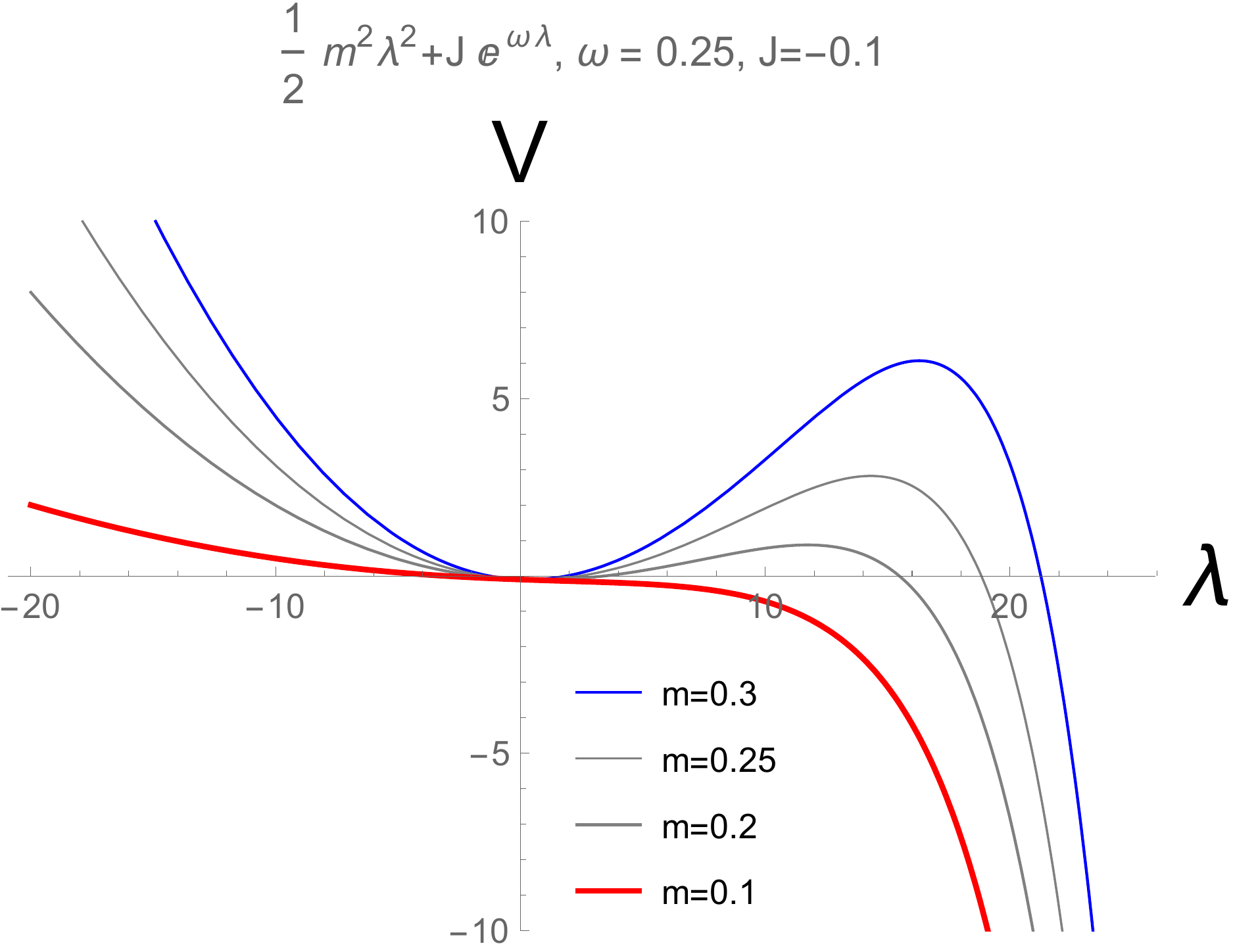}$\,\,\,\,\,\,\,\,\,\,$
\includegraphics[width=0.45\textwidth]{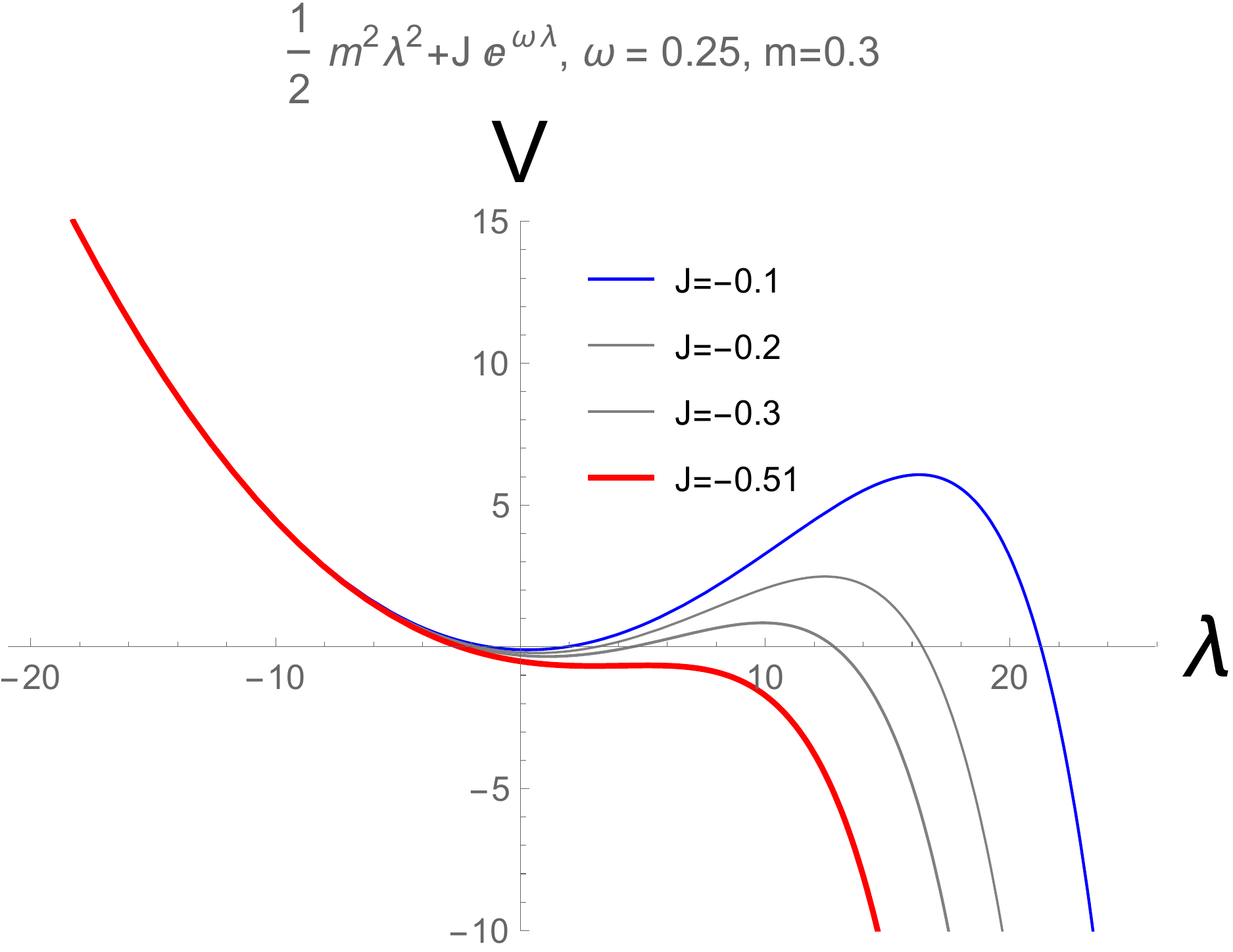}
\\
(a) $\qquad \qquad \qquad \qquad\qquad \qquad\qquad  \qquad$ (b)\\ $\,$\\
\caption {The appearance of phase transition under perturbation of the gaussian model by the  exponential potential  $\Delta V=Je^{\omega \lambda}$ taken with positive  $J$ and $w=0.25$. }\label{Fig:exp-PT}
%{\bf File: Deform-Gauss-model}
\end{center}
\end{figure}

Fig.\ref{Fig:exp-PT}  shows an appearance of a phase transition under a perturbation of the gaussian model by the  exponential potential  $V_{\exp}=Je^{\omega \lambda}$ taken with positive  $J$ and $w=0.25$.  We see that  the minimum of the potential disappears when $m$  decreases, Fig.\ref{Fig:exp-PT}.a, or $|J|$ increases for negative $J$, Fig.\ref{Fig:exp-PT}.b.

The shift of $\lambda\to \lambda+\delta \lambda$ in the potential \eqref{2-exp} produces the linear term in the potential and multiplies the current $J$ on the constant $J\to J\e^{\delta \lambda} $.
We parametrize our potential as
\be V(x)=m^2\Big(\frac12  x^2+C x+ J\, \e^{\omega x}\Big)
\label{Gauss-def}
\ee
and fix  the parameters in \eqref{Gauss-def} in an agreement with the measure localization on the segment $[0,\Lambda]$. To this purpose we first find the non-normalized measure  as a sum of non-normalized ones corresponding to the shifted Wigner and exponential potential distributions $\rho_{W}(\lambda,\Lambda)$ and $\rho_{e^{w\lambda}}(\lambda,\Lambda,\omega)$.

\begin{figure}[h!]
\begin{center}
\includegraphics[width=0.45\textwidth]{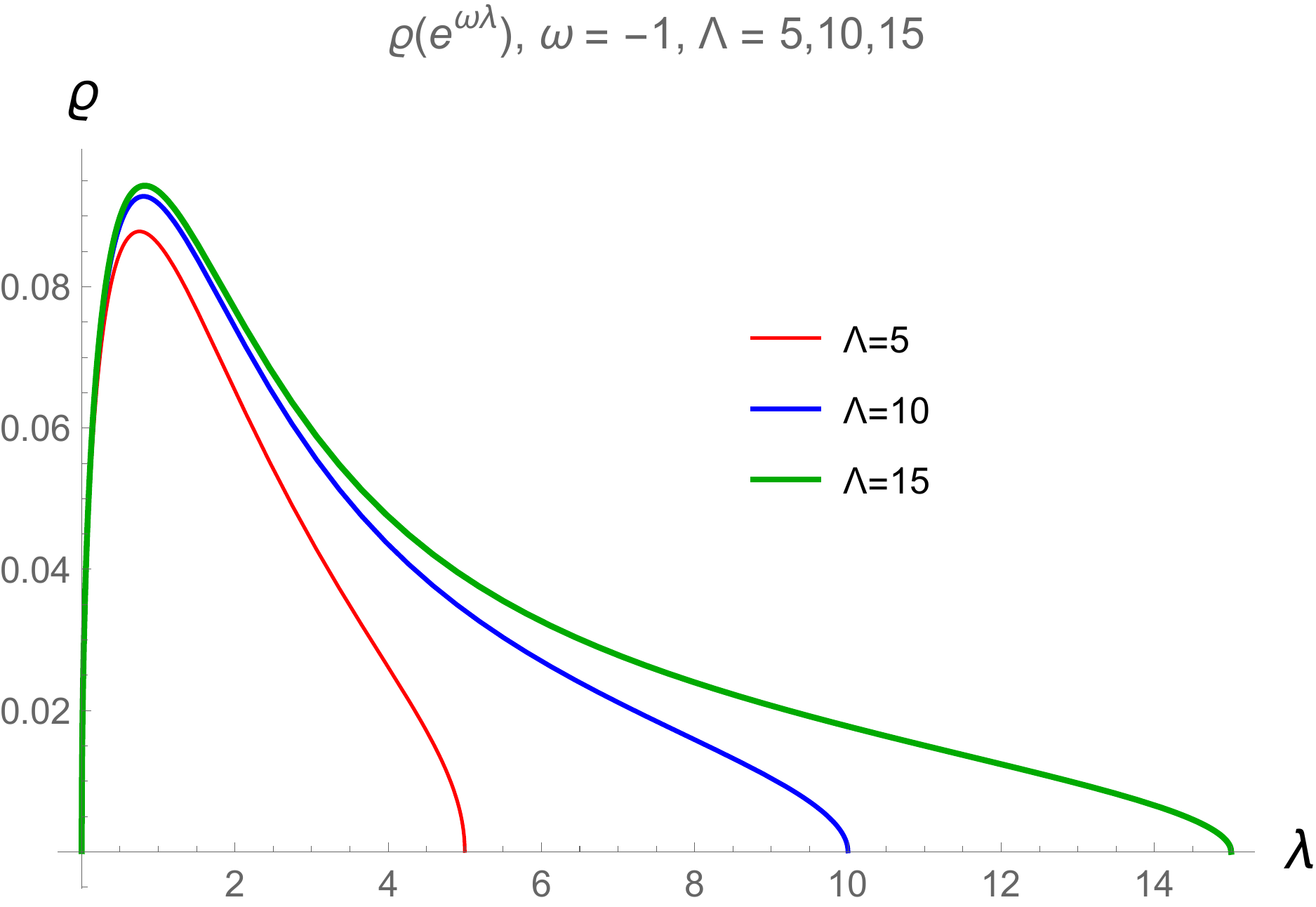}
\includegraphics[width=0.45\textwidth]{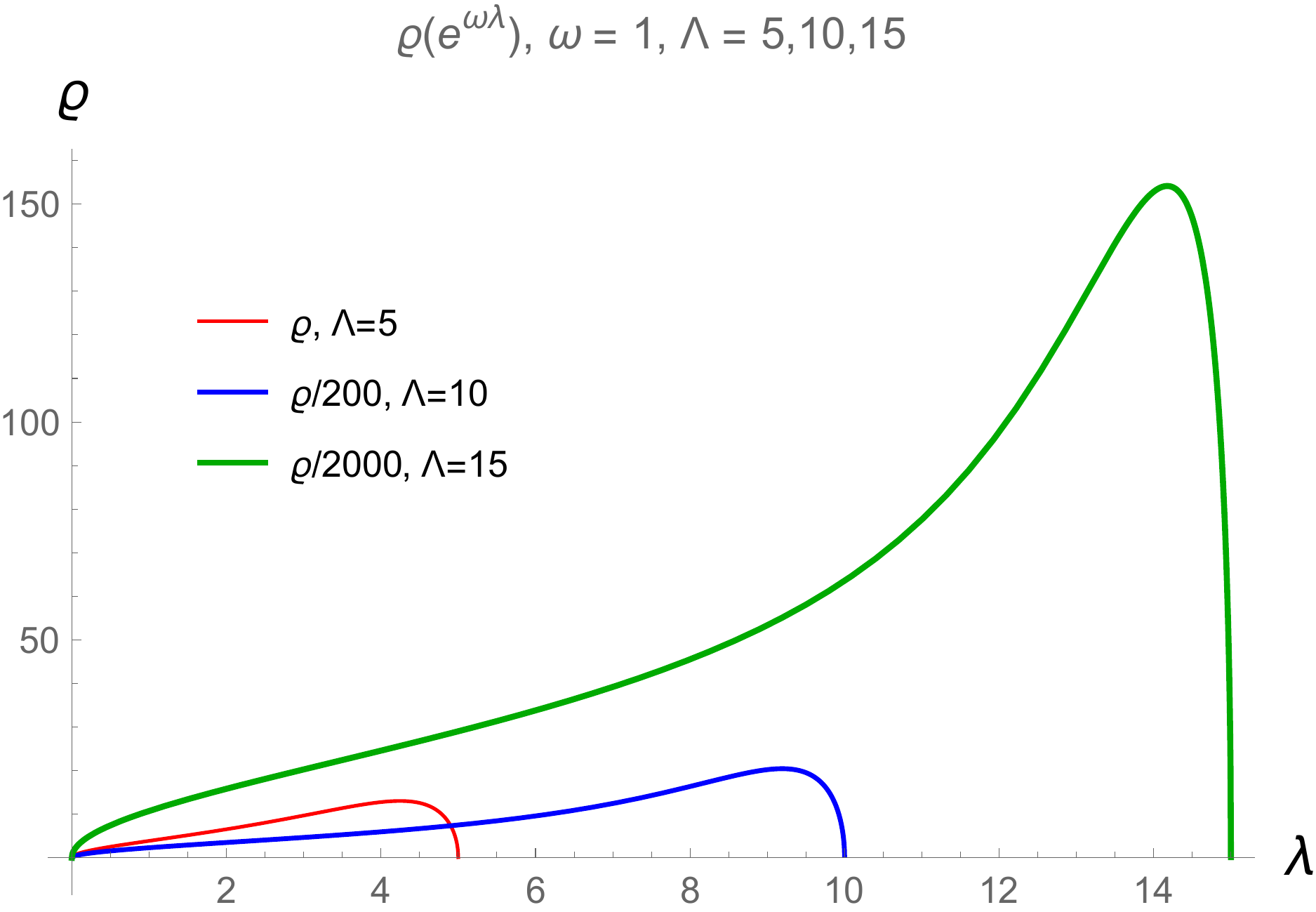}\\
(a) $\qquad \qquad \qquad \qquad\qquad \qquad\qquad  \qquad$ (b)
\caption {Non--normalized density plot for the exponential potential with negative $\omega=-1$ (a) and positive $\omega=1$ (b) and  different regularization parameter  $\Lambda$.}\label{Fig:neg-pos-exp}
%{\bf File: density-for-exp-pot-test-IA}
\end{center}
\end{figure}

The  forms of non-normalized measures $\rho_{e^{w\lambda}}(\lambda )$ for positive and negative $w$ presented are presented in Fig.\ref{Fig:neg-pos-exp}.
We can compare the contribution to the non-normalized density from the Wigner semi-circle for mass equal to 1 and the 
exponential potential taken with arbitrary current $J$. We see that this sum always defines the positive density for $J>0$ and becomes negative for $J<J_{cr}(\Lambda,\omega)<0$. Then for $J>J_{cr}(\Lambda,\omega)$ we find relation between $m^2$ and $\Lambda,\omega$ and $J$
from normalization condition.

\begin{figure}[h!]
\begin{center}
\includegraphics[width=0.33\textwidth]{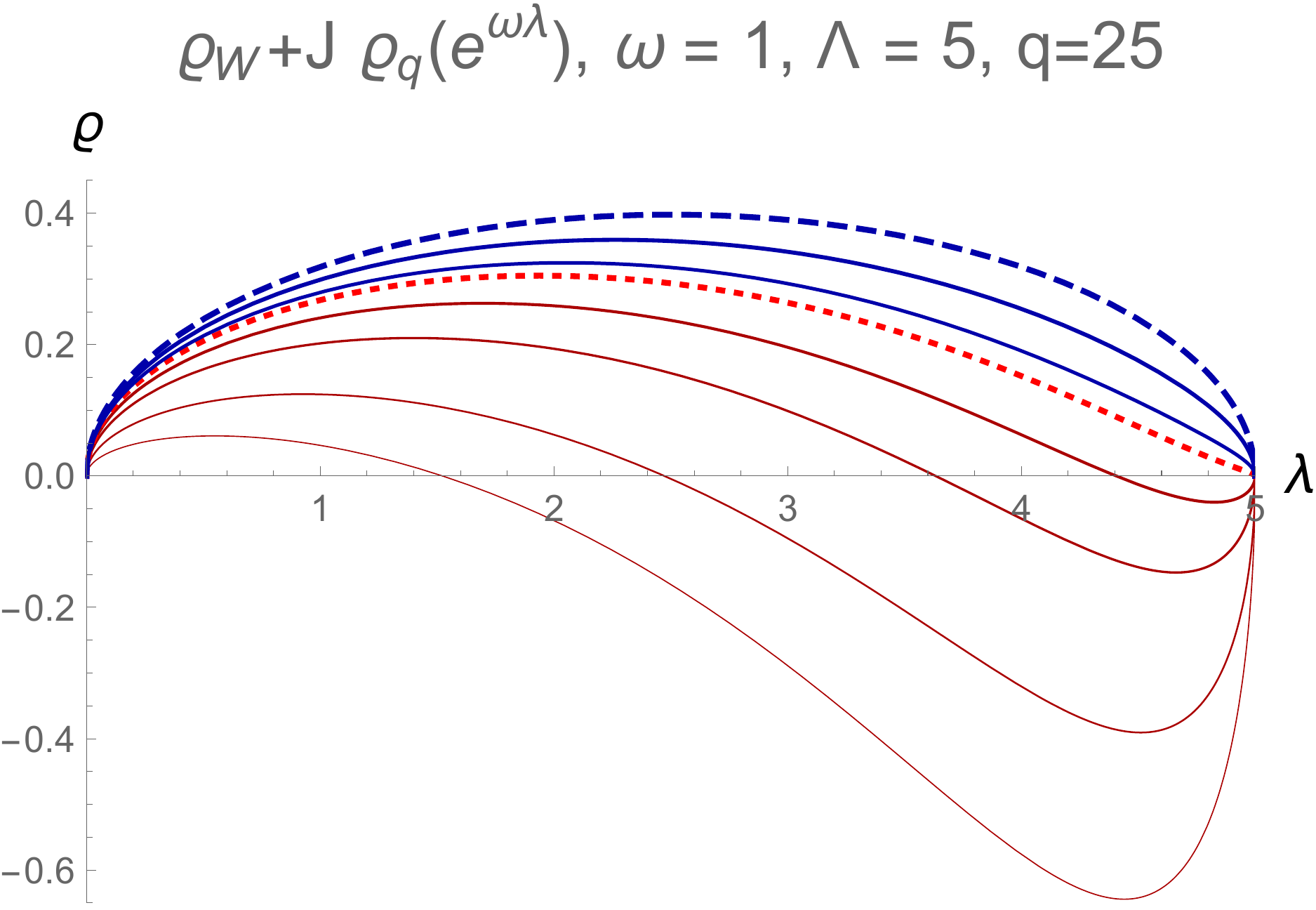}
\includegraphics[width=0.15\textwidth]{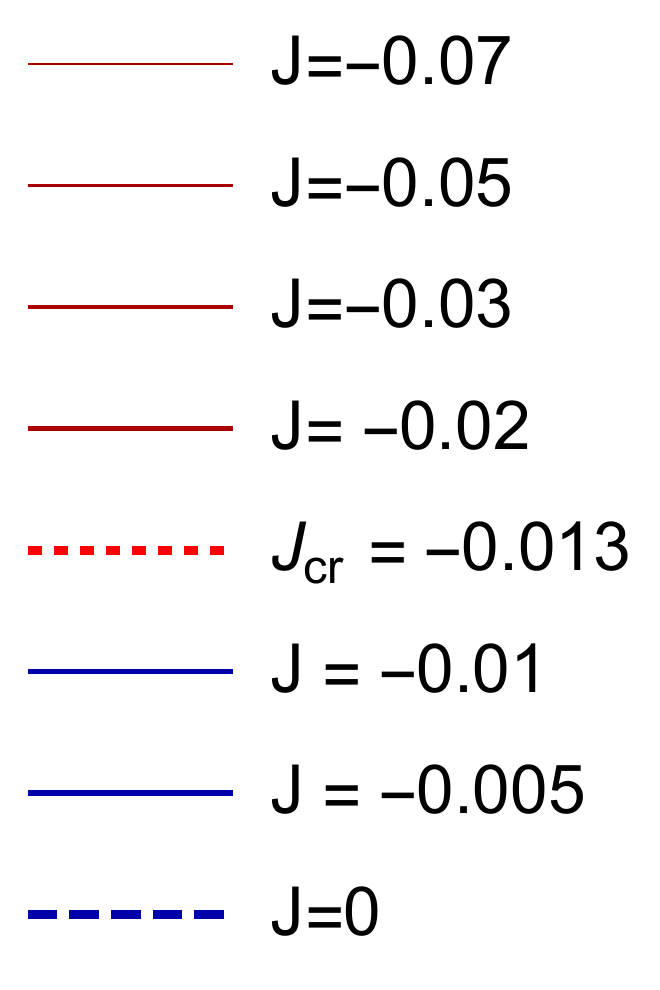}
\includegraphics[width=0.33\textwidth]{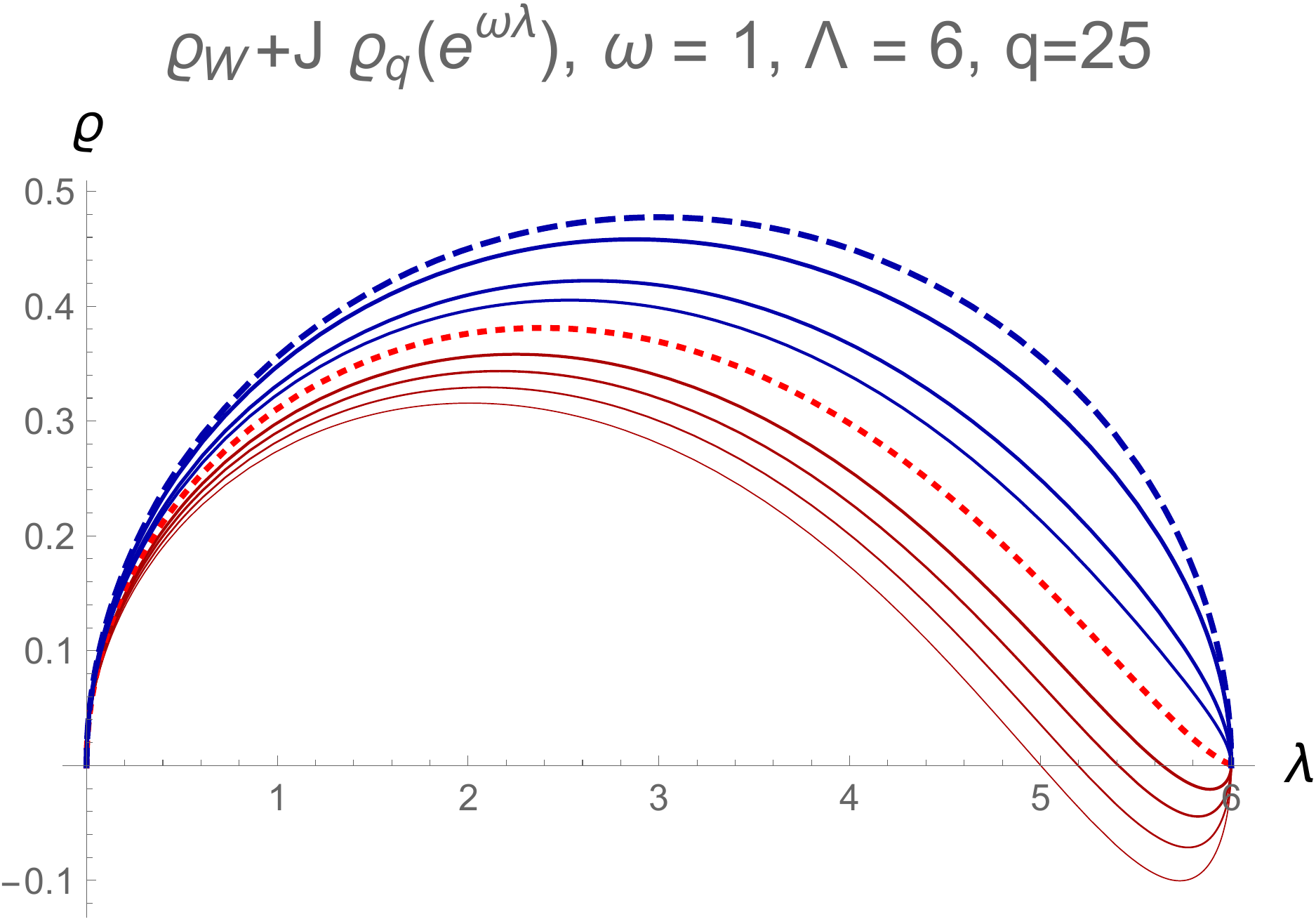}
\includegraphics[width=0.15\textwidth]{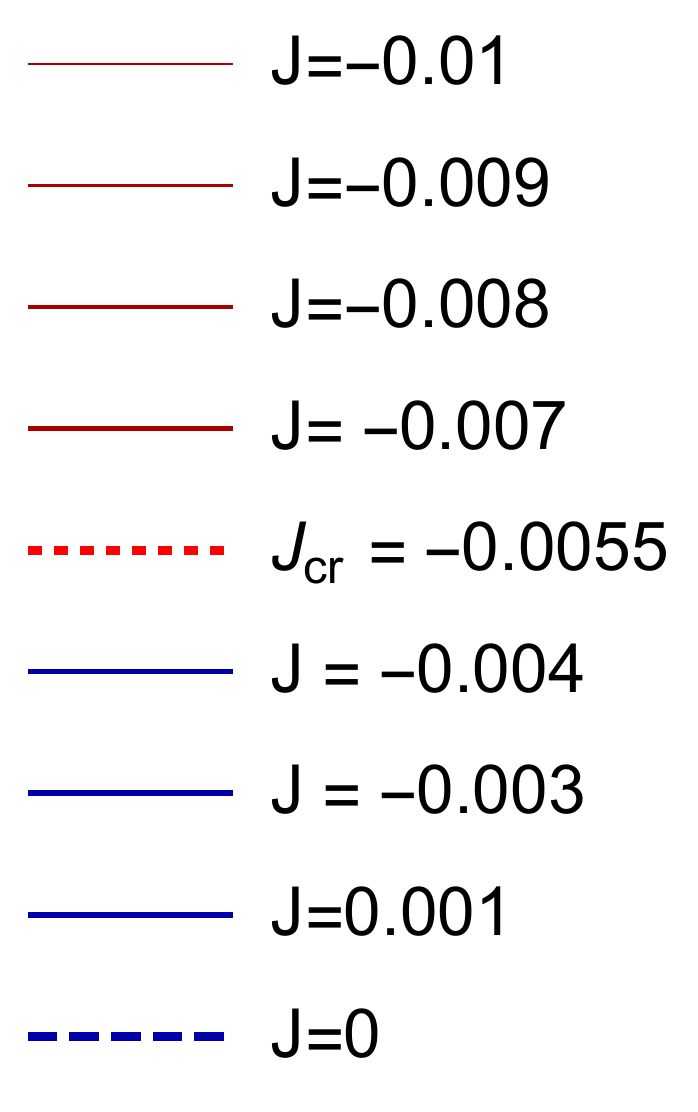}
\\
(a) $\qquad \qquad \qquad \qquad\qquad \qquad\qquad  \qquad$ (b)\\ $\,$\\$\,$\\
\includegraphics[width=0.33\textwidth]{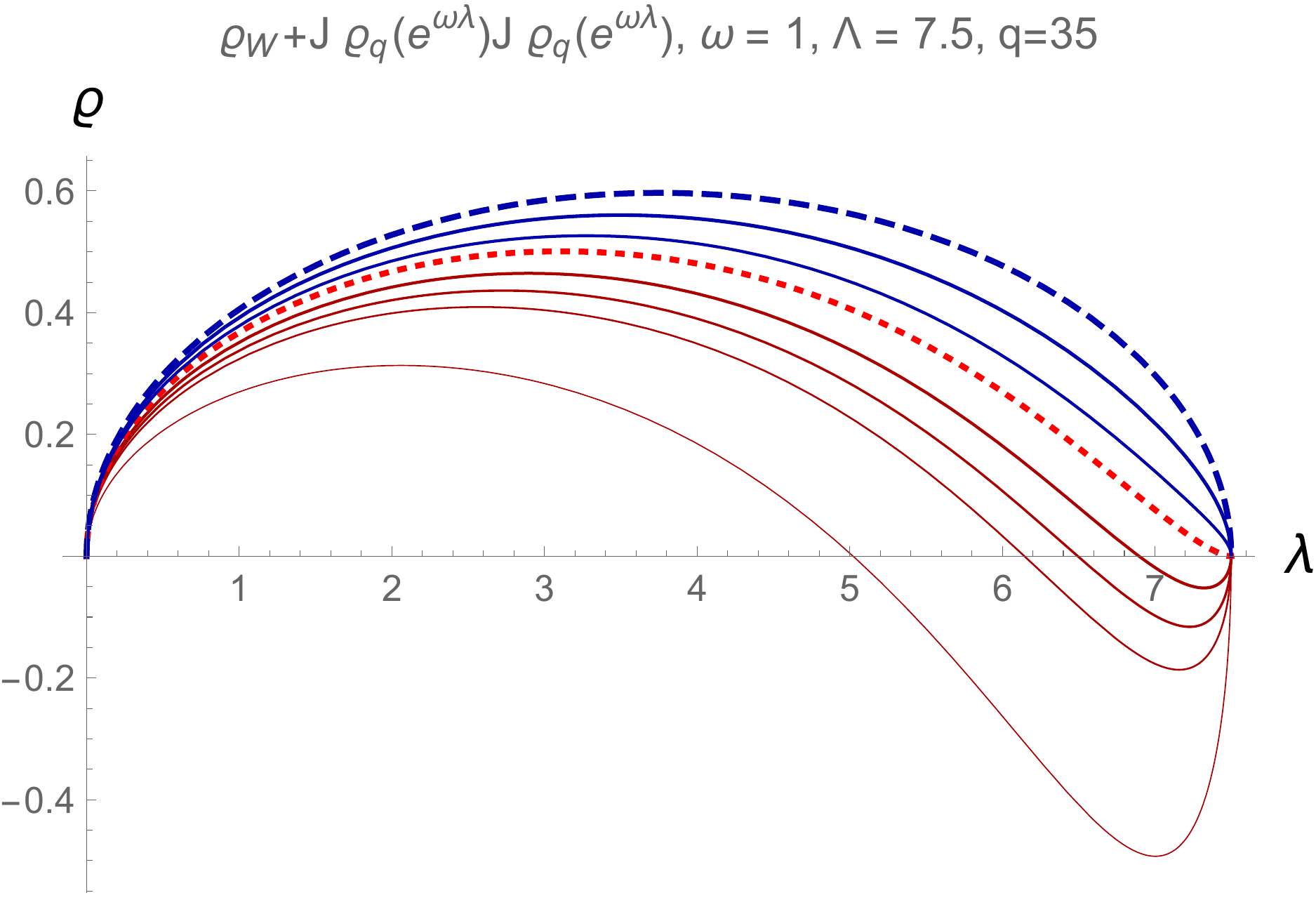}
\includegraphics[width=0.15\textwidth]{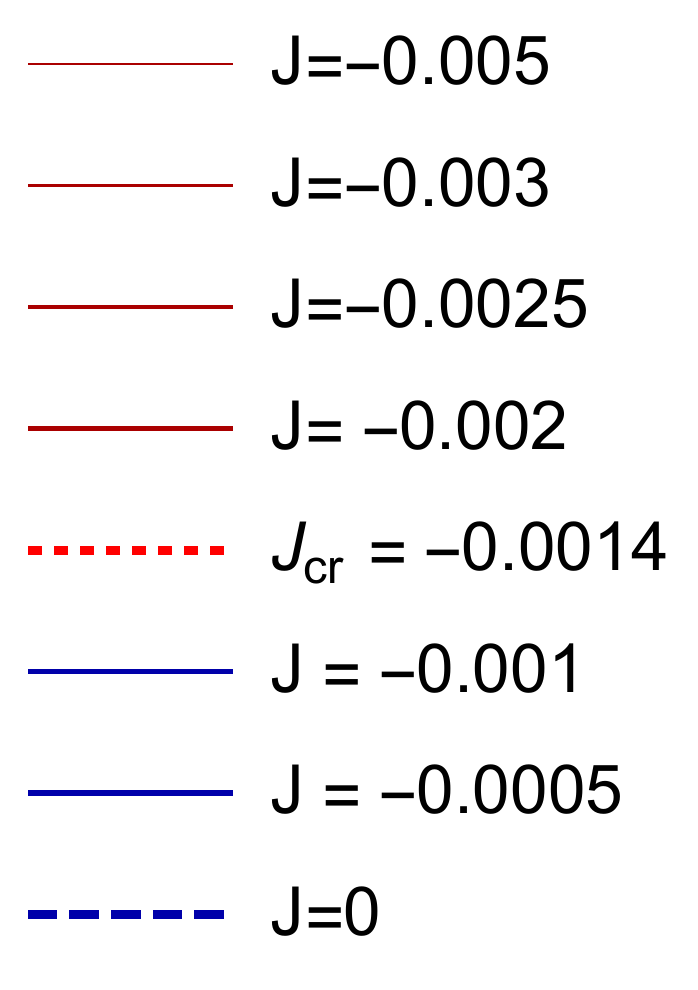}
\includegraphics[width=0.33\textwidth]{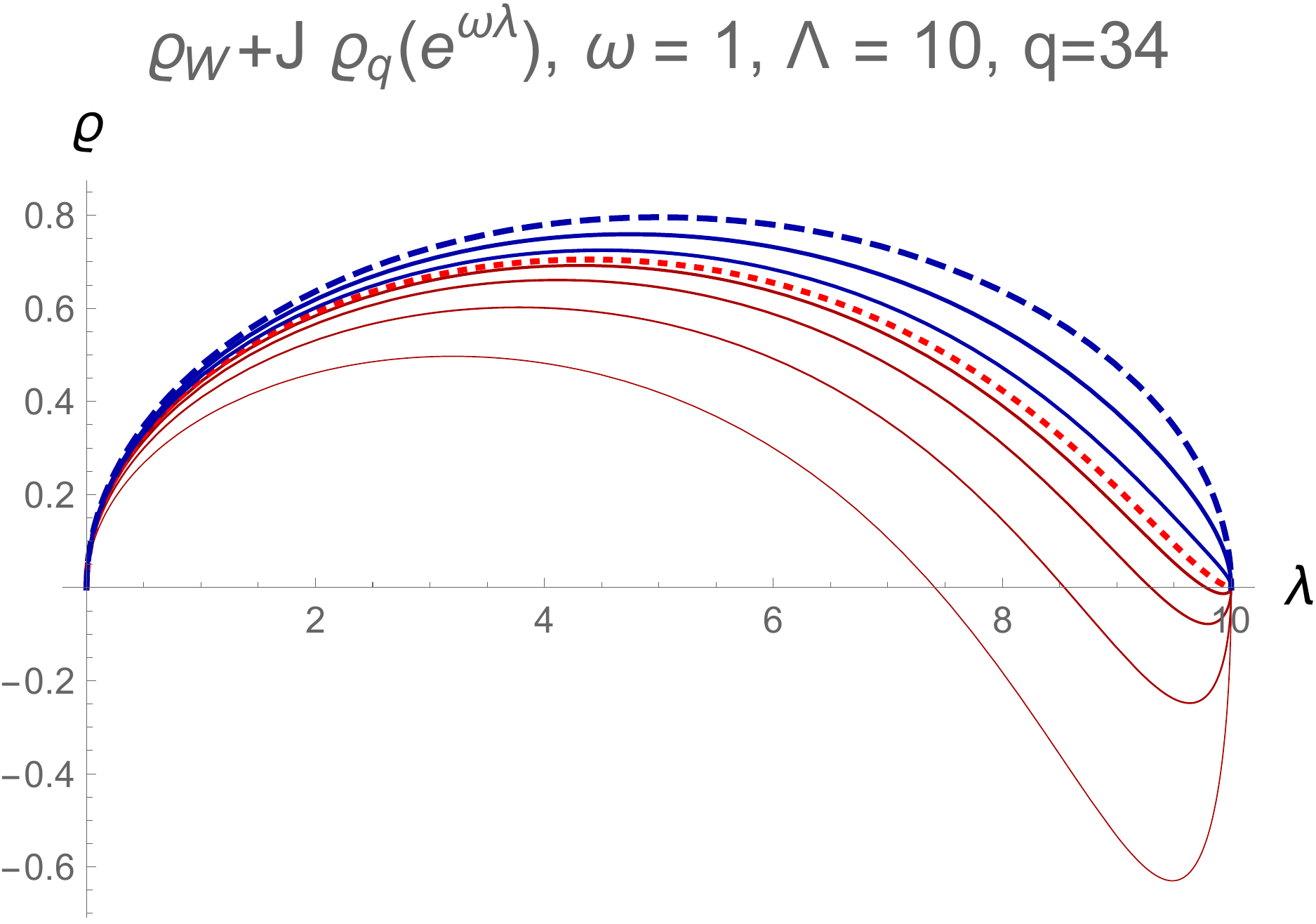}
\includegraphics[width=0.16\textwidth]{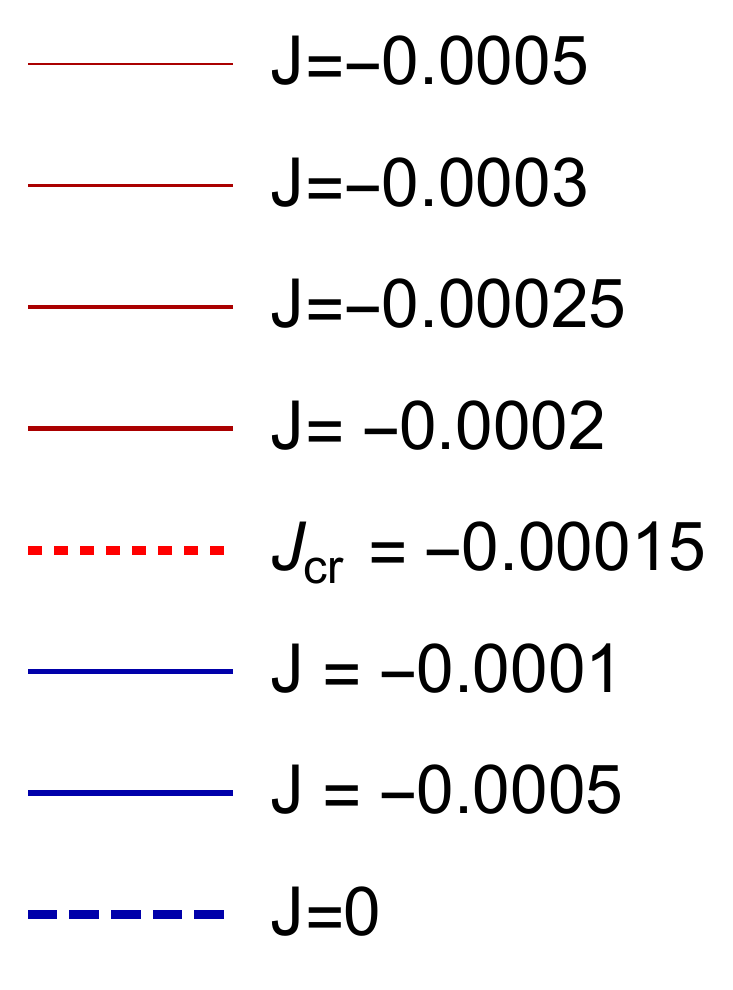}
\\
(c) $\qquad \qquad \qquad \qquad\qquad \qquad\qquad  \qquad$ (d)\\ $\,$\\$\,$\\
\includegraphics[width=0.33\textwidth]{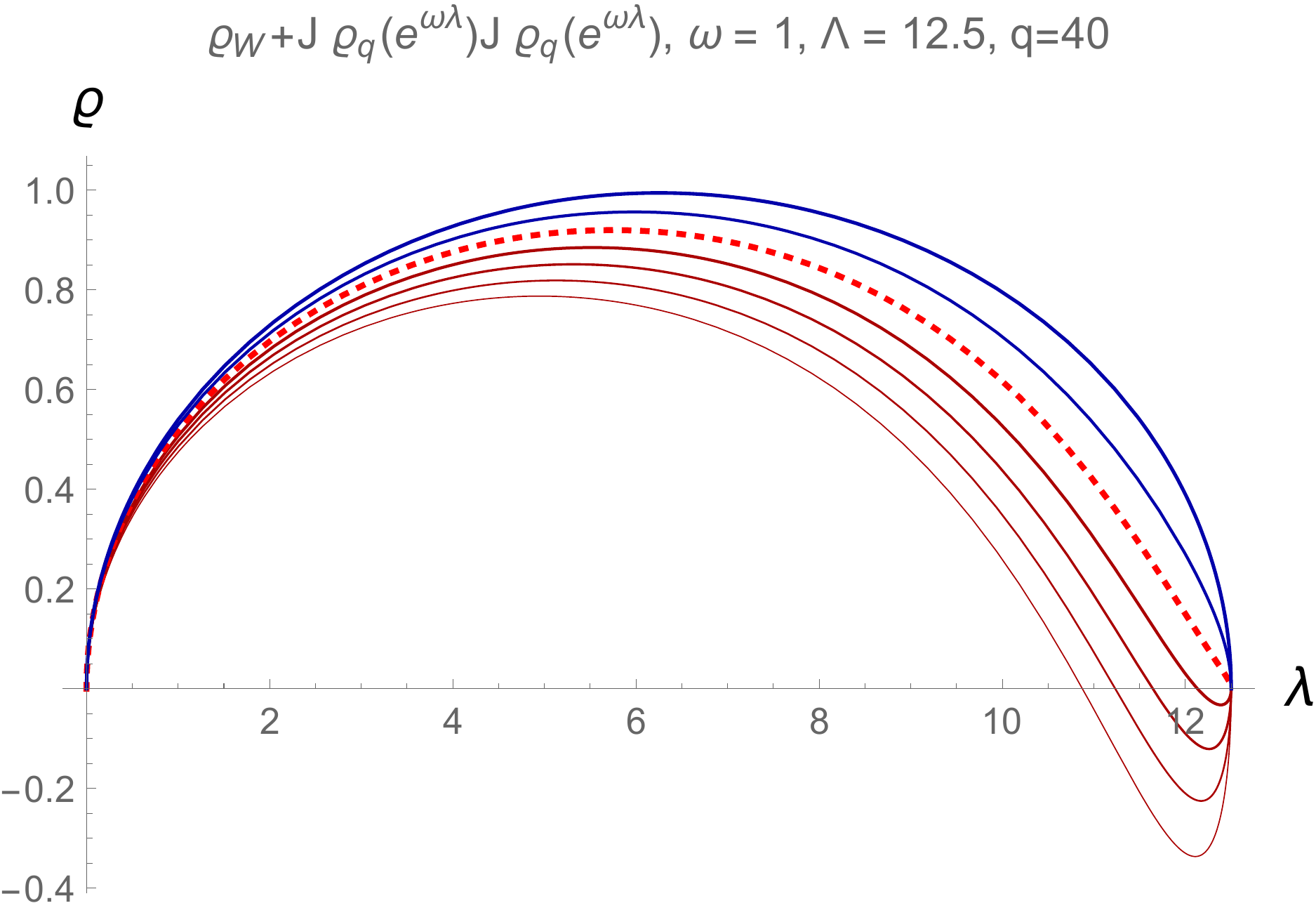}
\includegraphics[width=0.15\textwidth]{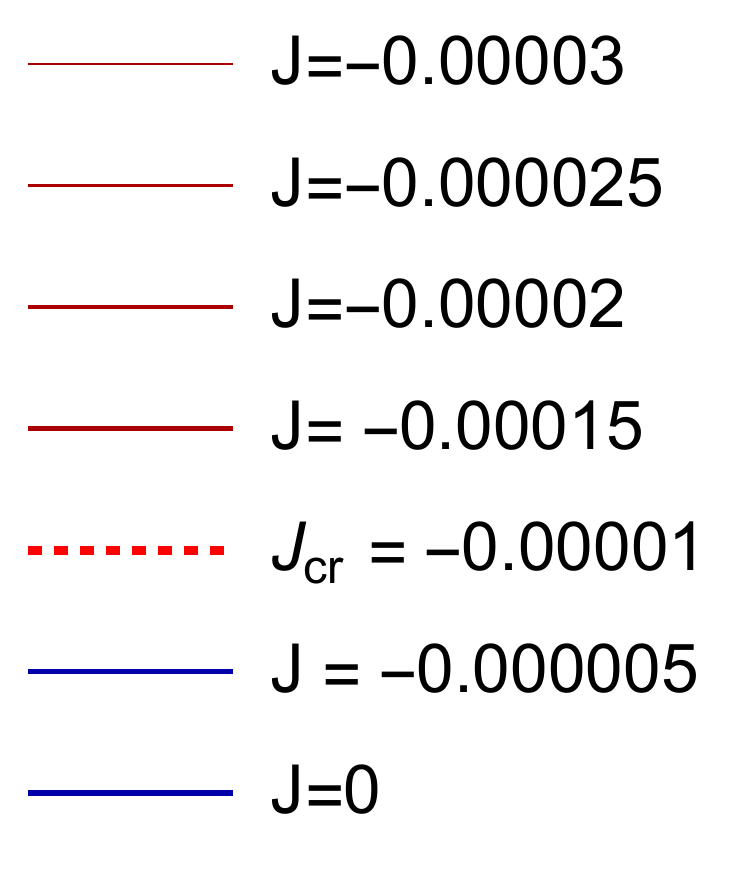}
\includegraphics[width=0.33\textwidth]{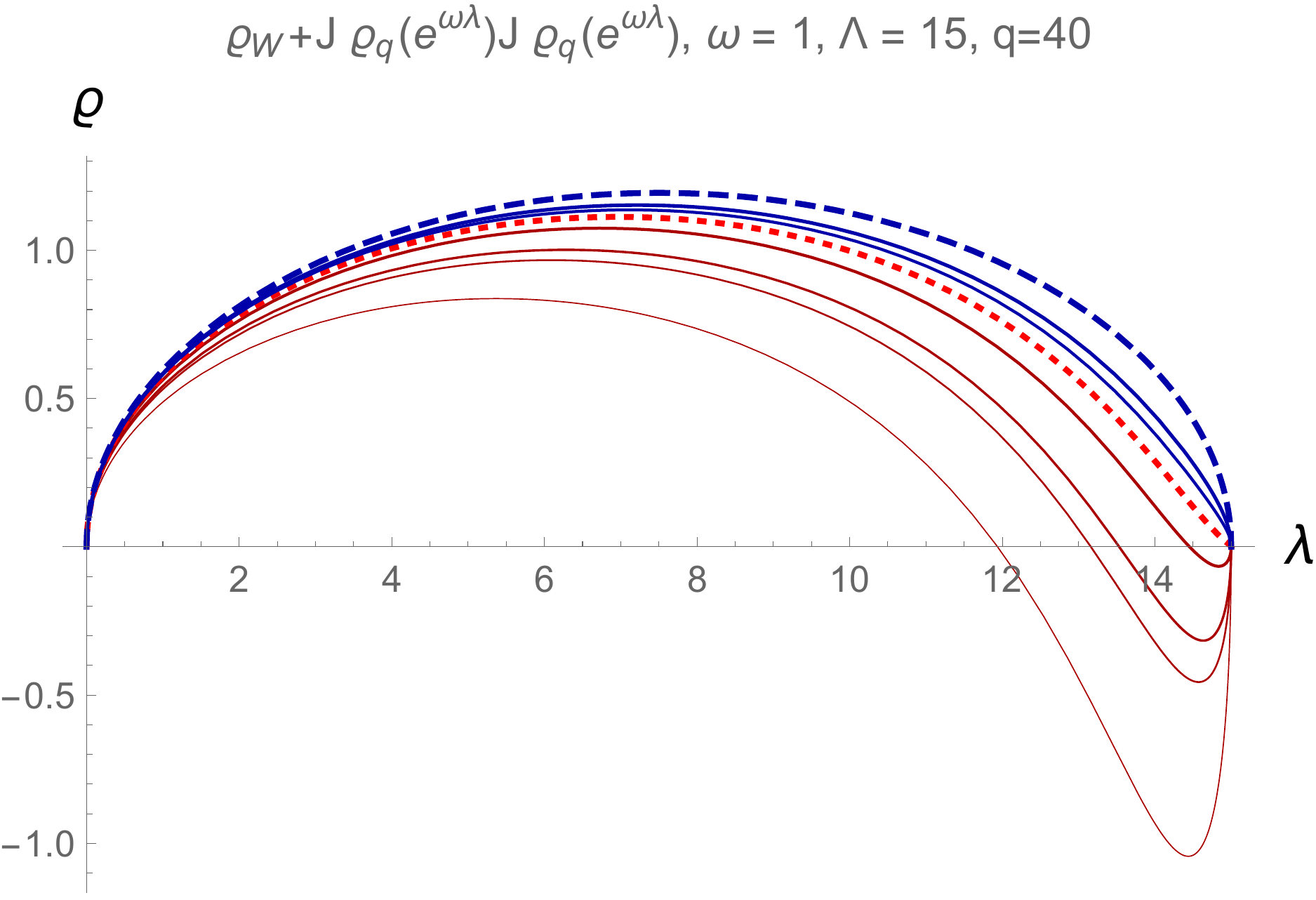}
\includegraphics[width=0.15\textwidth]{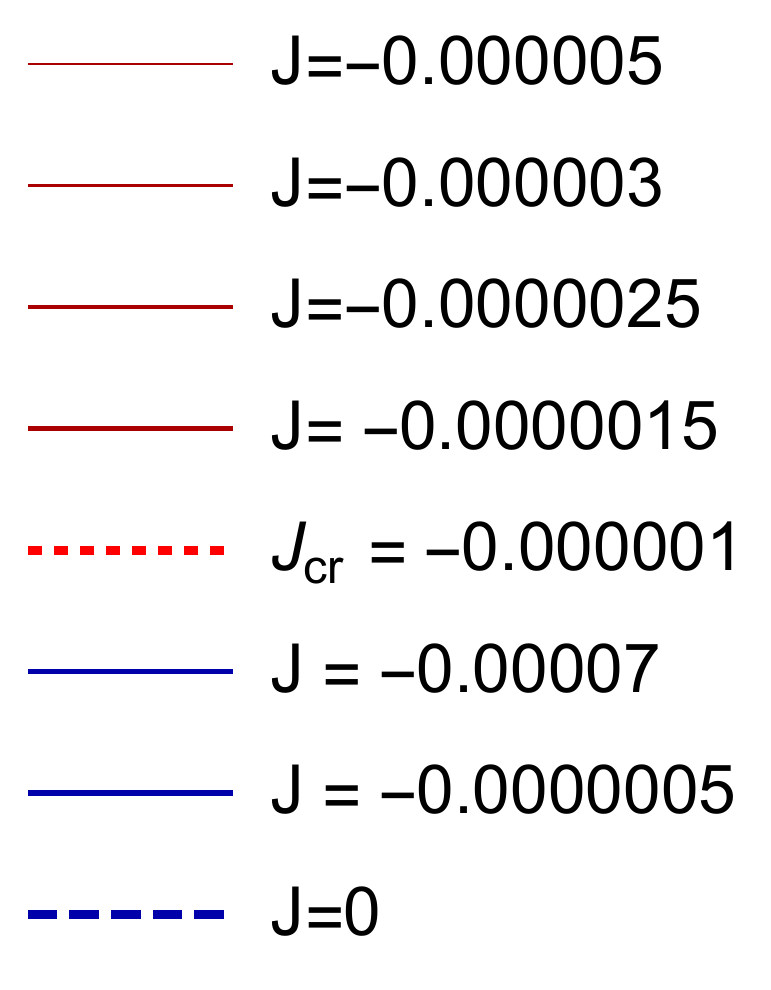}\\

(e) $\qquad \qquad \qquad \qquad\qquad \qquad\qquad  \qquad$ (f)\\ 

\caption {The plot of non--normalized density  for the quadratic  potential deformed by the exponential potential for different values of the regularization parameter  $\Lambda$ and  $\omega=1$.
}\label{Fig:PT-W-exp}
%{\bf File: density-for-exp-pot-test-IA}
\end{center}
\end{figure}

\begin{figure}[h!]
\begin{center}
\includegraphics[width=0.4\textwidth]{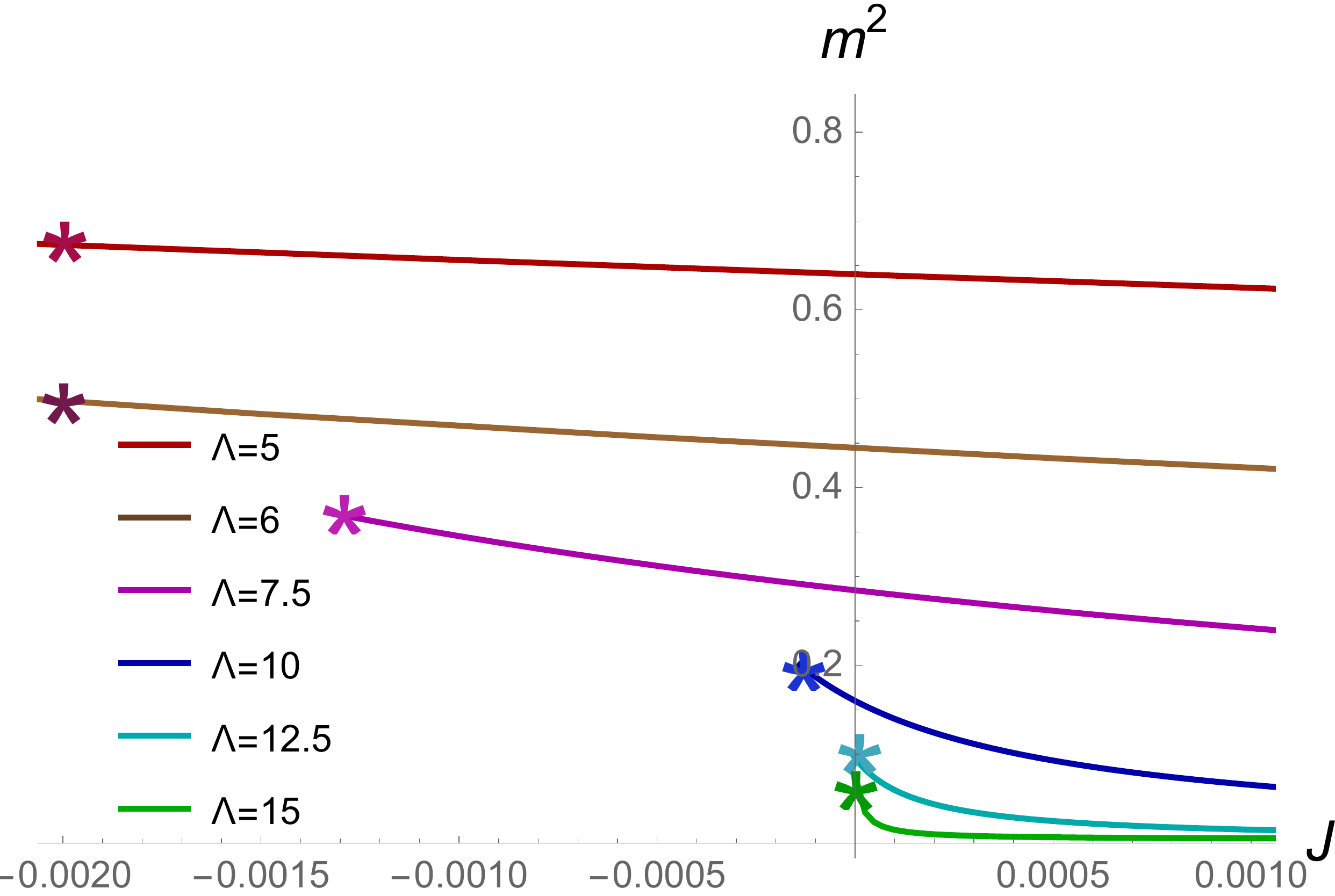}$\,\,\,\,\,\,\,$$\,\,\,\,\,\,\,$$\,\,\,\,\,\,\,$
\includegraphics[width=0.4\textwidth]{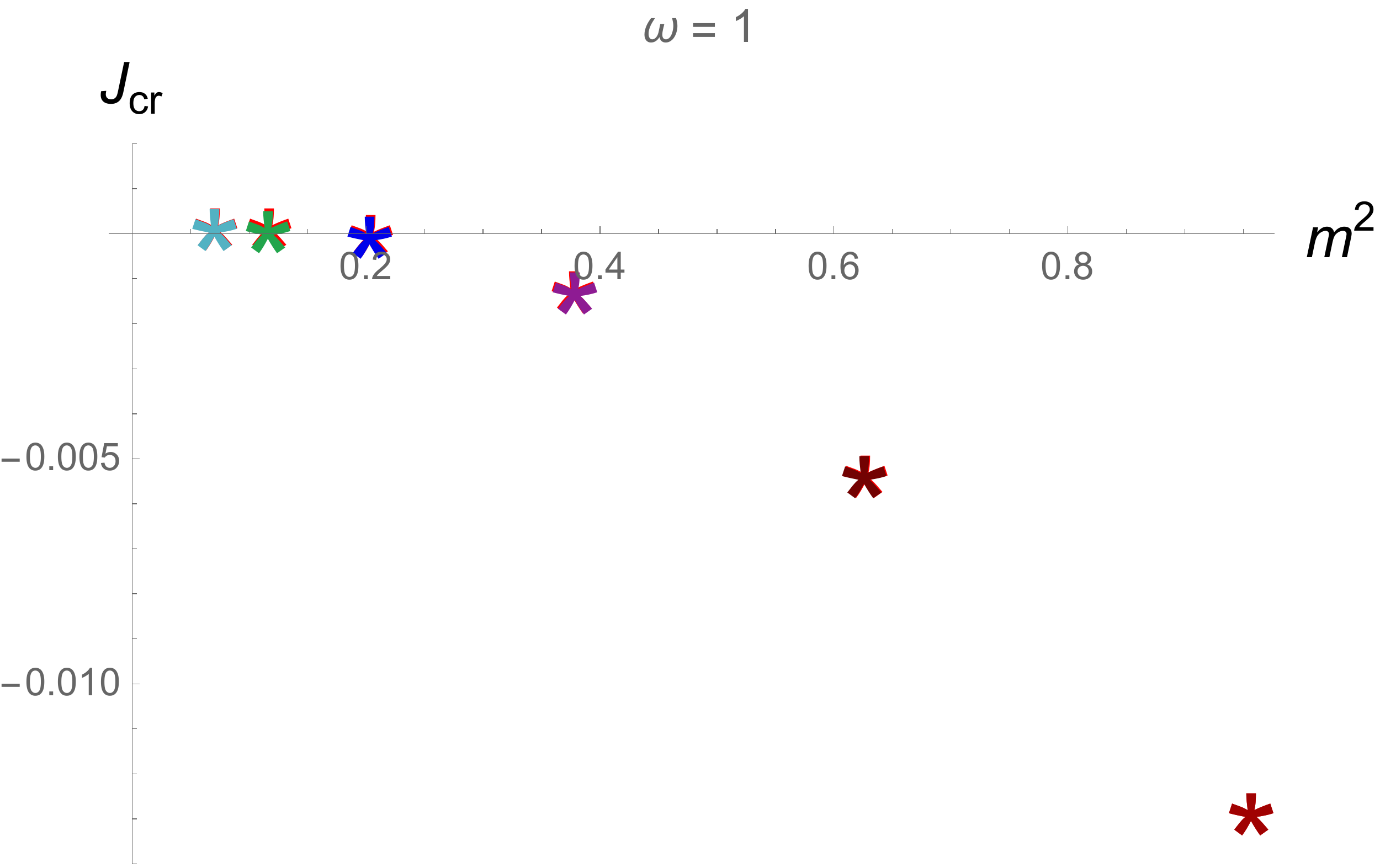}
\\
(a) $\qquad \qquad \qquad \qquad\qquad \qquad\qquad  \qquad$ (b)
\caption {(a) Relations between $m^2$ and $J$ for fixed $\Lambda$ and $\omega$. 
(b) The mass square at $J_{cr}$ for $\omega=1$. The legend  is the same as at (a).
}\label{Fig:Critical-points-exp}
%{\bf File: density-for-exp-pot-test-IA}
\end{center}
\end{figure}

In Fig.\ref{Fig:PT-W-exp}   the appearance of the phase transition at negative $J$ is presented.
In Fig. \ref{Fig:Critical-points-exp}.a relations between $m^2$ and $J$ for fixed $\Lambda$ and $\omega$ are shown for different values of $\Lambda$
and in Fig.\ref{Fig:Critical-points-exp}.b.

We can present the eigenvalues distribution corresponding to the potential 
\eqref{Gauss-def} as a sum
\footnote{Note that the potential $ \e^{\omega x}$ does not have a solution to the singular equation and does not itself defines the eigenvalues distribition, but $V=jx +J\e^{\omega x}$ does in the case $jJ\omega <0$. By $\rho_{e^{w\lambda}}$ we mean this distribution with corresponding choice of the linear term.}

\bea
\rho_{nn}(\lambda)=\rho_{W}(\lambda,\Lambda)+J\rho_{e^{w\lambda}}(\lambda,\Lambda,\omega)\label{sum-rho}
\eea
where
\bea
\rho_W(\lambda )&=&\frac{1}{2\pi^2}\sqrt{\lambda(\Lambda-\lambda)}\int^\Lambda_0\,\frac{d\mu}{\sqrt{(b-\mu)(\mu-a)}}\\
\rho_{e^{w\lambda}}(\lambda )&=&\frac{\omega}{2\pi^2}\sqrt{\lambda(\Lambda-\lambda)}\int^\Lambda_0\,
\frac{ (e^{\omega\lambda}-e^{\omega\mu})d\mu}{(\lambda-\mu)\sqrt{(b-\mu)(\mu-a)}}\label{exp-sol}
\eea
and fix constant $C$ and $m^2$ from the consistency condition and normalization, respectively,
\bea
0&=&\int_0^\Lambda \frac{\lambda+C+J\,\omega e^{\omega \lambda}}{S_{\Lambda}}\,d\lambda
\label{CC-w-exp}\\
\frac1{m^2}&=&\int_0^\Lambda\Big(\rho_{W}(\lambda,\Lambda)+J\rho_{e^{w\lambda}}(\lambda,\Lambda,\omega)\Big)\,d\lambda\label{NN-w-exp}
\eea

The  forms of non-normalized measures $\rho_{e^{w\lambda}}(\lambda )$ for positive and negative $w$ presented are presented in Fig.\ref{Fig:neg-pos-exp}. We see that these two segment distributions are nonsymmetric under the centre of the 
segment. The distribution of eigenvalues for the case of negative $\omega$ is pressed to the left boundary of the segment, and the for the case of positive $\omega$  it is pressed to the right one. Applying this deformation with a positive  $J$ to the GUE we "activate" 
the left or right part of of eigenvalues. Applying the same with negative $J$ we can destroy the constructed solution.
We  compare the contribution to the non-normalized density from the Wigner semi-circle for mass equal to 1 and the 
exponential potential taken with arbitrary current $J$. We see the this sum always define the positive density for $J>0$ and 
becomes negative for $J<J_{cr}(\Lambda,\omega)<0$. Then $J>J_{cr}(\Lambda,\omega)$ we find relation between $m^2$ and $\Lambda,\omega$, and $J$
from normalization condition.

In Fig.\ref{Fig:PT-W-exp}  is presented the appearance of the phase transition at negative $J$ for different $\Lambda$.
We see that for chosen parameters, $m^2=1,\omega=1$ the critical $J_{cr}$ decreases with increasing $\Lambda$.
To find the real mass that supports the normalized  solution,
\be
\rho (\lambda)=n(\lambda)\rho_{nn} (\lambda)\ee
we find $n(\lambda)$ from the normalization condition, so 
\be
n(\Lambda)^{-1}=\int _0^\Lambda \rho_{nn} (\lambda)d\lambda\ee 
and assume that the  mass is given my 
\be
m^2=n(\Lambda)\ee

In Fig.\ref{Fig:Critical-points-exp}.a the dependence of  $m^2$ on $J$ for fixed $\Lambda$ and $\omega$ is shown for different values of $\Lambda$. 
Here $J>J_{cr}(\Lambda)$. We see that mass (in our parametrization  of the potential of the model) decreases 
with increasing $J$.  The slow of the mass is more  fast for larger $\Lambda$.

In Fig.\ref{Fig:Critical-points-exp}.b the dependence of the critical current $J_{cr}$ on mass is shown. We see that $J_{cr}$
goes to zero when $m^2$ goes to zero, that corresponds to increasing $\Lambda$.

$$\,$$\newpage
\subsection{Fine tuning}
It is obvious that  fixing from the beginning the location of the eigenvalue one  immediately gives  restrictions on parameters of the potential of the matrix model. If we want to shift the location of the eigenvalues, $\cS_{[a,b]}\to\cS_{[0,b-a]}$ we have to make a shift in the potential,
$V(x)\to V(x+a)$. For the quadratic potential this shift produces  the linear term $\Delta V(x)=jx$ and $j$ can be determined  from the location of  the left point of the cut. For higher polynomial interaction the shift produces the linear term in the LHS of singular equation, as well as change of coupling constants. The shift in the exponential
potential produces just a multiplication on positive constant.

 We have also seen that if we want to deform a given distribution by a  new potential, that has the same locations of eigenvalues,  we can just take the sum of the given two distributions and  multiply all coupling constants of two initial model on the same parameter to fix the normalization condition for the distributions that is the some of given two distributions. As to the consistency condition it follows from consistency conditions of individual distributions. More precisely, if we know that
\bea
V'_1(\lambda)&=&2\fint _\cS\frac{\rho_{V_1}(\lambda ')}{\lambda -\lambda '}d\lambda'\,, \,\,\,\,\,\,\,\int _\cS\rho_{V_1}(\lambda)d\lambda=1\\
V'_2(\lambda)&=&2\fint _\cS \frac{\rho_{V_2}(\lambda ')}{\lambda -\lambda '}d\lambda '\,, \,\,\,\,\,\,\,\int _\cS\rho_{V_2}(\lambda)d\lambda=1,
\eea
Note that in both integrals the segment is the same. Taking
\be
\rho_{T}=\frac12(\rho_{V_1}+\rho_{V_2})\ee
we can claim that $\rho_{T}$ solves equations
\bea
V'_T(\lambda)\equiv \frac12(V'_1(\lambda)+V'_2(\lambda))&=&2\fint _\cS\frac{\rho_{T}(\lambda ')}{\lambda -\lambda '}d\lambda '\,, \,\,\,\,\,\,\,\int _\cS \rho_{T}(\lambda)d\lambda=1\eea
 The consistency condition is automatically satisfied. 

One can put a coupling constant in front of  the second potential, say $J$.  In our previous example this were the coupling constant $g$
in the perturbation of Gaussian model by qubic term, or $J$ in the case of the exponential potential. For positive $J$ 
we keep the positivity condition for the sum of two distribution, meanwhile we can lost it for the case of big negative current.
This loss of positivity leads to the destruction of the large $N$ expansion of the model and can be  interpreted as a phase transition. 

As to  double scaling limit, we can consider it  in three steps. First we bring the support of the eigenvalue distribution on the interval $[0,\Lambda]$ by fine tuning the linear term in the potential. 
Then one goes to the limit $N\to\infty$ and after that $\Lambda\to\infty$.

\newpage
\newpage

\section{Matrix model for JT gravity}
\subsection{Potentials for non-normalized density  distribution}

For fixed $\Lambda $ we  consider the eigenvalue distribution  normalized   to $1$
\be\label{hjk}
\rho_{norm,1}(E) = n(\Lambda)\rho_{nn}(E),\,\,\,\,\,\rho_{nn}(E)=\frac{1}{(2\pi)^2}\sinh(2\pi\sqrt{E}), \hspace{20pt} E>0.
\ee
This form of the eigenvalue distribution in the SYK model has been obtained in \cite{Bagrets,StanfordWitten} and is nothing but the 
 Bethe formula for the nuclear level density \cite{Bethe}.
 For large $\Lambda$ one has
$n(\Lambda)\approx 8 \pi ^3 e^{-2 \pi  \sqrt{\Lambda }}/\sqrt{\Lambda }$.
The distribution normalized to $N$  has the form
\be\label{hjk}
\rho_{d.s.}^{\text{nnorm}}(E)\equiv\rho_{norm,N}(E) = \frac{\e^{S_0}}{(2\pi)^2}\sinh(2\pi\sqrt{E}), \,\,\,\,\,\,\,\,\,\, E>0.
\ee
compare with \eqref{dsl}. It is evident that $\e^{S_0}=Nn(\Lambda)$

To  recover the potential that supports  the distribution  $\rho_{norm,1}(E)$,  we write
\bea
 V(\mu)&=&n(\Lambda)V_{nn}(\mu),\eea
 where $ V_{nn}'(\mu)$ is defined by
 \bea
 V_{nn}'(\mu)&=&\frac{1}{(2\pi)^2}\int_0^\Lambda  \frac{\sinh 2\pi\sqrt{\lambda}}{\mu-\lambda}d\lambda.
\eea
This gives
\bea
 V_{nn}'(\mu) &=&\frac{1}{(2\pi)^2}\Big[e^{-2 \pi  \sqrt{\mu }} \left(e^{4 \pi  \sqrt{\mu }}
   \text{Ei}\left(2 \pi  \left(\sqrt{\Lambda }-\sqrt{\mu
   }\right)\right)+\text{Ei}\left(2 \pi  \left(\sqrt{\Lambda
   }+\sqrt{\mu }\right)\right)\right)\nn\\&-&e^{-2 \pi  \sqrt{\mu }}
   \left(\text{Ei}\left(2 \pi  \left(\sqrt{\mu }-\sqrt{\Lambda
   }\right)\right)+e^{4 \pi  \sqrt{\mu }} \text{Ei}\left(-2 \pi 
   \left(\sqrt{\Lambda }+\sqrt{\mu }\right)\right)\right)  \Big] \eea
  Here $\text{Ei}$ is  the exponential integral that for real non zero values of x
    is defined as  
    \be
   \operatorname{Ei}(x)=-\fint_{-x}^{\infty}\frac{e^{-t}}t\,dt.
    \ee
    $ \operatorname{Ei}$ has an expansion 
    \bea
 \operatorname {Ei} (x)=\gamma +\ln x+\sum _{k=1}^{\infty }{\frac {x^{k}}{k\;k!}},\eea
 and  faster converging Ramanujan's series has the form 
\be  \operatorname {Ei} (x) = \gamma + \ln x + \exp{(x/2)} \sum_{n=1}^\infty \frac{ (-1)^{n-1} x^n} {n! \, 2^{n-1}} \sum_{k=0}^{\lfloor (n-1)/2 \rfloor} \frac{1}{2k+1}
\ee
$\operatorname {Ei}$ can be bounded by elementary functions as follows
\be
 \frac {1}{2}\e^{-x}\,\ln \!\left(1+{\frac {2}{x}}\right)<\operatorname {Ei}(x)<\e^{-x}\,\ln \!\left(1+{\frac {1}{x}}\right)\qquad x>0
\ee

\begin{figure}[h!]
\begin{center}
\includegraphics[width=0.75\textwidth]{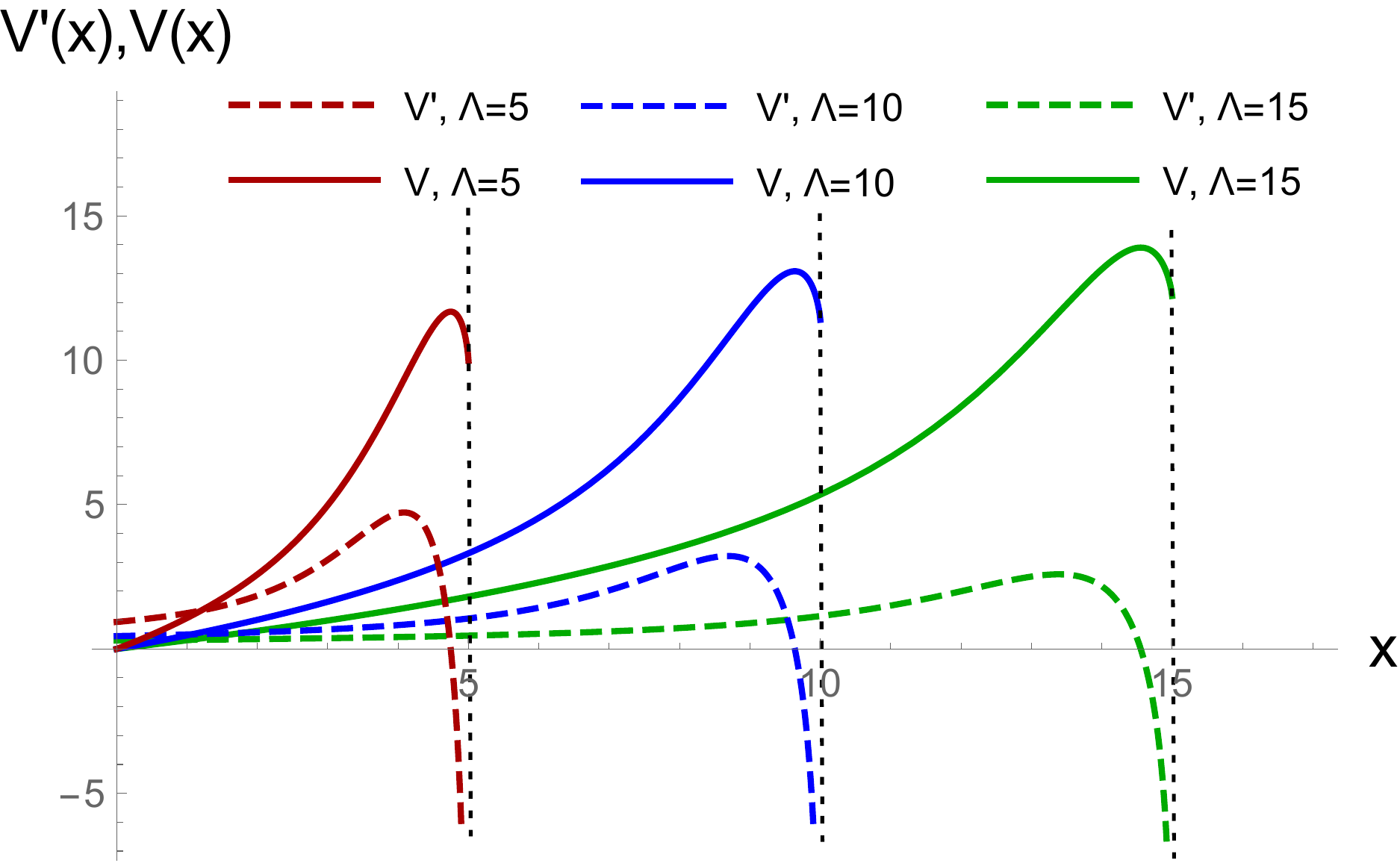}
\caption {The potential supported the density $ \rho_{norm,0}(E)$ for different  parameter  $\Lambda$ 
}\label{Fig:pot-sinh}
 %[[see math. file: {\bf pot-to-sinh.nb}]]
\end{center}
\end{figure}

The potential up to a constant is \be
  V(\mu) =\int ^\mu V'(\lambda)d\lambda\ee
 and it is presented in Fig.\ref{Fig:pot-sinh}.
\\

\subsection{Effective energy}
 
The effective energy $ E_{eff}$, \cite{BIPZ}, is evaluated on the normalized density  and it is given by the following formula
\bea
 E_{eff}&=&n^2(\Lambda)\, {\cal E}_{eff}=\frac{\e^{2S_0}}{N^2}\, {\cal E}_{eff},\nn\\
 \cE_{eff}&=&
 \int d\lambda  \rho_{nn}(\lambda)\,V_{nn}( \lambda)-\int\!\!\!\!\!\int d \lambda d \lambda'\, \rho_{nn}(\lambda)  
 \rho_{nn}(\lambda')\ln (\lambda-\lambda')^2
 \label{normV}
\eea

\begin{figure}[h!]
\begin{center}
\includegraphics[width=0.5\textwidth]{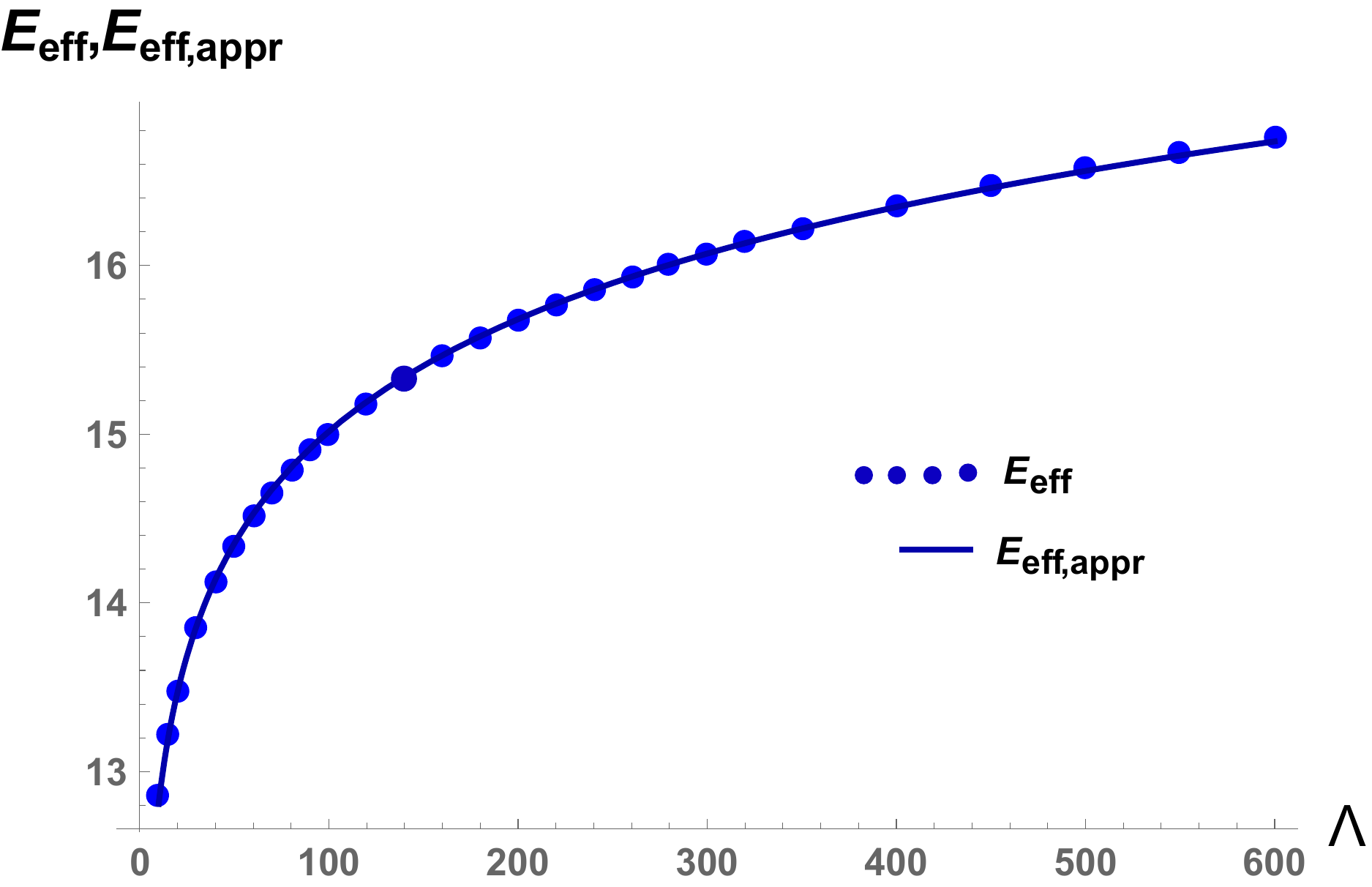}
\caption {$E_{eff}$  as function of $\Lambda$ and its  approximation
$10.6 0 + 0.959 \log x$. }\label{Fig:effective-action}
%{\bf Math.: pot-to-sinh-24-04-mian-test.nb}
\end{center}
\end{figure}

In Fig.\ref{Fig:effective-action} the effective action $E_{eff}$  as function of $\Lambda$ is shown.
We see that it can be approximated by  a  log $10.6 0 + 0.959 \log x$ (the next approximation is given by the double-log, $10.76 + 1.050 \log x- 0.393 \log \log x$).

\newpage
\newpage
\subsection{Phase transition}
To study the phase transition we consider  the deformation of $\rho_{\sinh \sqrt{\lambda}}$ by  $\rho_{\exp}$
\bea
\rho_{nn}(\lambda)=\rho_{\sinh  \sqrt{\lambda}}(\lambda,\Lambda)+J\rho_{e^{w\lambda}}(\lambda,\Lambda,\omega)
\label{sinh-exp}\eea

It is interesting to compare this density with the density 
\bea
\rho_{nn,s}(\lambda)=\rho_{\sinh  \sqrt{\lambda(1-\lambda/\Lambda)}}(\lambda,\Lambda)+J\rho_{e^{w\lambda}}(\lambda,\Lambda,\omega),
\label{sinh-exp-mod}\eea
where
\be
\rho_{\sinh  \sqrt{\lambda(1-\lambda/\Lambda)}}(\lambda,\Lambda)=\frac{n_s(\Lambda)}{(2\pi)^2}\sinh 2\pi \sqrt{\lambda(1-\lambda/\Lambda)}.\ee

\begin{figure}[h!]
\begin{center}
\includegraphics[width=0.33\textwidth]{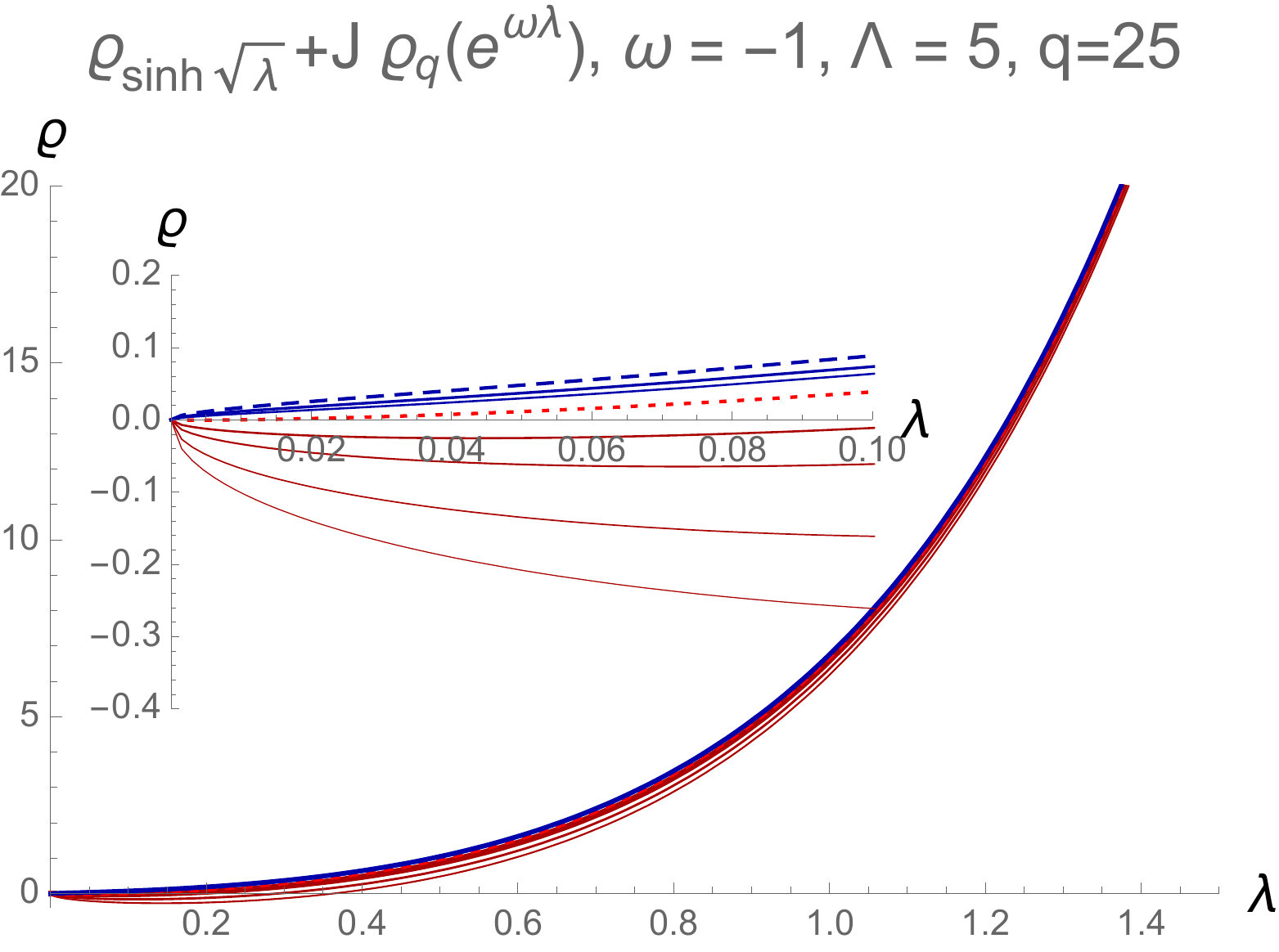}
\includegraphics[width=0.14\textwidth]{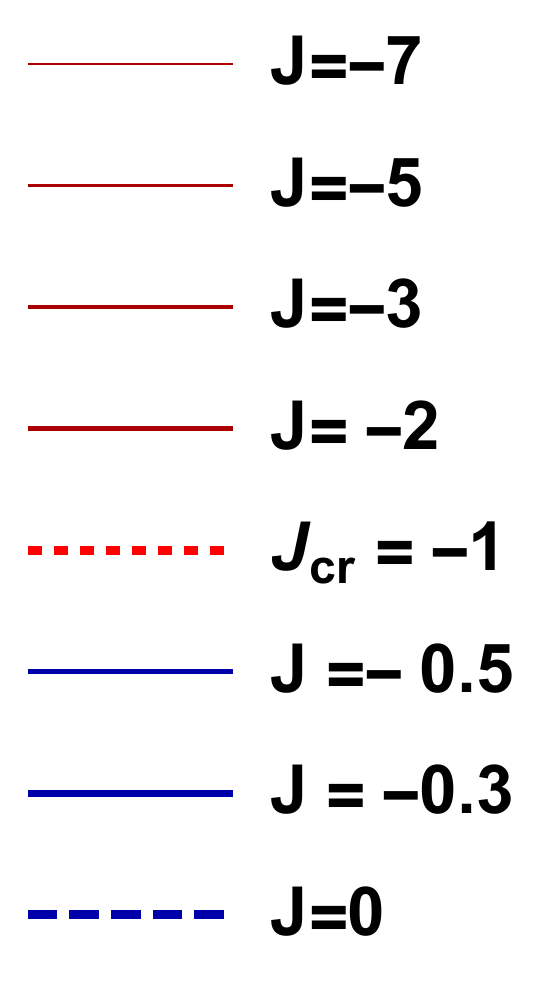}
\includegraphics[width=0.33\textwidth]{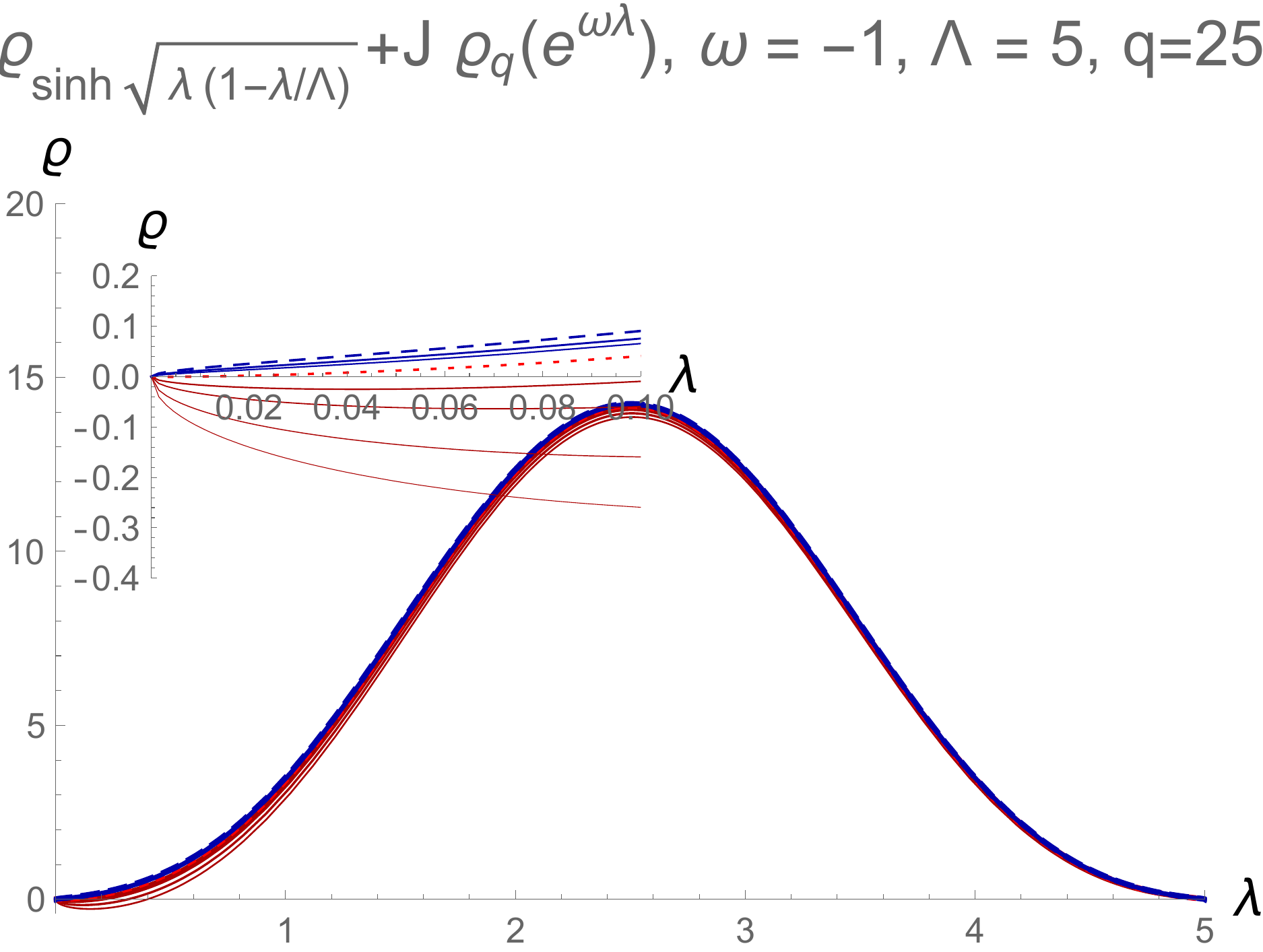}\\
(a) $\qquad \qquad \qquad \qquad\qquad \qquad\qquad  \qquad$ (b)\\ 
\includegraphics[width=0.33\textwidth]{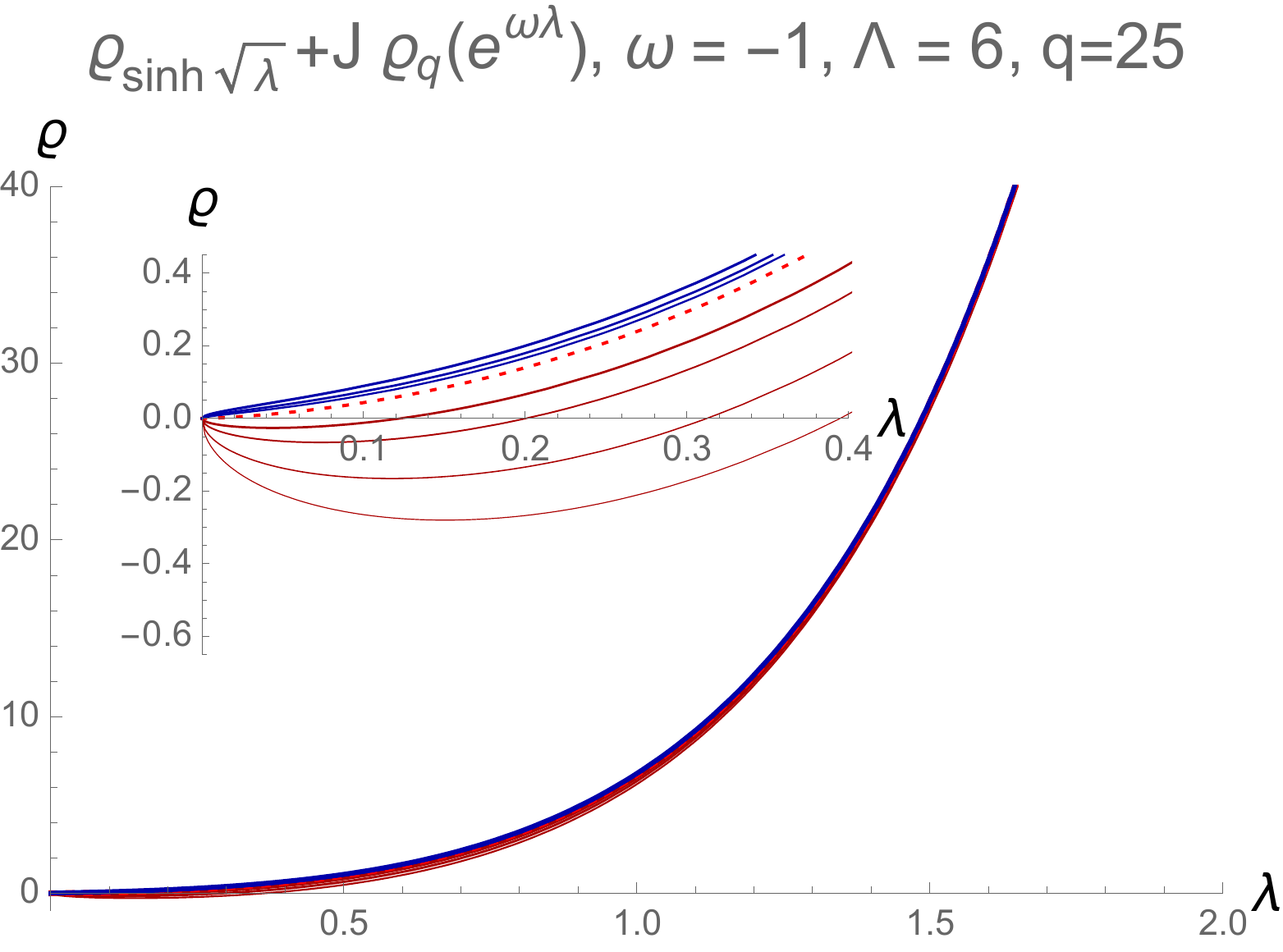}
\includegraphics[width=0.14\textwidth]{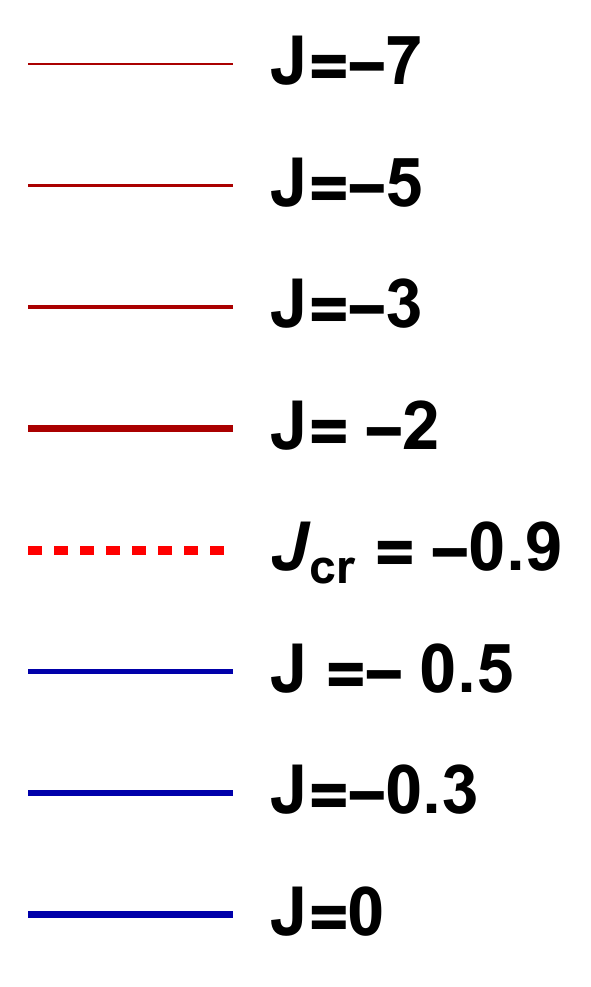}
\includegraphics[width=0.33\textwidth]{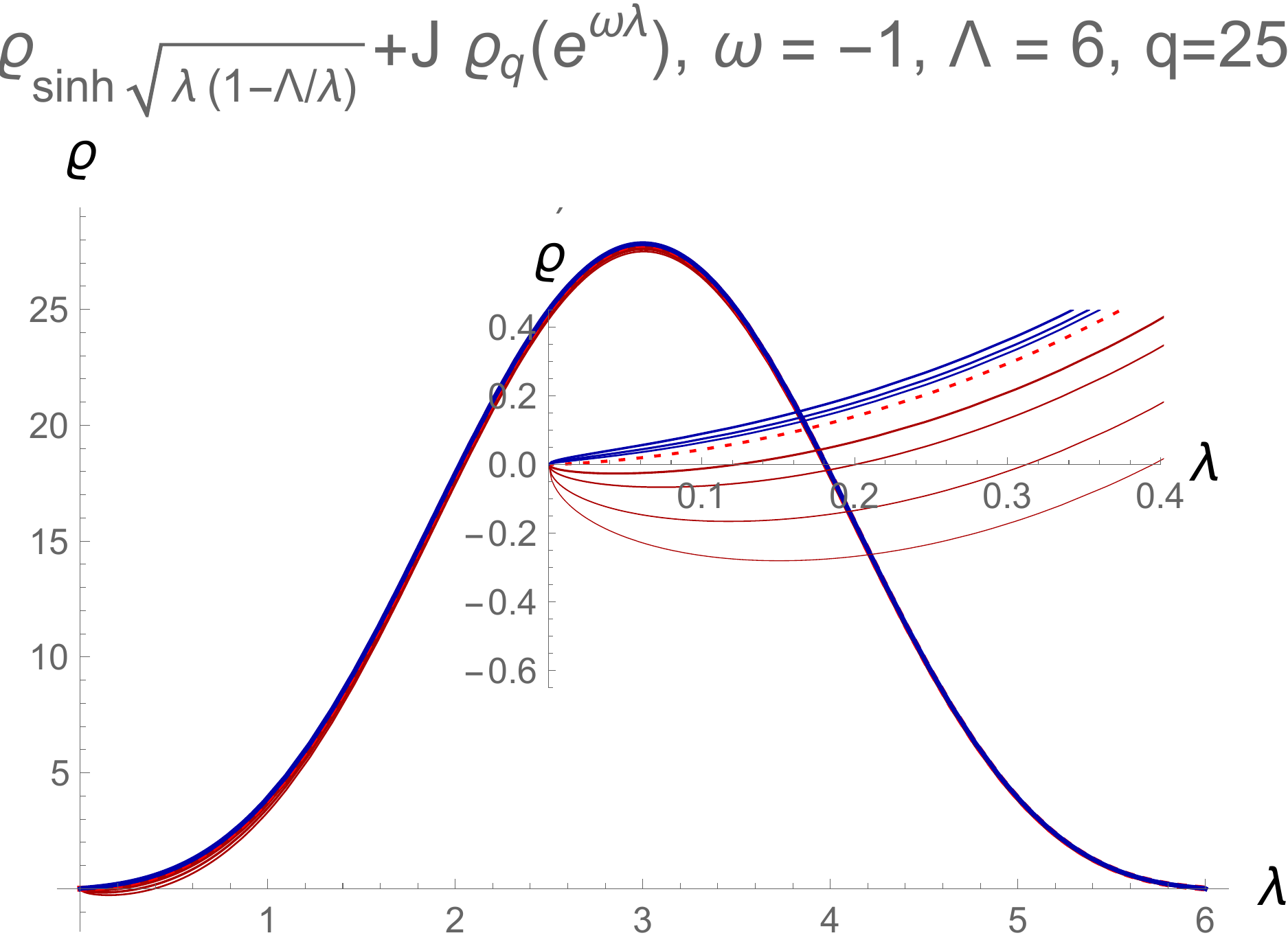}\\
(c) $\qquad \qquad \qquad \qquad\qquad \qquad\qquad  \qquad$ (d)\\
\includegraphics[width=0.33\textwidth]{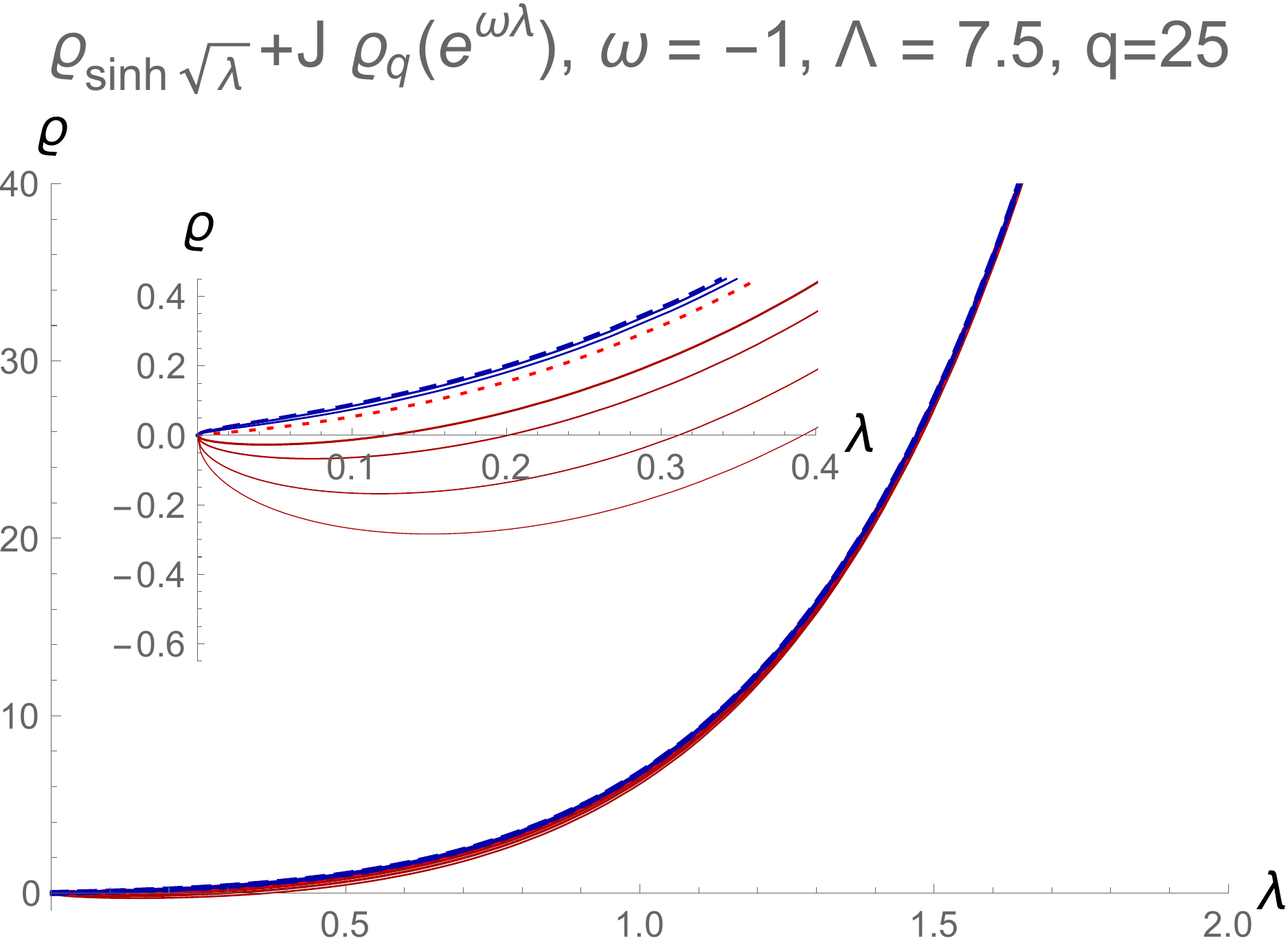}
\includegraphics[width=0.15\textwidth]{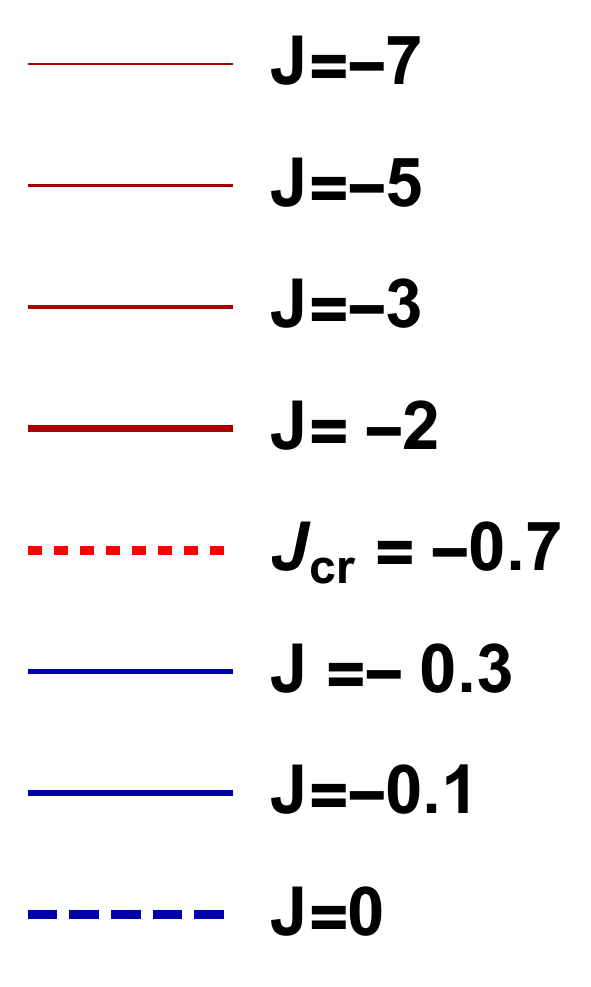}
\includegraphics[width=0.33\textwidth]{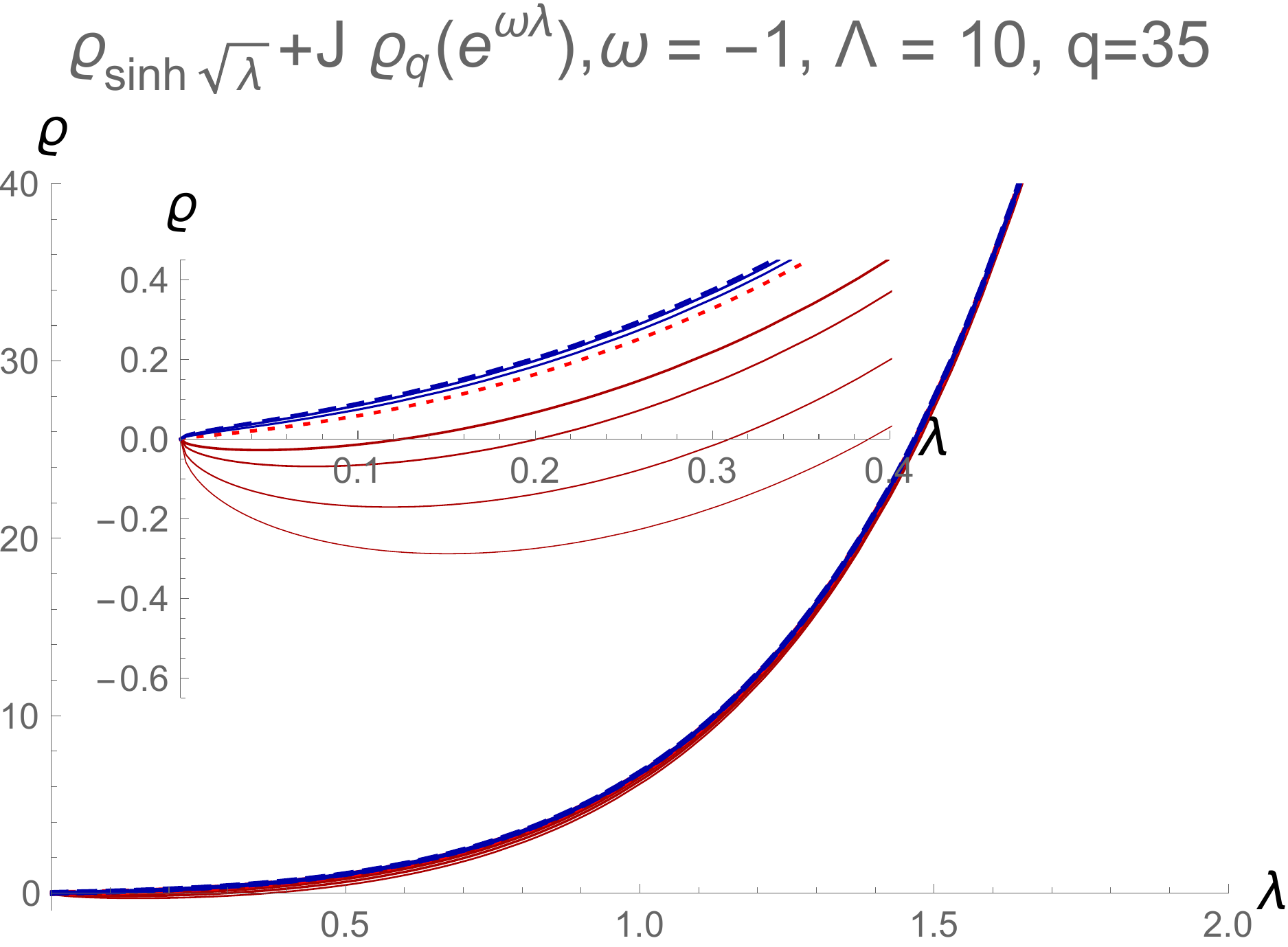}
\includegraphics[width=0.15\textwidth]{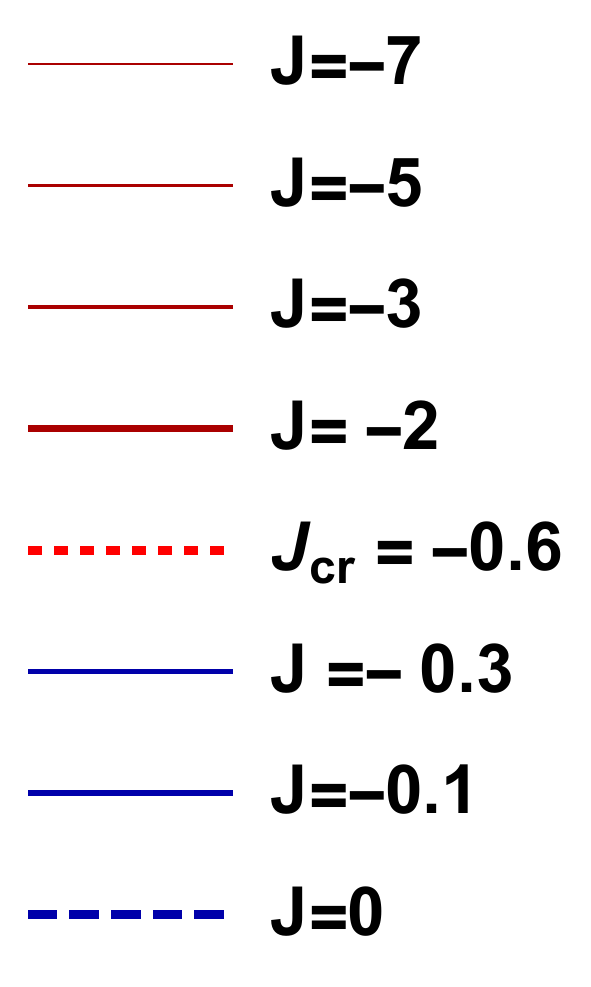}
\\
(e) $\qquad \qquad \qquad \qquad\qquad \qquad\qquad  \qquad$ (f)\\
\includegraphics[width=0.33\textwidth]{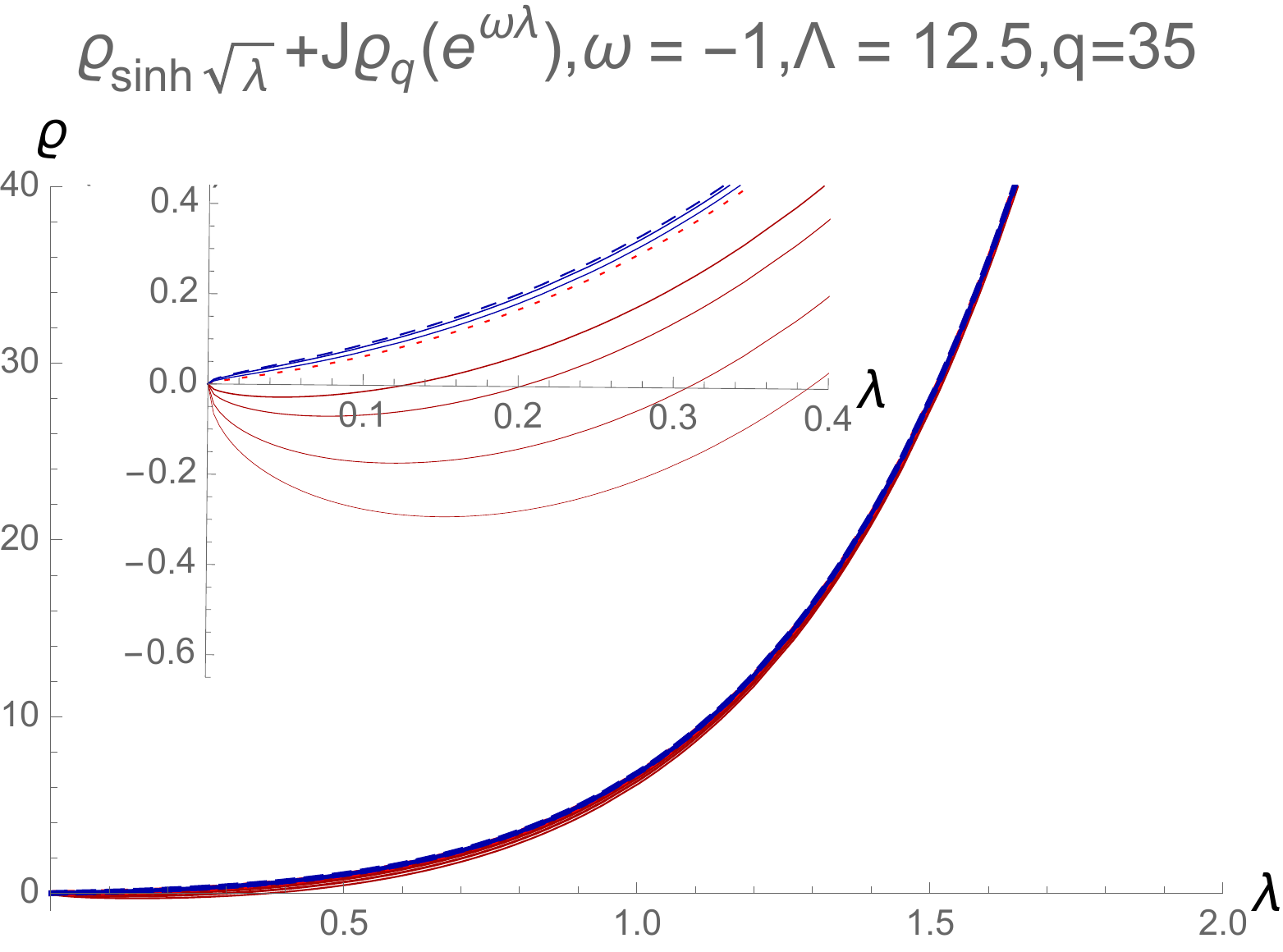}
\includegraphics[width=0.15\textwidth]{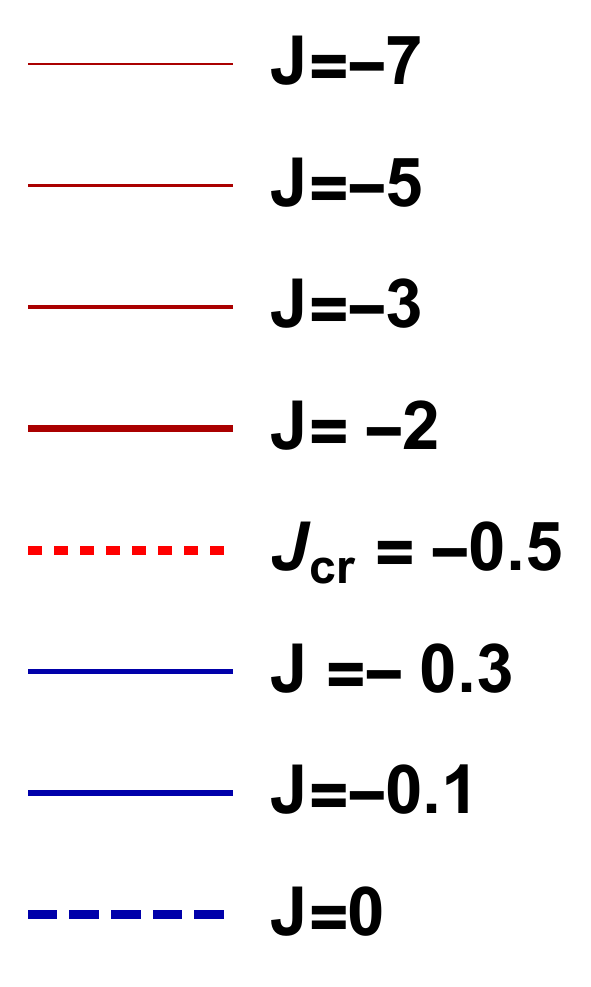}
\includegraphics[width=0.33\textwidth]{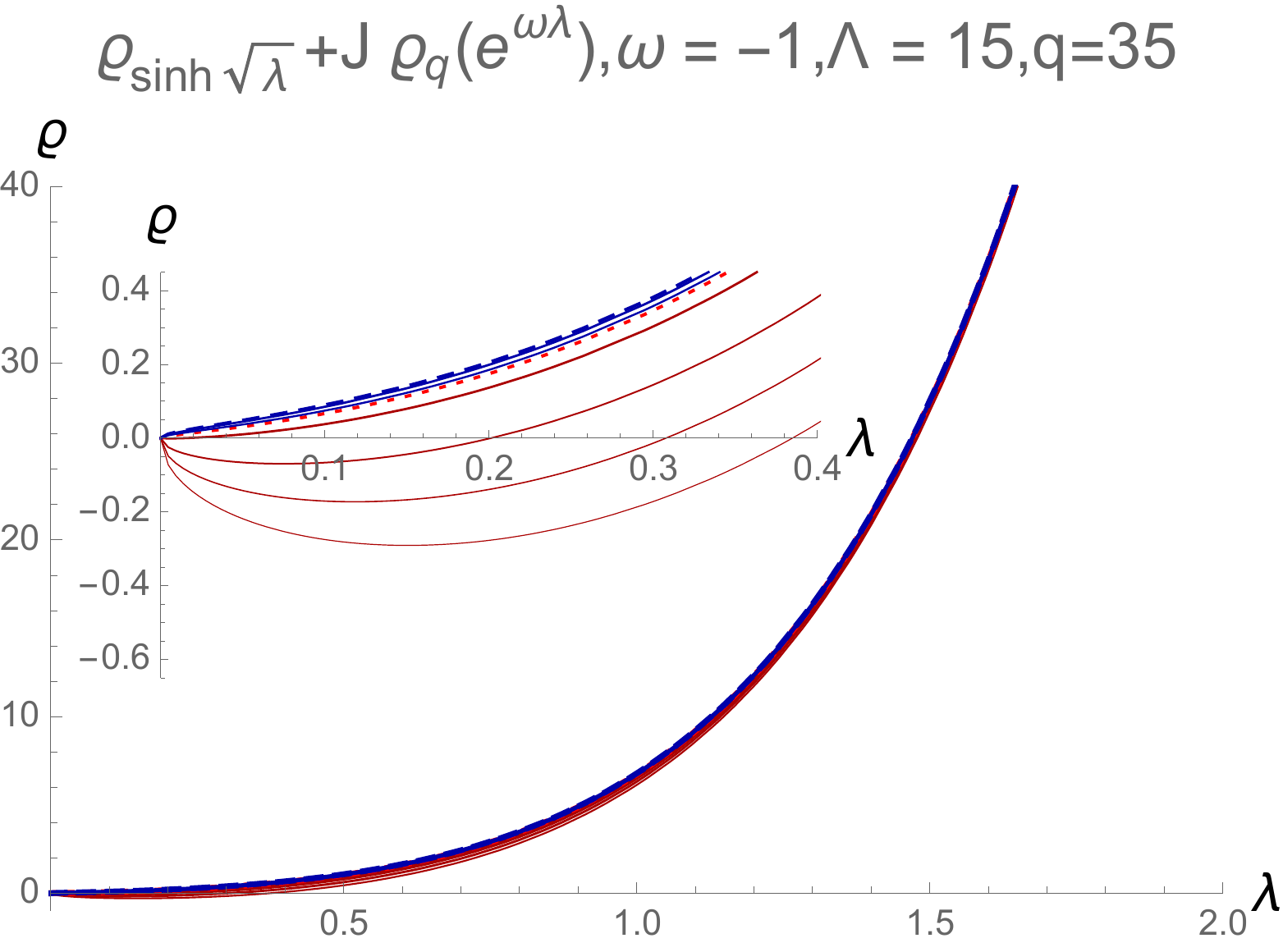}
\includegraphics[width=0.15\textwidth]{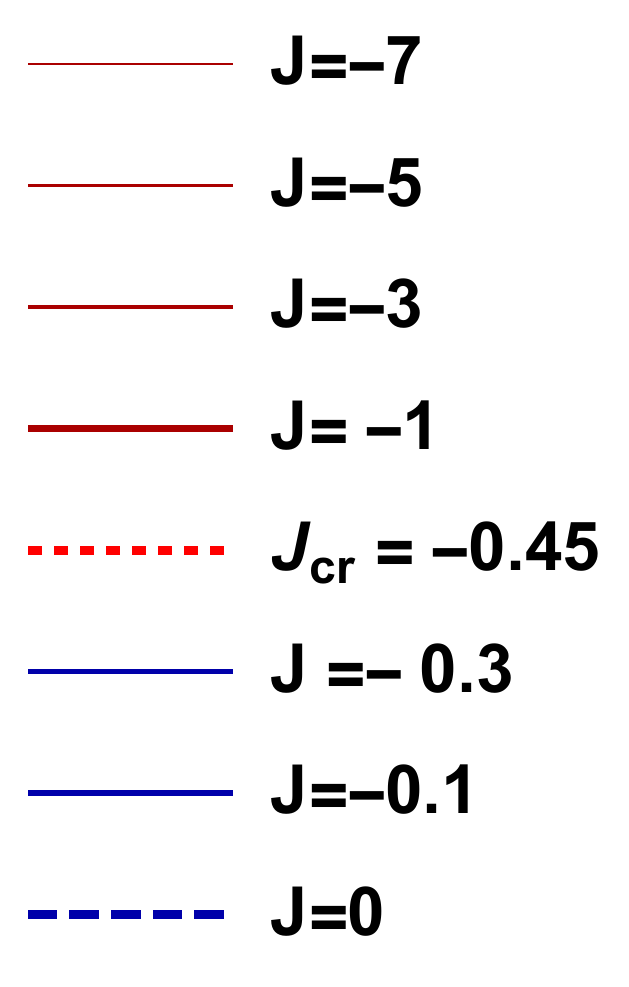}
\\
(i) $\qquad \qquad \qquad \qquad\qquad \qquad\qquad  \qquad$ (j)
\caption {The plots of  density \eqref{sinh-exp} for  $\Lambda=5,6,7.5,10,12.5,15$ (a,c,e,f,i,j)
and density  \eqref{sinh-exp-mod} for  $\Lambda=5,6$ (b,d). In all cases $\omega=-1$. Legends in (a,b), and (c,d) are the same.
}\label{Fig:PT-sinh-exp-more}
%{\bf File: density-for-exp-pot-test-IAA-one-again}
\end{center}
\end{figure}

\begin{figure}[h!]
\begin{center}
\includegraphics[width=0.45\textwidth]{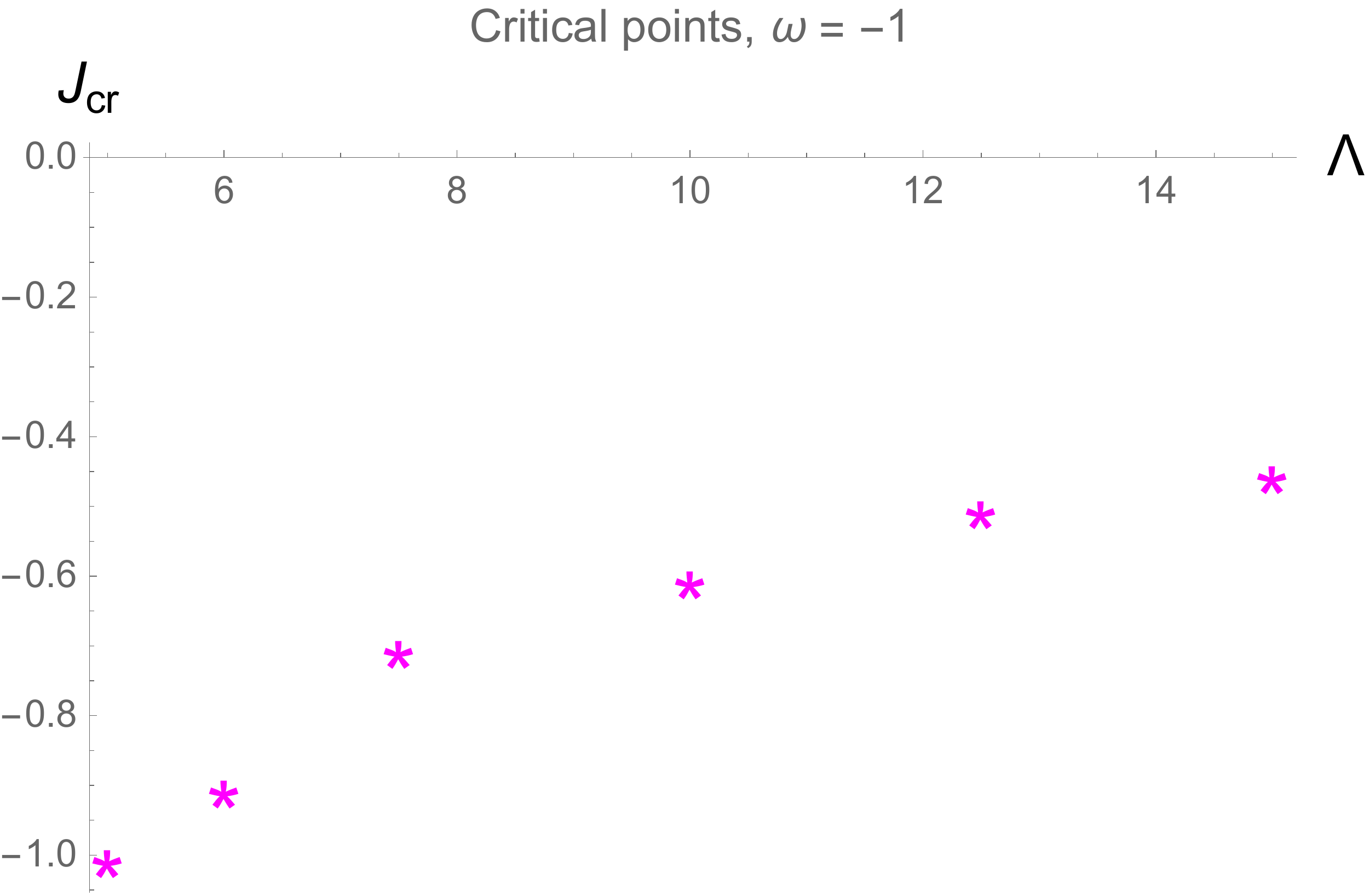}
\includegraphics[width=0.44\textwidth]{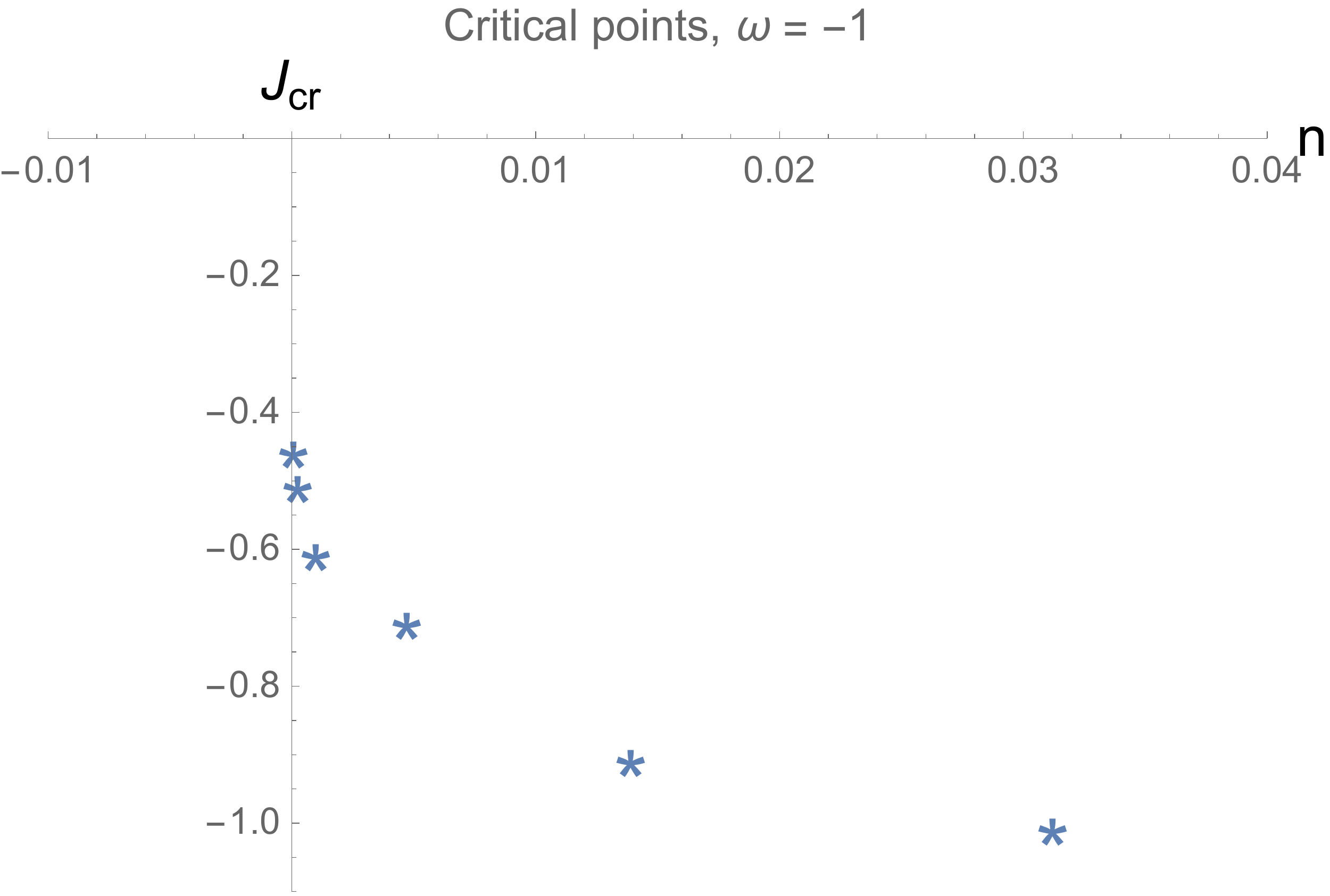}
\caption {(a) The points show the values of $J_{cr}$ for  $\rho(\lambda)$ given by  \eqref{sinh-exp} for different $\Lambda$. (b) $J_{cr}$ vs normalization $n(\Lambda)$ for the same $\Lambda$ as at (b).  $\omega=-1$.  
}\label{Fig:Critical-Lambda}
%{\bf File: density-for-exp-pot-test-IA-one-again}
\end{center}
\end{figure}

In  Fig.\ref{Fig:PT-sinh-exp-more} we plot  the density \eqref{sinh-exp} 
and \eqref{sinh-exp-mod} for  $\omega=-1$ and different $J$ and $\Lambda$.
We see that for negative $J>J_{cr}(\Lambda)$   domains on the segment $0,\Lambda$, where $\rho_{nn}$ and $\rho_{nn,s}$ become negative, appear. 
We interpret this as a destruction of the solution, or in other words, as an appearance of a  phase transition
at $J_{cr}$. It is interesting that the critical value  $J_{cr}$ are the same for both $\rho_{nn}$ and $\rho_{nn,s}$.
 In Fig.\ref{Fig:Critical-Lambda} the locations of critical points depending on $\Lambda$ are shown. 
 
\begin{figure}[h!]
\begin{center}
\includegraphics[width=0.3\textwidth]{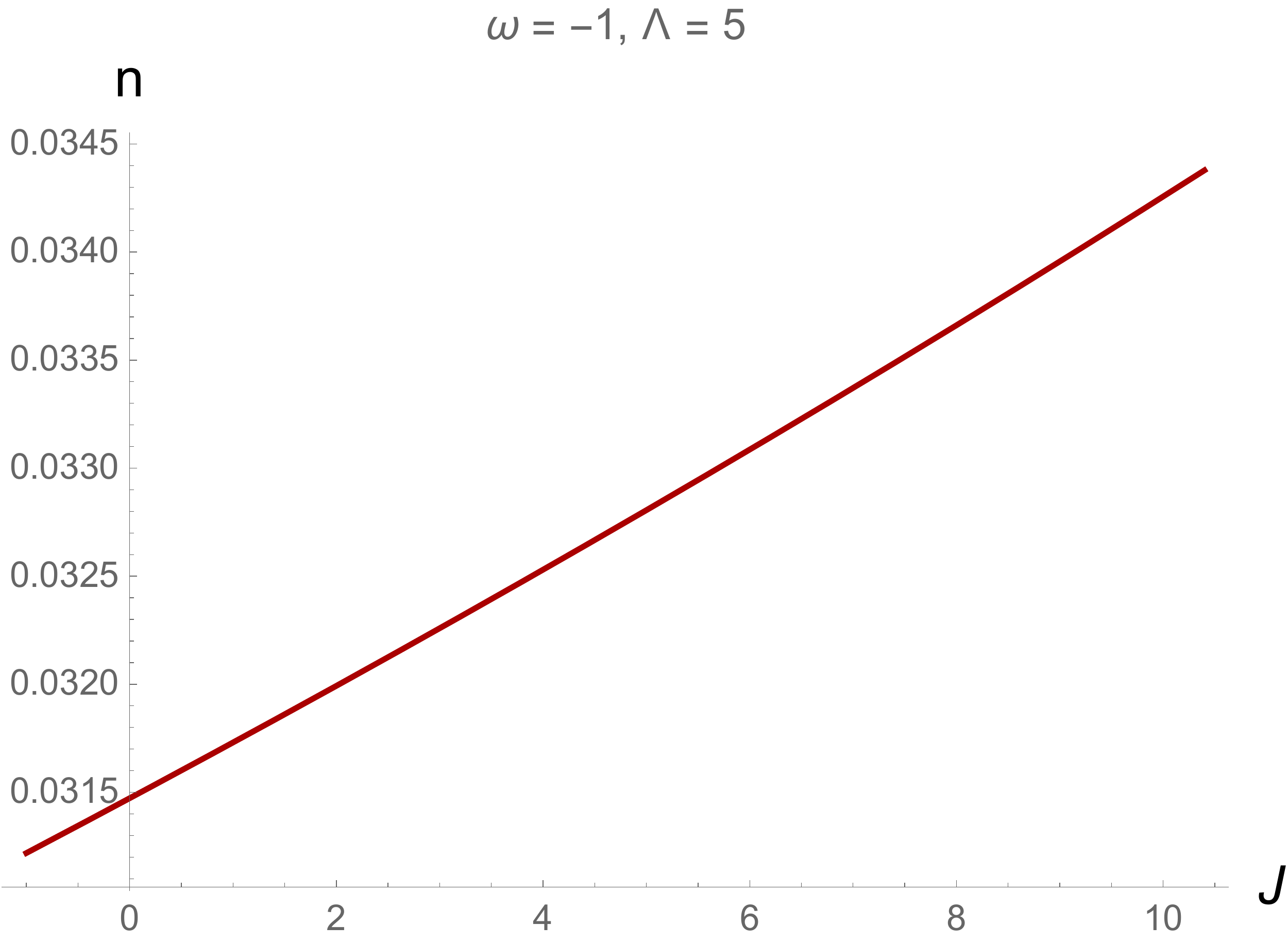}
\includegraphics[width=0.3\textwidth]{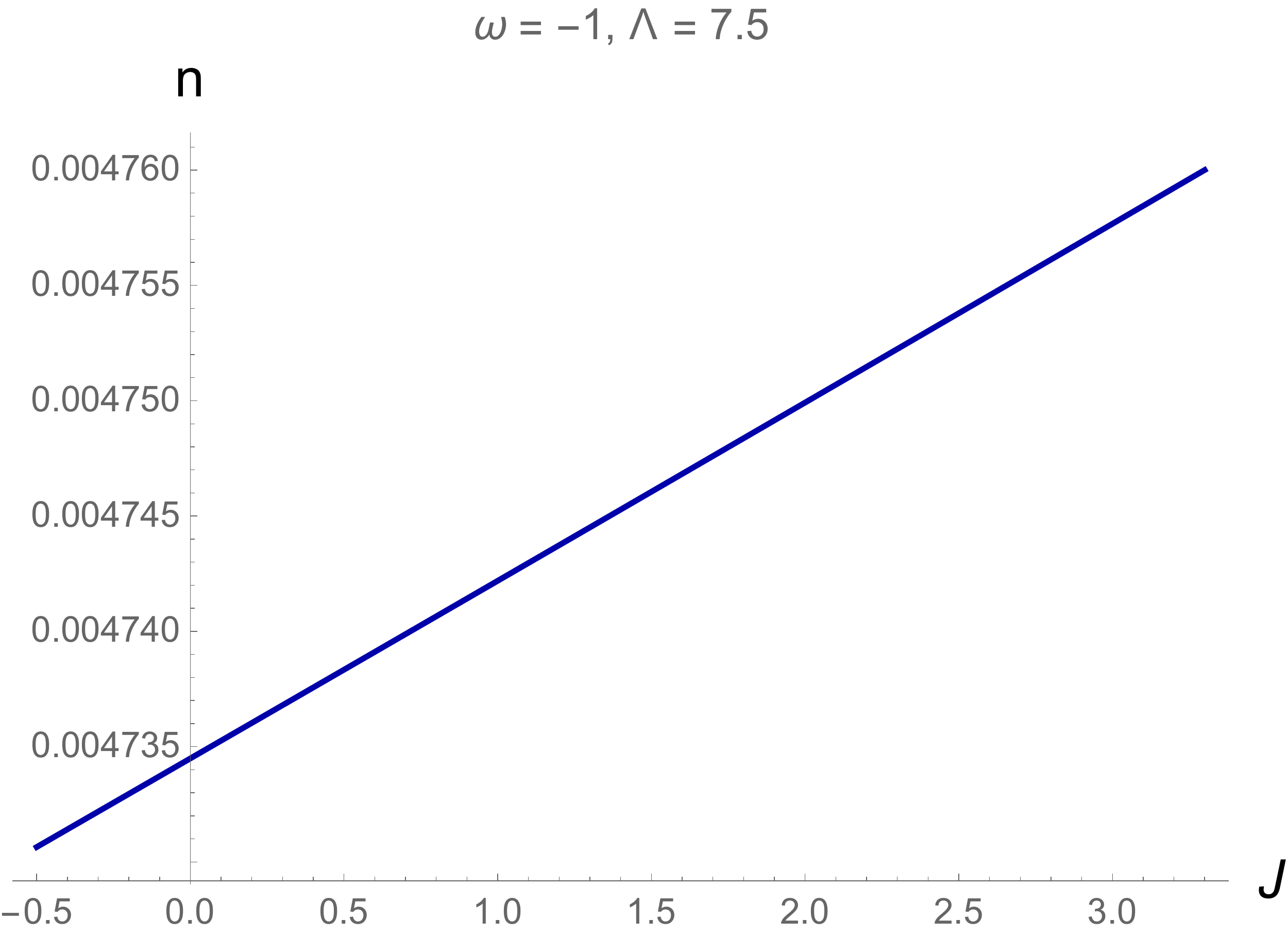}
\includegraphics[width=0.3\textwidth]{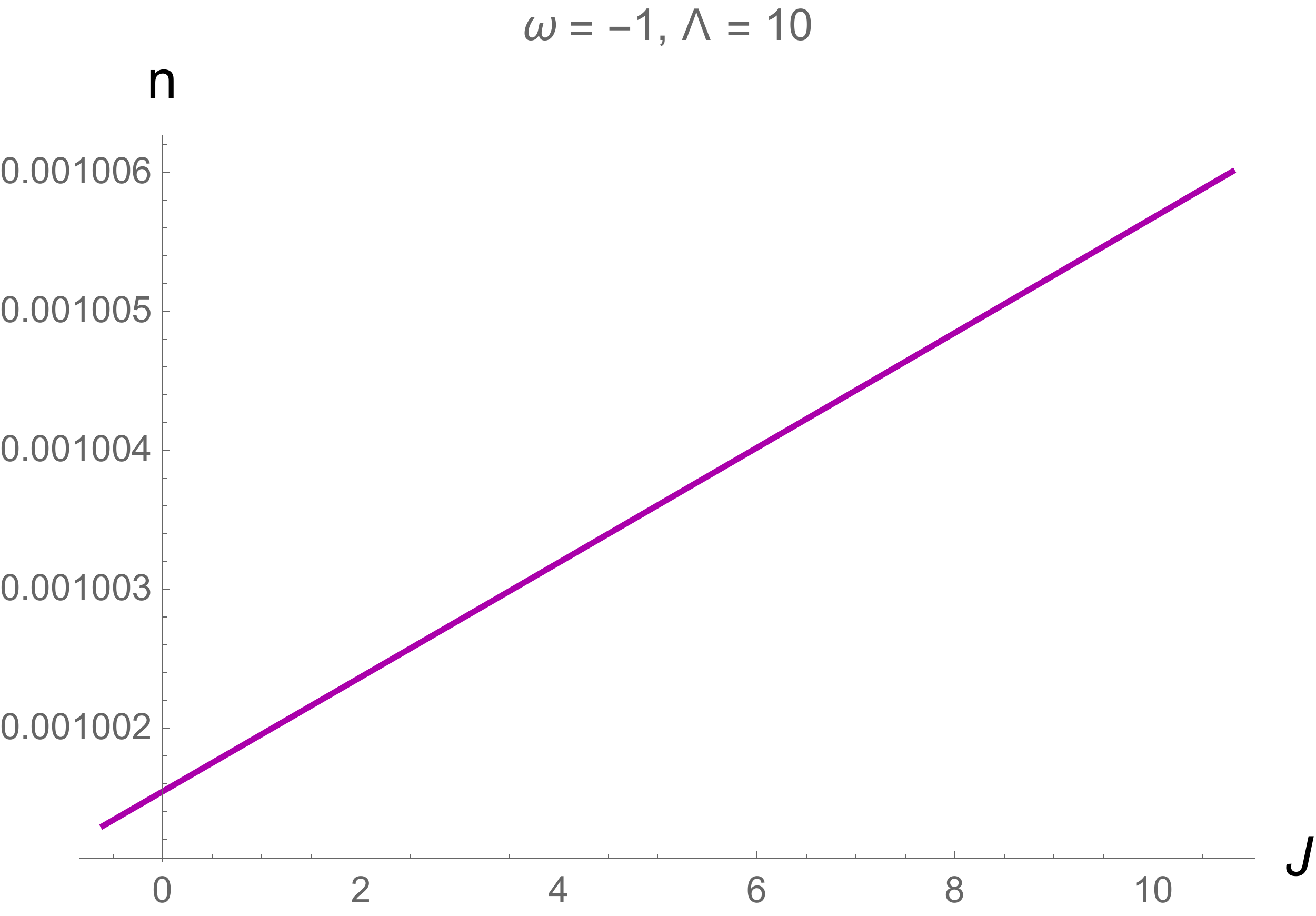}
\caption {Relations between $n(\Lambda)$, normalization factor, and $J$ for fixed $\Lambda$   for  $\rho(\lambda)$ given by  \eqref{sinh-exp}. The lines start at the  points presented at Fig.\ref{Fig:Critical-Lambda}.a.
}\label{Fig:m-J-SSS}
%{\bf File: density-for-exp-pot-test-IA-one-again}
\end{center}
\end{figure}

 In Fig.\ref{Fig:m-J-SSS} the dependence of  $n(\Lambda)$ on $J$ for fixed $\Lambda$ and $\omega$  for different values of $\Lambda$ is shown. 
For (a),(b) and (c) these plots are for $\Lambda =5,7.5,10$.
Here $J>J_{cr}(\Lambda)$. We see normalization factor  increases 
with increasing $J$ and this dependence is linear. This dependence differs from the dependence for the Gaussian perturbed
ensemble (see Fig.\ref{Fig:Critical-points-exp}.a), where it
is given by a decreasing nonlinear function.  The plot in  Fig.\ref{Fig:m-J-SSS}.b) shows the dependence of critical current $J_{cr}$ on $n(\Lambda)$.
$$\,$$

\newpage
\section{Discussion and conclusion}

The generating functional for the correlation functions  of the   boundaries of the Riemann serfaces is considered
in JT gravity and in matrix theory. The matrix integral \cite{SSS} provides a nonperturbative completion of  the genus expansion in  JT gravity on Riemann surfaces with  fixed  number of boundaries for all genus. The generating functional ${\fZ}(J)$ considered in this paper, 
gives  completion also for infinitely many  number of boundaries. \\

By using this formulation with  several matrix models,  including
 double scaling limit for the Gaussian model, relevant to topological gravity, cubic  model,
and JT gravity  have been  investigated.
In all these cases (here we have presented  only results for topological and JT gravities),  in  the corresponding  gas of baby universes  the phase transition is  observed. By analogy with states of matter one could expect that this phase change is condensation  from the gaseous state of JT
multiverse to liquid state.\\

To study this phase transition, we consider generating functional that we associate with baby universes.
The generating functional for a matrix model with potential $V(x)$  is obtained  from the partition function of the model just by
shift $V(x)\to V(x)-\tilde {J}(x)$ where $\tilde {J}(x)$ is the Laplace transform of $J(\beta)$.  Baby universes should look as point-like objects with very small length of the boundary. Therefore the distribution $J(\beta)$ should have a peak somewhere near zero.
In our model with $J(\beta)=J\delta (\beta -\omega)$ baby universes correspond to small  $\omega$. The shifted potential in this case is
$V(x)-Je^{-\omega x}$. Note that one could expect $<Z(\beta)^n>\approx <Z(\beta)>^n$ for large $n$ similarly
to the infinite replica number limit considered in \cite{AKTV}. \\

We have obtained that in all cases  there are critical negative $J_{cr}$, such that 
for $J<J_{cr}$ the corresponding solution is destroyed. The mass at the critical points decreases with increasing of $\Lambda$.  
In this paper we have considered the deformation of the density function by the exponential potential, but is also possible to study all momenta
deformations, as well as $\det(E-M)$, compare with \cite{García}.\\

The holographic boundary dual of the JT gravity in spacetime with $M$ boundaries is the $M$-replica SYK model in the low energy limit \cite{Jensen:2016pah,Maldacena:2016upp,Engelsoy:2016xyb,Harlow}. 
The studies of \cite{AV,AKTV,Kamenev,1902.09970,Okuyama:2019xvg, AKV} show that the nonperturbative completion of SYK involves nontrivial replica-nondiagonal saddle points. 
The replica-nondiagonal structures in SYK with replica interaction \cite{1804.00491,1901.06031,Klebanov,AKV}
 demonstrate nontrivial phase structures and symmetry breaking patterns. \\

Note that the deformation of potential by the linear term or nonlinear ones as a tool for investigation of phase transitions and spontaneous symmetry breaking in SYK-like models was used in \cite{AV,1902.09970, AKV}.  In particular, a nonlocal interacting of two-replica by a nonlocal term proportional to an external current permits to  
reveal  nonperturbative effects in the SYK model. This nonlocality in some sense is analogous to an external nonlocal source $J(x)$ (cosmological daemon) in cosmology \cite{1103.0273}, there it specified  boundary conditions. \\

There are numerous investigations of wormholes and baby universes in cosmology and particle physics, including the Giddings--Strominger wormhole solution \cite{GS}
and Coleman's approach to the cosmological constant problem \cite{Coleman}. There are many open questions in theory of wormholes and baby universes. In particular, it would be interesting to see whether  in JT gravity or its generalizations there is a mechanism 
of suppression \cite{CK} the probability of creation of giant wormholes and big baby universes.   
By using the wormhole/baby universe approach it was found that  the probability for the universe to undergo a spontaneous compactification down to a four-dimensional spacetime is greater than to remain in the original homogeneous multidimensional state \cite{IV}. It is interesting to find an analog of this  in the context of JT gravity.\\

In the study of SYK from gravity perspective, a crutial role plays consideration  of  wormholes in JT gravity 
\cite{1804.00491,1904.01911,1904.12820,1903.10532,1903.05732,1903.05658,1901.06031}.
A interplay between baby universes and wormholes could lead to nontrivial effects.\\

Using the recent result of  \cite{1905.03780} according which  JT gravity in $dS_2$ is an analytic continuation of JT gravity
in Euclidean $AdS_2$ it would be interesting to understand the meaning of the phase transition considered here in the dS case.\\

It would be also interesting to to compare the  Hartle-Hawking construction of string baby universes related with $AdS_2\times S^2$ geometry, using free fermionic formulation \cite{Dijkgraaf:2005bp}
with the  fermionic interpretation of determinant in matrix models, 
and also find a slot for baby universes in an quasi-classical \cite{1812.00918,1902.11194} or  exact quantization of JT proposed  in \cite{1905.02726}.

\section*{Acknowledgments}
The authors are grateful to M.~Khramtsov  for
useful discussions.
This work is supported by the Russian Science Foundation (project
19-11-00320, Steklov Mathematical Institute).
%\newpage
%%%%%%%%
%\input{appendix}
%\input{appendicBC}


\begin{thebibliography}{99}
\bibitem{SSS} 
 P.~Saad, S.~H.~Shenker and D.~Stanford,
  ``JT gravity as a matrix integral,''
  arXiv:1903.11115.
  %%CITATION = ARXIV:1903.11115;%%
  \bibitem{Jackiw:1984je} 
R.~Jackiw, ``Lower Dimensional Gravity,''
 Nucl. Phys.
 B252 (1985) 343--356.
%%CITATION = NUPHA,B252,343;%%.

\bibitem{Teitelboim:1983ux}
C.~Teitelboim, ``Gravitation and Hamiltonian Structure in Two Space-Time
  Dimensions,''
 Phys. Lett.
B126 (1983) 41--45.
%%CITATION = PHLTA,B126,41;%%.
 
\bibitem{Mir}
M.~Mirzakhani, ``Growth of Weil-Petersson volumes and random hyperbolic
  surface of large genus,'' Journal of Differential Geometry, 
  94 (2013) 267--300.
  \bibitem{eynard2007invariants}
B.~Eynard and N.~Orantin, ``Invariants of algebraic curves and topological
  expansion,'' 
  arXiv:math-ph/0702045.
  \bibitem{EO}
B.~Eynard and N.~Orantin, ``Weil-Petersson volume of moduli spaces,
  Mirzakhani's recursion and matrix models,''
 arXiv:0705.3600 [math-ph].
 
%%%%%%%Witten-2gr,MK,ManinZograf,DW

 \bibitem{Witten-2gr}
E.~Witten, ``Two-dimensional gravity and intersection theory on moduli
  space,''
Surveys Diff. Geom.
 1 (1991) 243--310.
%%CITATION = 00078,1,243;%%.

\bibitem{MK}
M.~Kontsevich, ``Intersection theory on the moduli space of curves and the
  matrix airy function,'' Comm. Math. Phys.147 (1992)1-23. 


\bibitem{ManinZograf}
Y.~I. Manin and P.~Zograf, ``Invertible Cohomological Field Theories and
  Weil-Petersson volumes", arXiv: math/9902051

  
  \bibitem{DW}
R.~Dijkgraaf and E.~Witten, ``Developments in Topological Gravity,''
arXiv:1804.03275
  
   \bibitem{Cotler:2016fpe}
J.~S. Cotler, G.~Gur-Ari, M.~Hanada, J.~Polchinski, P.~Saad, S.~H. Shenker,
  D.~Stanford, A.~Streicher, and M.~Tezuka, ``Black Holes and Random
  Matrices,'' JHEP 05 (2017)118, arXiv:1611.04650.
  %%CITATION = ARXIV:1611.04650;%%.
   \bibitem{Saad:2018bqo}
P.~Saad, S.~H. Shenker, and D.~Stanford, ``A semiclassical ramp in SYK and in
  gravity,''
 arXiv:1806.06840.
 
%%CITATION = ARXIV:1806.06840;%%.

  \bibitem{BookSFT} M.B. Green, J.H. Schwarz, E. Witten, {\it Superstring Theory,} Cambridge, UK: Univ. Press. ( 1987) 469 pp
  \bibitem{W-SFT}   E.~Witten,
  ``Noncommutative Geometry and String Field Theory,''
  Nucl.\ Phys.\ B 268  (1986) 253.
  %doi:10.1016/0550-3213(86)90155-0
  %%CITATION = doi:10.1016/0550-3213(86)90155-0;%%E. Witten, Nucl. Phys. B 268 (1986) 253
  \bibitem{AV-MatrixSFT} I.Ya. Aref'eva and I.V. Volovich, Two-dimensional gravity, string field theory and spin glasses, Phys.Lett. 255 (1991) 197-201
  
  %%%%%%%%%%%%%%%%Wigner,Dyson,BIPZ,mehta
     \bibitem{Wigner} 
     E.P. Wigner, Proc. Cambridge Philos. Soc. 47 (1951) 790, reprinted in C.E. Porter, Statistical theories of spectra: fluctuations (Academic Press, New York, 1965)

  \bibitem{Dyson}  F. J. Dyson,  A Class of Matrix Ensembles, J. Math. Phys. 13  (1972) 90.
%doi: 10.1063/1.1665857 
\bibitem{BIPZ} E. Brezin, C. Itzykson, G. Parisi and J.-B. Zuber, Commun. Math. Phys. 50 (1978) 35
\bibitem{mehta}
M.L. Mehta, {\it Random matrices}, 2nd edition, Academic Press, New York, 1991.
   \bibitem{georgia}%Gakhov
N.I. Muskhelishvili, {\it Singular integral equations}, Noordhoff, 1953.
 \bibitem{Gakhov} F.D. Gakhov, {\it Boundary problems}, Fizmatgiz, Moscow, 1977 [in Russian].    
  \bibitem{doublescaling}
E.~Br\'ezin and V.~A.~Kazakov,
``Exactly Solvable Field Theories Of Closed Strings,''
Phys.\ Lett.\ B 236  (1990) 144.
%%CITATION = PHLTA,B236,144;%%

\bibitem{DS} M.~R.~Douglas and S.~H.~Shenker, ``Strings In Less Than One-Dimension,''
Nucl.\ Phys.\ B  335  (1990) 635.
%%CITATION = NUPHA,B335,635;%%

\bibitem{GM} D.~J.~Gross and A.~A.~Migdal,
``Nonperturbative Two-Dimensional Quantum Gravity,''
Phys.\ Rev.\ Lett.\   64 (1990) 127 .
%%CITATION = PRLTA,64,127;%%

\bibitem{DFGZ}
P.~Di Francesco, P.~Ginsparg and J.~Zinn-Justin, ``2-D Gravity and random matrices,''
Phys.\ Rept.\   254 (1995) 1,
arXiv:hep-th/9306153.
%%CITATION = HEP-TH 9306153;%%
\bibitem{MM}
  M.~Marino,
``Les Houches lectures on matrix models and topological strings,''
  hep-th/0410165.
  %%CITATION = HEP-TH/0410165;%%


\bibitem{Ey5Lectures} 
  B.~Eynard, T.~Kimura and S.~Ribault,
  ``Random matrices,''
  arXiv:1510.04430 [math-ph].
  %%CITATION = ARXIV:1510.04430;%%
  
\bibitem{migdal}
A. Migdal, ``Loop equations and $1/N$ expansion,''
Phys.\ Rept.\   102  (1983) 199.
%%CITATION = PRPLC,102,199;%%
\bibitem{AJM} J.~ Ambjorn, J. ~Jurkiewicz and Yu. M.~ Makeenko, "Multiloop correlators for two-dimensional quantum gravity,"  Phys. Lett. B251 (1990) 517

\bibitem{eynard}
B.~Eynard,  ``Topological expansion for the 1-Hermitian matrix model correlation functions,''
  JHEP {\bf 0411}, 031 (2004),
 % doi:10.1088/1126-6708/2004/11/031;
  arXiv:hep-th/0407261.
%%CITATION = HEP-TH 0407261;%%
%v1 B.~Eynard, ``All genus correlation functions for the hermitian 1-matrix model,''   arXiv:hep-th/0407261.
\bibitem{Almheiri:2014cka}
A.~Almheiri and J.~Polchinski, ``Models of AdS$_{2}$ backreaction and
  holography,''  JHEP,
 11 (2015) 014,
 arXiv:1402.6334 [hep-th].
%%CITATION = ARXIV:1402.6334;%%} 

 \bibitem{Hawking} S W Hawking, Phys Lett B195 (1987) 337
  
  \bibitem{HL} S W Hawkang and R Laflamme, Baby universes and the non-renormalizability of gravity,
  Phys.Lett.209 (1988), 39-41

    \bibitem{LRT}  
  G.~V. Lavrelashvili, V.~A. Rubakov, and P.~G. Tinyakov, ``Disruption of
  Quantum Coherence upon a Change in Spatial Topology in Quantum Gravity,''
JETP Lett 46 (1987) 167 
%%CITATION = JTPLA,46,167;%%.
\\
G.~V. Lavrelashvili, V.~A. Rubakov, and P.~G. Tinyakov,  Nuel. Phys. B 290 (1988) 757 
  \bibitem{GS}
  S.~B. Giddings and A.~Strominger, ``Axion Induced Topology Change in Quantum
  Gravity and String Theory,''
Nucl. Phys. B306 (1988) 890--907.
 %http://dx.doi.org/10.1016/0550-3213(88)90446-4
%%CITATION = NUPHA,B306,890;%%.
  \bibitem{GS-BU}
  S.~B. Giddings and A.~Strominger, ``Baby Universes, Third Quantization and the
  Cosmological Constant,''
 B321 (1989) 481--508.
%%CITATION = NUPHA,B321,481;%%.
\bibitem{AS} A.~ Strominger, Baby Universes, In: "Quantum Cosmology and Baby Universes", (1991)  pp. 269-346.
%https: doi.org/10.1142/9789814503501$\_$0005

    \bibitem{Coleman} 
      S.~R.~Coleman,
  ``Why There Is Nothing Rather Than Something: A Theory of the Cosmological Constant,''
  Nucl.\ Phys.\ B 310 (1988) 643.
  %doi:10.1016/0550-3213(88)90097-1
  %%CITATION = doi:10.1016/0550-3213(88)90097-1;%%
       \bibitem{IV} I. V. Volovich, "Baby universes and the dimensionality of spacetime", Phys. Lett. B, 219 (1989), 66-70.  
   \bibitem{1807.00824} 
  A.~Hebecker, T.~Mikhail and P.~Soler,
  ``Euclidean wormholes, baby universes, and their impact on particle physics and cosmology,''
  Front.\ Astron.\ Space Sci.\   5 (2018) 35 ,
 % doi:10.3389/fspas.2018.00035
 arXiv:1807.00824.
  %%CITATION = doi:10.3389/fspas.2018.00035;%%Arthur Hebecker, Thomas Mikhail, Pablo Soler, arXiv:1807.00824. 

\bibitem{Mathur} 
  S.~Jain and S.~D.~Mathur,
  ``World sheet geometry and baby universes in 2-D quantum gravity,''
  Phys.\ Lett.\ B  286 (1992) 239,
  %doi:10.1016/0370-2693(92)91769-6
  hep-th/9204017.
  %%CITATION = doi:10.1016/0370-2693(92)91769-6;%%
\bibitem{Renata} 
  J.~Ambjorn, J.~Barkley, T.~Budd and R.~Loll,
  ``Baby Universes Revisited,''
  Phys.\ Lett.\ B 706 (2011) 86,
 % doi:10.1016/j.physletb.2011.10.062
  arXiv:1110.3998.
  %%CITATION = doi:10.1016/j.physletb.2011.10.062;%%

%%%%%bz1,BB,BI,Pastur,Widom,KM,DKMVZ
\bibitem{bz1}
  E.~Brezin and A.~Zee,
  ``Universality of the correlations between eigenvalues of large random matrices,''
  Nucl.\ Phys.\ B 402  (1993) 613.

\bibitem{BB}  M. Bowick and E. Brezin, "Universal scaling of the tail of the density of eigenvalues in random matrix models," Phys. Lett B268 (1991) 21
\bibitem{BI} Pavel Bleher, Alexander Its, "Double scaling limit in the random matrix model: the Riemann-Hilbert approach",
Comm. on Pure and Applied Math., Vol.56, Issue 4 (2003), doi.org/10.1002/cpa.10065;
arXiv:math-ph/0201003
\bibitem{Pastur} A. Boutet de Monvel, L. Pastur, and M. Shcherbina. "On the statistical mechanics approach in the random matrix theory: Integrated density of state", J. Stat. Phys. 79 (1995) 585.\\

L. Pastur, and M. Shcherbina, "Universality of the Local Eigenvalue Statistics for a Class of Unitary Invariant Random Matrix Ensembles", J. Stat. Phys.
 Vol. 86, (1997), 109


\bibitem{Widom}  C. Tracy and H. Widom, "Level-Spacing Distributions and the Airy Kernel", Commun.Math. Phys. 159  (1994) 151-174.\\
C. Tracy and H. Widom, "Fredholm Determinants, Differential Equations and Matrix Models", Commun. Math. Phys. 163:35 (1994) 33-72 




\bibitem {KM} A. B. J. Kuijlaars and K. T-R McLauglin, Generic behavior
of the density of states in random matrix theory and equilibrium
problems in the presence of real analytic
external field,  Commun. Pure Appl. Math.  53  (2000) 736-785.


\bibitem{DKMVZ} P. Deift, T. Kriecherbauer, K. T-R. McLaughlin,
S. Venakides, and X. Zhou,
Uniform asymptotics for polynomials orthogonal with respect
to varying exponential weights and applications to universality
questions in random matrix theory,
{\it Commun. Pure Appl. Math.},  52 (1999) 1335-1425 .


 \bibitem{BH} 
  E.~Brezin and S.~Hikami,
  {\it Random Matrix Theory with an External Source},
  doi:10.1007/978-981-10-3316-2
  %%CITATION = doi:10.1007/978-981-10-3316-2;%%






\bibitem{AIM} I.Ya.~Arefeva, A.C.~Ilchev and
B.K.~Mitruchkin, "Phase structure of matrix NxN
Goldstoune model in the large
N limit", In: Proceedings of "III
international Symposium on
selected topics in statistical
mechanics", Dubna 22-26 August,
1984; preprint JINR D17-84-850, p.20-26.

\bibitem {CCM} G. M. Cicuta, L. Molinari, and Montaldi,
Large $N$ phase transition in low dimensions,
Mod. Phys. Lett. A1 (1986) 125.

\bibitem {CM} C. Crnkovic and G. Moore,
Multicritical multi-cut matrix models, 
{\it Phys. Lett. B}  257 (1991).

\bibitem{VK} V.A. Kazakov, A simple solvable model of quantum field theory of open strings, Phys.Lett.B. 237 (1990), 212-216

\bibitem{Bagrets} D. Bagrets, A. Altland, A. Kamenev, "Sachdev-Ye-Kitaev Model as Liouville Quantum Mechanics", Nucl. Phys. B 911 (2016) 191-205, arXiv:1607.00694 
\bibitem{StanfordWitten}
D.~Stanford and E.~Witten, ``Fermionic Localization of the Schwarzian
  Theory,''  JHEP 10 (2017) 008,  arXiv:1703.04612.
%%CITATION = ARXIV:1703.04612;%%.



%\bibitem{DKM} {\bf CHECK} P. Deift, T. Kriecherbauer, K. T-R. McLaughlin,
%New results on the equilibrium measure for logarithmic
%potentials in the presence of an external field,
%{\it J. Appr. Theory} {\bf 95}, 399-475 (1998). 


\bibitem{Bethe} H. A. Bethe, "An Attempt to Calculate the Number Energy Levels of a Heavy Nucleus", Phys.Rev., 50 (1936) 332

 
 \bibitem{García} 
  A.~M.~Garcia-Garcia, Y.~Jia and J.~J.~M.~Verbaarschot,
  ``Exact moments of the Sachdev-Ye-Kitaev model up to order $1/N^2$,''
  JHEP 1804 (2018) 146,
  %doi:10.1007/JHEP04(2018)146
  arXiv:1801.02696.
  %%CITATION = doi:10.1007/JHEP04(2018)146;%%
  
  %%%%%%Jensen:2016pah,Maldacena:2016upp,Engelsoy:2016xyb,Harlow
  \bibitem{Jensen:2016pah} 
  K.~Jensen,
  ``Chaos in AdS$_2$ Holography,''
  Phys.\ Rev.\ Lett.\  {\bf 117}, no. 11, 111601 (2016)
 % doi:10.1103/PhysRevLett.117.111601
  arXiv:1605.06098 
  %%CITATION = doi:10.1103/PhysRevLett.117.111601;%%\
\bibitem{Engelsoy:2016xyb}
J.~Engelsöy, T.~G. Mertens, and H.~Verlinde, ``An investigation of AdS$_{2}$
  backreaction and holography,''
JHEP 07 (2016) 139,  arXiv:1606.03438 [hep-th].
%%CITATION = ARXIV:1606.03438;%%.


\bibitem{Maldacena:2016upp}
J.~Maldacena, D.~Stanford, and Z.~Yang, ``Conformal symmetry and its breaking
  in two dimensional Nearly Anti-de-Sitter space,''
PTEP 12, (2016) 12C10, arXiv:1606.01857.
%%CITATION = ARXIV:1606.01857;%%.

\bibitem{Harlow}
D.~Harlow and D.~Jafferis, ``{The Factorization Problem in Jackiw-Teitelboim
  Gravity},''
arXiv:1804.01081.
 \bibitem{1804.00491}  
J. Maldacena, X.-L. Qi, "Eternal traversable wormhole",  arXiv:1804.00491
 \bibitem{AV} 
  I.~Aref'eva and I.~Volovich,
  ``Notes on the SYK model in real time,''
  Theor.\ Math.\ Phys.\ 197   (2018) 1650
%doi:10.1134/S0040577918110090, 10.4213/tmf9533
  arXiv:1801.08118.


   \bibitem{1902.09970}  I.~Aref'eva and I.~Volovich,
  ``Spontaneous symmetry breaking in fermionic random matrix model,''
  arXiv:1902.09970.
  %%CITATION = ARXIV:1902.09970;%%
  \bibitem{Okuyama:2019xvg} 
  K.~Okuyama,
  ``Replica symmetry breaking in random matrix model: a toy model of wormhole networks,''
  arXiv:1903.11776 [hep-th].

    \bibitem{AKTV}  I.~Aref'eva, M.~Khramtsov, M.~Tikhanovskaya and I.~Volovich,
  ``Replica-nondiagonal solutions in the SYK model,''
  arXiv:1811.04831 [hep-th].
  %%CITATION = ARXIV:1811.04831;%%
  \bibitem{Kamenev} H. Wang, D. Bagrets, A. L. Chudnovskiy and A. Kamenev, "On the replica structure
of Sachdev-Ye-Kitaev model," arXiv:1812.02666 [hep-th].

  \bibitem{1103.0273} I.Ya. Aref'eva, I.V. Volovich, Cosmological daemon, JHEP, 8 (2011) 102, arXiv: 1103.0273  

\bibitem{CK}  S.R. Coleman, Ki-Myeong Lee,
  Escape From the Menace of the Giant Wormholes,
Phys.Lett. B221 (1989) 242-249.
 


\bibitem{Klebanov} J. Kim, I. R. Klebanov, G. Tarnopolsky and W. Zhao, "Symmetry Breaking in
Coupled SYK or Tensor Models," arXiv:1902.02287 .

   \bibitem{AKV}   I.~Aref'eva, M.~Khramtsov and I.~Volovich,
  ``Revealing nonperturbative effects in the SYK model,''
  arXiv:1905.04203 [hep-th].
   %%CITATION = ARXIV:1905.04203;%%

 \bibitem{1904.01911}
J. Maldacena, G. J. Turiaci, Zh. Yang, 
"Two dimensional Nearly de Sitter gravity", arXiv:1904.01911

\bibitem{1904.12820} H. W. Lin, J. Maldacena, Ying Zhao, "Symmetries Near the Horizon", arXiv:1904.12820

\bibitem{1903.10532} Y. Chen, P. Zhang, "Entanglement Entropy of Two Coupled SYK Models and Eternal Traversable Wormhole",
arXiv:1903.10532  

\bibitem{1903.05732}
 B. Freivogel, V. Godet, Ed. Morvan, J. F. Pedraza, A. Rotundo
"Lessons on Eternal Traversable Wormholes in AdS", arXiv:1903.05732 


\bibitem{1903.05658}
P. Betzios, E. Kiritsis, O. Papadoulaki, Euclidean Wormholes and Holography, arXiv:1903.05658 

\bibitem{1901.06031}
A. M. Garci­a-Garcia, T. Nosaka, D. Rosa, J. J. M. Verbaarschot,
Quantum chaos transition in a two-site SYK model dual to an eternal traversable wormhole,
arXiv:1901.06031

 \bibitem{1905.03780} 
J. Cotler, K. Jensen, A. Maloney, Low-dimensional de Sitter quantum gravity, arXiv:1905.03780

\bibitem{Dijkgraaf:2005bp} 
  R.~Dijkgraaf, R.~Gopakumar, H.~Ooguri and C.~Vafa,
  ``Baby universes in string theory,''
  Phys.\ Rev.\ D 73 (2006)  066002,
 % doi:10.1103/PhysRevD.73.066002
  hep-th/0504221.
  %%CITATION = doi:10.1103/PhysRevD.73.066002;%%
  %1812.00918,1902.11194
 \bibitem{1812.00918}  A. Blommaert, T. G. Mertens and H. Verschelde, "Fine Structure of Jackiw-Teitelboim
Quantum Gravity," arXiv:1812.00918 .
\bibitem{1902.11194} A. Blommaert, T. G. Mertens and H. Verschelde, "Clocks and Rods in
Jackiw-Teitelboim Quantum Gravity," arXiv:1902.11194.
 \bibitem{1905.02726} L. V. Iliesiu, S. S. Pufu, H. Verlinde, Y. Wang, 
An exact quantization of Jackiw-Teitelboim gravity,
arXiv:1905.02726


\end{thebibliography}
\end{document}